\documentclass[iop]{emulateapj}
\newcommand{\kms}{km\,s$^{-1}$}\newcommand{\vsini}{$v\sin i$}
\newcommand{\msun}{$M_{\sun}$}

\newcommand{\lli}{$A$(Li)}
\newcommand{\ac}{$A$(C)}
\newcommand{\cratio}{$^{12}$C/$^{13}$C}
\newcommand{\rsun}{$R_{\sun}$}
\newcommand{\teff}{$T_{\rm eff}$}
\newcommand{\logg}{$\log g$}
\newcommand{\eqref}[1]{(\ref{#1})}

\shorttitle{Rapid Rotation and Light Element Replenishment}
\shortauthors{Carlberg et al.}

\begin{document}

\title{	OBSERVABLE SIGNATURES OF PLANET ACCRETION IN RED GIANT STARS I: \\ RAPID ROTATION AND LIGHT ELEMENT REPLENISHMENT}

\author{Joleen K. Carlberg\altaffilmark{1,2}, Katia Cunha\altaffilmark{3,4}, Verne V. Smith\altaffilmark{3}, and Steven R. Majewski\altaffilmark{2}}
\altaffiltext{1}{Department  of Terrestrial Magnetism,
Carnegie Institution of Washington,
5241 Broad Branch Road NW,
Washington, DC 20015, USA
 jcarlberg@dtm.ciw.edu}
\altaffiltext{2}{Department of Astronomy,
University of Virginia,
Charlottesville, VA 22904, USA
srm4n@virginia.edu}
 \altaffiltext{3}{NOAO
 950 North Cherry Avenue,
Tucson, AZ 85719, USA
 vsmith@noao.edu, cunha@noao.edu}
  \altaffiltext{4}{ Observat\a'{o}rio Nacional, R. Gal. Jos\a'{e} Cristino, 77, 20921-400
S\~{a}o Crist\a'{o}v\~{a}o, Rio de Janeiro, RJ, Brazil}

\begin{abstract}
The orbital angular momentum of a close-orbiting giant planet can be sufficiently large that, if transferred to the envelope of the host star during the red giant branch (RGB) evolution, it can spin-up the star's rotation to unusually large speeds. This spin-up mechanism is one possible explanation for the rapid rotators detected among the population of generally slow-rotating red giant stars. These rapid rotators thus comprise a unique stellar sample suitable for searching for signatures of planet accretion in the form of unusual stellar abundances due to the dissemination of the accreted planet in the stellar envelope.  In this study, we look for signatures of replenishment in the Li abundances and (to a lesser extent) \cratio, which are both normally lowered  during
  RGB evolution.
Accurate abundances were measured from high signal-to-noise echelle spectra for samples of both slow and rapid rotator red giant stars. We  find that the rapid rotators are on average enriched in lithium compared to the slow rotators, but both groups of stars have identical  distributions of \cratio\ within our measurement precision.
Both of these abundance  results are consistent with the accretion of planets of only a few Jupiter masses.  
We also explore alternative scenarios for understanding the most Li-rich stars in our sample---particularly Li regeneration during various stages of stellar evolution. 
Finally, we  find that our stellar samples show non-standard abundances even at early RGB stages, suggesting that initial protostellar Li abundances and \cratio\ may be more variable than originally thought. 
\end{abstract}

\keywords{stars: abundances -- stars: atmospheres -- stars: chemically peculiar -- stars: late-type -- stars: rotation}

\section{Introduction}   
\label{sec:intro}
Composition studies of red giant stars have yielded insightful clues into the internal structure and history of low mass stars.  Some elemental abundances do not change much over the entire lifetime of the star, and these abundances can encode information about the star's birthplace and the chemical evolution history of the gas that became the star. 
Other abundances, particularly those of light elements, are altered by nuclear processes throughout the star's life, and measurements of these abundances and isotopic ratios have  provided observational constraints on stellar evolution models. Light abundance
alterations also provide a means of studying planet accretion in red giant stars. The abundances of some light element in red giants are significantly depleted from the stars' initial stellar abundances and, by extension, from the abundances of planets that formed from the same
protostellar material.  Therefore, accreted planets can replenish the stellar surface abundances of depleted elements. In this work, we focus on the 
 lithium abundance,  \lli \footnote{The abundance of Li is defined as $A({\rm Li})\equiv\log N_{\rm Li}-\log N_{\rm H} +12.00$}, and the ratio of $^{12}$C to $^{13}$C. The initial stellar and planetary abundances of these elements in solar metallicity stars, as measured in the solar system, 
are \lli$\sim3.3$\, dex \citep[see, e.g., ][]{bosegaard88,lodders98} and \cratio$\sim$90 \citep{anders89}. 

Lithium is  destroyed at relatively low temperatures  in a star by proton
capture reactions, and its abundance both in the stellar interior and at the stellar surface changes considerably throughout the lifetime of most stars.
During the main sequence (MS) phase, Li is preserved only in the coolest, outermost regions of the star that comprises only a few percent of the total stellar mass.  In the stellar interior,
the star is essentially devoid of Li.  
For stars more massive than $\sim1.4$\,\msun, the surface \lli\ stays constant throughout  the MS lifetime, while \lli\ slowly decreases in lower mass stars. 
Therefore, by  the end of the MS, the surface \lli\  in the outer convection zone can
range from near the original stellar abundance to depleted by factors of 100, depending on the star's mass.
 Near the beginning of the post-MS evolutionary phases, the star undergoes first dredge-up (FDU), during which the convection zone first deepens into the star and then recedes.
The Li that had been preserved  in the outer $\sim1$\% of the stellar mass on the MS gets mixed into the Li-poor interior, diluting the surface abundance by a factor of $\sim 60$
\citep{iben67}.
  Similar processes affect the \cratio\ ratio. The main chain of the CNO cycle reduces \cratio\ in the stellar interior towards the equilibrium value of 3.5, and mixing during FDU will lower the overall  \cratio\ of a red giant's envelope by a factor of $\sim 3$ \citep{sweigart89}. 
 These are the ``standard model'' abundance changes, which have been observationally verified by, e.g.,  \cite{lambert81} and \cite{brown89}. 
 
 Stars less massive than  $\sim$2.3\,\msun\ will experience additional surface abundance changes when the 
outward moving H-burning shell reaches the chemical discontinuity left behind by the convective envelope during FDU, and this evolutionary stage creates in the color-magnitude diagram what is known as the luminosity bump or RGB bump.  
In stars more massive than $\sim$2.3\,\msun, the H-burning shell does not reach the chemical discontinuity before the star evolves off of the red giant branch (RGB). The ``erasing'' of the chemical discontinuity in low mass stars allows for additional  non-convective mixing methods to further decrease the surface \lli\ and \cratio\ by varying amounts.
\cite{denissenkov03} coined the term ``canonical extra-mixing'' to refer to the prevalent processing
of light elements beyond standard dilution predictions for low-mass, post-RGB bump
giant stars that have been observed  in open clusters \citep{gilroy89,gilroy91,luck94,smiljanic09}, globular clusters \citep{shetrone03,pilachowski03,reicoblanco07}, and field red giants \citep{sneden86,charbonnel98,gratton00,kumar09}. The study by \cite{denissenkov03} described  a two-parameter turbulent diffusion model \citep[following the work by ][]{wasserburg95,boothroyd99} and explored rotation-driven mixing  \citep[as previously employed by ][]{sweigart79,charbonnel98} as a physical source of the turbulent diffusion.
They concluded that most of the non-standard abundances seen in red giants 
could be reproduced by a relatively simple two-parameter diffusion model.

However, within this framework of standard dilution and canonical extra-mixing models,
 inexplicable light-element  abundances can remain.
  Of particular interest to the present study are the  few percent of red giants that are ``Li-rich,'' with  \lli\,$>1.5$\,dex \citep{dasilva95,balach00,drake02,reddy05,kumar09,gonzalez09Li,carlberg10b,kumar11,monaco11,ruchti11,lebzelter12}.  The Li abundances of these stars are larger than that expected from stellar evolution and sometimes even exceed the assumed undiluted abundances of these stars \citep[e.g.,][]{balach00,kumar09,carlberg10b,monaco11,ruchti11}.
For massive, luminous red giant stars (i.e., $M_{\star}>1.5$\,\msun and $\log L/L_{\sun}>4$), temperatures at the base of the convection
envelopes are hot enough for the nucleosynthesis of $^7$Li through the Cameron--Fowler
chain \citep{cameron71}.   If this freshly synthesized Li (or the $^7$Be that decays to Li) is quickly transported to the cooler  regions of the convective envelope, then the surface \lli\ can 
substantially increase \citep{scalo75}. This mechanism is known as hot bottom burning (HBB). 
However, many of the Li-rich giants do not fit this picture because their convective envelopes are too cool, even at the hottest depths, for Li synthesis. Both \cite{charbonnel00} and
\cite{reddy05} found that many of these inexplicably Li-rich giants tend to cluster near the luminosity bump.   It is thought that the erasing of the chemical discontinuity at the bump may allow non-convective mixing processes to connect the cool convective envelope to  hotter depths, where Li can be regenerated. 
Like the HBB mechanism, the mixing must be rapid to replenish  the surface abundance. 
For example, \cite{sack99}
used a parameterized ``conveyor belt'' model and could reproduce the abundances of
Li-rich giants with certain mixing geometries; however, they did not provide a physical
mechanism for their successful models. Rotation was thought to be a likely mechanism
\citep{denissenkov04} until \cite{palacios06} found that a self-consistent
model of rotational mixing could not generate enough circulation to build up Li in
the stellar envelope. Recently, \cite{palmerini10} summarized
various mixing mechanisms capable of providing the needed conveyor belt 
and noted that the two current contenders are thermohaline mixing \citep{eggleton06,charbonnel07} and magnetic buoyancy \citep{busso07,nordhaus08,guandalini09,palmerini10}.
Because thermohaline mixing is a relatively slow process, it may only be an alternative
model for the canonical extra-mixing that destroys Li. The magnetic buoyancy models
circulate material fast enough to replenish Li in the stellar envelope; however, both the  \cite{guandalini09} and \cite{palmerini10} models predict maximum lithium 
enrichments of \lli$\sim$\,2.5 dex, which  are an order of magnitude smaller than the abundances observed in some of the most Li-rich
red giants.

Although there is no consensus on the physical mixing mechanisms capable of
transporting freshly regenerated Li to the stellar envelope, the intrinsically nuclear
origin of the Li can be observationally distinguished from Li supplied by an accreted planet.
First, any extra mixing should dredge-up  material with low \cratio, which will further decrease
the surface \cratio.  The ``enhanced extra-mixing" models of \cite{denissenkov04} yield \cratio\ between 8.2 and 17.4 compared to the 
expected values of 23 without the enhanced mixing.  Planet accretion, by contrast, should raise both \lli\ and \cratio.
Second, because only a small fraction of stars associated with the RGB bump
are Li-rich, then either only a small fraction of stars evolving through the bump regenerate
Li, which is then long-lived in the envelope, or all stars evolving through the bump
phase experience a short-lived phase of high Li abundances---a ``Li flash.'' \cite{charbonnel00} considered the latter scenario more likely because of the paucity
of Li-rich giant stars between the luminosity bump and the RGB tip.  
Therefore, Li-rich giants at other evolutionary stages (particularly pre-bump stages) require an alternative explanation,
such as planet accretion.
Finally, accreted planets should contribute angular momentum to the stellar envelope in addition to altering the stellar abundances of light elements. 
\cite{carlberg09} showed that some known exoplanets are expected to experience tidal orbital decay, and the orbital angular momentum that is transferred to the 
stellar envelope is sufficiently large to measurably increase the host stars'  rotational velocities.     Because
red giant stars are generally slow rotators \citep{deMed96b,gray81,gray82}, red giants with enhanced rotation and replenished light-element abundances
are candidates for stars that have accreted a planetary companion.

In our project, we select  samples of both rapid rotators and 
slow rotators (as a control sample) to look for multiple signatures of planet accretion simultaneously. In addition to a Li enhancement, we are  looking
for evidence of enhancements of \cratio\  and (in a future paper) refractory elements in the atmospheres of rapid rotators, which should also be indicative of planet 
accretion. Interpretation of these signatures will likely be open for debate  for any individual star; however, trends
that differ significantly between the two main samples (defined only by their relative \vsini) should make a
much stronger case to either defend or refute the proposition that some rapid rotators gain their angular momentum from a former 
planet. 
Ours is not the first attempt to correlate Li-rich giants and rapid rotation nor the first to invoke planet accretion as the explanation for the correlation. 
Planet accretion was first put forward by \cite{alexander67} to explain Li enrichment in giant stars, while \cite{pete83} were the first to consider planets as sources of 
angular momentum in evolved stars.  \cite{drake02} noted that while both slow and rapid rotators could be Li-rich, only 2\% of the slow rotators were Li-rich compared to nearly 
50\% of the rapid rotators.  \cite{siess99} modeled the  accretion of sub-stellar companions by red giant stars and calculated observational signatures of planet  accretion in 
those stars, such as rapid rotation and $^7$Li enrichment. From the actual occurrence of these predicted observational signatures in the red giant population, \cite{siess99}  
estimated that  4\%--8\% of Sun-like stars host giant planets at orbital radii small enough for 
   significant interactions to take place between the stars and  their planets while  the stars are on the RGB.   
Others who have considered  planet accretion as the solution to both Li enhancement and rapid rotation in giants include \cite{wallerstein82}, \cite{reddy02a}, \cite{carney03}, and \cite{denissenkov04}.

We describe the selection and observation of our red giant sample in Section \ref{sec:sample_obs}.
 In Section \ref{sec:stellparams}, we detail the derivation of the atmospheric properties, rotational velocity, and abundances.
We present our combined abundance and rotation
results in Section \ref{sec:placc}, where we  find  Li enhancement in the rapid rotators that is consistent with planet accretion. We find an unmeasurable difference in \cratio\ between the slow and rapid rotators; 
however, this \cratio\ result is still consistent with planet accretion given our measurement uncertainties and the modest \cratio\ enhancement expected to accompany the observed level of Li enrichment.  
We divide our sample into groups of similar evolutionary stages and find that the Li enrichment of the  rapid rotators persists within these subgroups. 
In Section \ref{sec:discuss}, we discuss alternative internal and external mechanisms that could explain enriched Li and how the enhanced angular momentum of the rapid rotators might affect these mechanisms. We select a small group of Li rich rapid rotators that are the best candidates for planet accretion to test whether the rotation and Li enrichment is consistent with planet engulfment. 
Our conclusions are presented in Section \ref{sec:summary}.

\section{Observations and Data Reduction}
\label{sec:sample_obs}
Our observational sample was built from three different primary sources.  Most of the stars, both rapid and slow rotators,  come from the sample of {\it Space Interferometry Mission} Astrometric Grid candidates described in \citet[][hereafter C11]{carlberg11}. Twenty-eight candidate rapid rotators were identified from the $\sim 1300$ stars studied in C11.  However, 
 because of the paucity of the total number of rapid rotators as well as the faintness of the Grid Giant Star Survey (GGSS) and {\it Tycho} stars that constitute the Astrometric Grid sample, we supplemented our observing list with brighter K giants from the catalog of bright F, G, and K  giant stars by \citet{demed99}. 
We included northern hemisphere G9-K9~III stars  from this catalog that have  \vsini\,$\ge$\,10\,\kms \ and that were not suspected to be spectroscopic 
binaries. (Only four stars met all of  these selection criteria and only two were  ultimately observed.)   Four more of our observed rapid rotators were selected from \cite{glebocki00}, which is a compilation
of rotational velocities for all types of stars. We selected  bright K~III stars, and  their primary  \vsini\  measurements come from  \cite{strassmeier93}, \cite{henry95}, \cite{fekel97}, and/or \cite{strassmeier00}.  A comparison sample of slow rotators was also selected from these literature sources. In total, we observed 15 rapid rotators and 58 slow rotators.

 In anticipation of measuring the lithium abundance  and \cratio, we added a sample of known Li-rich giants from \cite{drake02} and stars with previous determinations of \cratio\ from \cite{dearborn75}, \cite{smith85}, and \cite{fekel93} as a control sample to which we can compare our abundance determinations.
We also found published \lli\  for two rapid rotators (HD31993 and HD34198) and four slow rotators (Arcturus, HD108255, HD115478, and HD116010) after their inclusion in our main sample. 
Finally, we added 13 stars from the literature to act as  special control samples.  Ten of these stars are red giants with known planetary companions; they are listed in Table \ref{tab:swps} together with the  associated planetary orbital parameters.  
These stars with planets (SWPs) are all slow rotators and are generally expected to show abundances similar to those of the other slow rotators (since neither sample is expected to have accreted planets). 
The other three stars (HIP35253, HIP36896, and HIP81437) come from \citet[][hereafter M08]{massarotti08a},  who found that these stars' enhanced rotation is most plausibly explained by planet accretion.
These stars form an independent sample of stars  expected to show planet accretion  signatures. 
\tabletypesize{\scriptsize}
\begin{deluxetable}{lrrrrr}
\tablewidth{\columnwidth}
\tabletypesize{\scriptsize}
\tablecaption{Orbital Parameters for Companions of Stars with Known Planets\label{tab:swps}}
\tablehead{
  \colhead{Star Name} &
  \colhead{$M_{\rm p}\sin i$} &
  \colhead{$P$} &
  \colhead{$a_{\rm p}$} &
  \colhead{$e$}&
    \colhead{Ref.} \\
    \colhead{ } &
    \colhead{($M_{\rm Jup}$) } &
    \colhead{(days) } &
    \colhead{(AU) } &
    \colhead{ } &
    \colhead{ } }
\startdata 
HD 104985 & 6.3 & 198.2 & 0.78 & 0.03 & 1\\  
HD 122430 & 3.71 & 344.95 & 1.02 & 0.68 & 2 \\
HD 13189 & 14 &471.6 & 1.85 & 0.27 & 3\\ 
HD 137759 &  9 & 511.10 & 1.3 & 0.712& 4, 5 \\ 
HD 177830 & 1.5& 410 & 1.23 & 0.10 & 5, 6\\ 
HD 219449 A & 2.9 & 182 & 2.9 & \nodata & 7\\
HD 47536  & 5, 7 &43, 2500 &1.61, \nodata & 0.2, \nodata & 8 \\ 
HD 59686 & 5.25 & 202 & 0.911 & & 9 \\ 
HD 73108 & 7.1 & 269 & 0.88 & 0.43 & 10  \\ 
Pollux     & 2.7 & 589.6 & 1.73 & 0.02 &11, 12
\enddata
\tablerefs{(1) \citealt{sato03}; (2) \citealt{setiawan04}; (3)  \citealt{hatzes05}; (4)  \citealt[][discovery]{frink02}; (5)  \citealt[][orbit]{butler06}; (6)  \citealt[][discovery]{vogt00}; 
(7)  \citealt{raghavan06}; (8)  \citealt{setiawan03}; (9)  \citealt{mitchell03}; (10)  \citealt{dollinger07}; (11)  \citealt[][discovery]{hatzes93}; (12)  \citealt[][orbit]{hatzes06}}
\end{deluxetable}

To meet the high signal-to-noise ratio (S/N) and spectral resolution requirements needed for chemical abundance work, we obtained spectra with S/N$>100$ per pixel
for  our K giant sample using the echelle spectrographs on the Kitt Peak (KPNO) 4-m Mayall telescope and the Apache Point Observatory (APO) 3.5-m telescope.
At KPNO, we chose the 31.6 line~mm$^{-1}$ grating and used a $1.5''$ slit, yielding a spectral resolution  of $R=\lambda/\Delta\lambda\approx$~22,000 when using the long red camera. 
Observations at APO used the default $1.6''$ slit, yielding $R\approx$~31,500.
The observations were taken between 2007 March and 2010 March. Thorium--argon (Th--Ar) calibration lamp images, which are used for the wavelength calibration, were taken at each telescope position either immediately preceding or following each target star's observation. For stars requiring very long integration times, a Th--Ar comparison image was taken both
preceding and following the target star observation.  
Between 2010 March and 2011 August, we  obtained additional low S/N echelle spectra of the rapid rotators in an effort to identify radial velocity (RV) variations that could signify a stellar binary companion as the source of the
rapid rotation.
We detail our observations in Table \ref{tab:obs}, which lists each star along with its right ascension ($\alpha$), declination ($\delta$),  $V$ magnitude, total exposure time (given as the product of the number of exposures and individual exposure time), observatory where the data were obtained, and the date of observation.   
The table is organized in groups. The slow rotators  are  presented first followed by the rapid rotators, the M08 accretion candidates, SWPs, and control stars.
 Each of the control stars was used as a comparison for previous determinations of either \lli\ or \cratio.
 
The echelle spectra were reduced using standard IRAF procedures for overscan correction, bias removal, two-dimensional (2D) flat-fielding with quartz lamp images (KPNO data only;  see below for special flat-fielding procedures for the APO data), order extraction, scattered light removal, and wavelength calibration. The wavelength calibrations of the target star spectra come from the associated Th--Ar lamp image(s) taken at the same telescope position as the target spectrum.  
Although our targets were observed during bright time,  they  are bright enough that the expected sky flux at full moon is less than 1\% of the star flux even for the faintest stars ($V=12.6$).  Therefore, we did not perform sky subtraction. 
  The APO data required  non-standard flat-fielding procedures because of how closely spaced the echelle orders are on the detector. 
Following the recommendations of the  APO echelle reduction manual,\footnote{http://www.apo.nmsu.edu/arc35m/Instruments/ARCES/\\images/echelle\_data\_reduction\_guide.pdf} we  digitally magnify the 2D images by a factor of  four in the spatial direction before extracting the spectra.  We then  create an average flat-field image, magnify the average
flat-field in the same manner as the target star images, and extract a spectrum from the magnified image. The target stars are  flat-fielded by dividing the extracted target star spectra  by the normalized flat-field spectrum. 

The extracted, wavelength-calibrated echelle orders are then combined and continuum corrected to create normalized, continuous, one-dimensional (1D) spectra. These 1D spectra were cross-correlated with the 
\cite{hinkle00} atlas Arcturus spectrum to measure the observed stellar RVs, and these velocities were used to shift the wavelength solution to the stellar rest frame. 
The APO spectra span a large wavelength range: there is near-continuous coverage between 3200\,\AA\ and 1\,$\mu$m.  The KPNO data, on the other hand, span only 5210--8250\,\AA.

\begin{deluxetable*}{lrrrrrr}
\tablewidth{\textwidth}
\tabletypesize{\scriptsize}
\tablecaption{Observation Log \label{tab:obs}}
\tablecolumns{7}
\tablehead{
       \colhead{Star}&
       \colhead{$\alpha$}&
       \colhead{$\delta$}&
       \colhead{$V$}&
       \colhead{$n_{\rm ex}\times t_{\rm exp}$\tablenotemark{a}}&
       \colhead{Observatory}&
       \colhead{DATE-OBS} \\
       \colhead{ } &
       \colhead{(hh:mm:ss.ss)}&
       \colhead{($^\circ:':''$)}&
       \colhead{(mag)}&
       \colhead{(s)}&
       \colhead{ }&
       \colhead{(mm/dd/yyyy)}
}
\startdata
\multicolumn{7}{c}{{\bf Slow Rotators}} \\ \hline
Arcturus           &14:15:39.70  &$+$19:10:57 & 0.0    &$1\times0.5$  & KPNO      & 01/12/2008         \\     
Arcturus           &14:15:39.70  &$+$19:10:57 & 0.0    &$1\times(5+2)$  & KPNO      & 03/06/2007     \\     
G0300+00.29        &03:02:04.51 &$+$00:01:00 & 12.6    &$6\times1900$  & KPNO      & 01/13-14/2008           \\
G0319+56.5830      &03:23:22.10 &$+$56:29:04  & 10.7   &$2\times900$  & KPNO      & 01/10/2008             \\   
G0319+56.6888      &03:26:27.55 &$+$56:25:55  & 10.6   &$2\times840$  & KPNO      & 01/12/2008             \\   
G0453+00.90        &04:56:22.09 &$+$00:12:00  & 12.5   &$6\times1620$  & KPNO      & 01/11/2008             \\
G0639+56.6179      &06:43:52.65   &$+$56:09:49   & 10.4&$2\times900$  & APO       & 11/15/2007             \\   
G0653+16.552       &06:55:50.93 &$+$16:39:52 & 12.3    &$5\times1600$  & KPNO      & 01/14/2008                    \\
G0654+16.235       &06:56:51.41 &$+$16:49:15 & 12.4    &$5\times1750$  & KPNO      & 01/12/2008                    \\
G0840+56.5839      &08:46:08.70   &$+$56:14:38   & 10.6 &$2\times1200$  &APO        & 03/31/2007              \\  
G0840+56.9122      &08:45:38.71   &$+$56:07:36   & 10.5 &$2\times1200$  & APO       & 03/30/2007             \\
G0909-05.211       &09:12:46.87   &$-$05:57:47 & 11.9  &$4\times1450$  & KPNO      & 01/10/2008                    \\
G0912-05.11       &09:15:05.97   &$-$05:58:12 & 12.2  &$4\times1800$  & KPNO      & 01/12/2008                   \\
G0935-05.152       &09:38:06.22 &$-$05:49:18 & 12.5    &$6\times1650$  & KPNO      & 01/14/2008       \\ 
G1053+00.15        &10:55:47.15 &$-$00:07:02   & 10.9  &$2\times1200$  & APO       & 03/31/2007            \\   
G1124-05.61        &11:27:28.73 &$-$05:58:53   & 12.3  &$5\times1700$ & KPNO      & 01/11/2008             \\   
G1127-11.60        &11:29:21.64 &$-$11:24:13 & 11.0    &$2\times1200$  & KPNO      & 01/12/2008             \\   
G1130+39.9414      &11:32:14.20 &$+$39:05:46   & 9.7   &$1\times4000$  & KPNO      & 03/05/2007             \\   
G1200+67.3882      &11:59:20.10 &$+$67:26:24   & 10.7  &$2\times1500$  & KPNO      & 03/06/2007             \\   
G1240+56.8464      &12:45:02.80 &$+$55:52:25 & 10.3    &$1\times1500$  & KPNO      & 03/06/2007            \\    
G1331+00.13        &13:33:25.57 &$-$00:03:46 & 10.8    &$2\times1060$  & KPNO      & 01/13/2008            \\   
G1421+28.4625      &14:24:19.10 &$+$28:03:07  & 10.5    &$2\times1200$  & APO       & 03/30/2007            \\    
G1551+22.9456      &15:54:22.70 &$+$22:24:18  & 10.1    &$1\times1050$  & KPNO      & 03/08/2007            \\
G1640+56.6327      &16:42:43.13 &$+$56:09:57  & 10.3    &$1\times1200$  & KPNO      & 03/08/2007            \\    
G1800+61.12976     &18:01:15.05 &$+$61:43:52  & 10.5    &$2\times1200$  & APO       & 05/02/2007            \\    
G1800+61.12976     &18:01:15.05 &$+$61:43:52  & 10.5    &$2\times1200$  & APO       & 05/30/2007            \\    
G1936+61.14369     &19:34:43.25 &$+$61:49:48  & 10.3    &$2\times960$  & APO       & 05/04/2007            \\    
G2200+56.3466      &21:59:05.90 &$+$56:44:43  & 10.6    &$2\times1200$  & KPNO      & 01/10/2008           \\    
HD108225           &12:25:50.90 &$+$39:01:07 & 5.0     &$1\times300$  & KPNO      & 03/06/2007            \\    
HD109742           &12:36:58.30 &$+$17:05:22 & 5.7     &$1\times300$  & KPNO      & 03/08/2007            \\    
HD115478           &13:17:15.60 &$+$13:40:33 & 5.3     &$1\times200$  & KPNO      & 03/07/2007            \\    
HD116010           &13:20:19.00 &$+$40:09:02 & 5.6     &$1\times250$  & KPNO      & 03/08/2007            \\    
HD118839           &13:39:02.30 &$+$18:15:55 & 6.5     &$1\times(540+465)$  & KPNO      & 03/05/2007            \\    
HD191277           &20:05:32.88 &$+$61:59:43 & 5.4     &$1\times120$  & APO       & 05/04/2007            \\    
HD206445           &21:42:10.12 &$+$01:17:06 & 5.7     &$1\times30$  & APO       & 07/28/2007            \\    
HD221862           &23:35:09.46 &$+$67:29:31 & 7.2     &$2\times300$  & KPNO      & 01/12/2008                 \\
HD26162            &04:09:10.00 &$+$19:36:33 & 5.5     &$1\times300$  & KPNO      & 03/06/2007      \\       
Tyc0195-02087-1    &08:10:33.60 &$+$00:01:57 & 9.3     &$1\times900$  & KPNO      & 03/06/2007            \\    
Tyc0205-01287-1    &08:23:39.31 &$+$04:30:49 & 9.8     &$1\times1200$  & APO       & 03/30/2007            \\    
Tyc0276-00327-1    &11:58:03.00 &$+$04:21:12 & 9.4     &$2\times600$  & APO      & 03/30/2007            \\   
Tyc0319-00231-1    &14:04:25.70 &$+$04:03:28 & 9.7     &$1\times900$  & KPNO      & 03/06/2007            \\    
Tyc0913-01248-1    &14:36:30.44 &$+$11:22:11 & 10.1      &$1\times1600$  & APO       & 04/08/2007            \\     
Tyc0914-00571-1    &14:46:21.20 &$+$11:11:05 & 9.4     &$1\times1620$  & KPNO      & 03/06/2007          \\
Tyc1469-01108-1    &14:23:16.62 &$+$15:31:38 & 9.7      & $2\times660$  & APO      & 05/02/2007        \\
Tyc1780-00654-1    &02:43:50.95 &$+$28:52:58 & 10.1    &$2\times1800$  & APO       & 11/15/2007            \\    
Tyc1780-00654-1    &02:43:50.95 &$+$28:52:58 & 10.1    &$2\times500$  & KPNO      & 01/12/2008           \\    
Tyc1890-01314-1    &06:20:21.29 &$+$28:43:49 & 10.0    &$2\times900$  & APO       & 11/15/2007                   \\
Tyc1938-00311-1    &07:59:14.21 &$+$28:45:54 & 9.8     &$2\times660$  & APO       & 03/30/2007            \\    
Tyc2043-00747-1    &16:24:05.00 &$+$22:32:30 & 9.5     &$2\times900$  & KPNO      & 03/07/2007           \\    
Tyc2521-01716-1    &10:56:32.22 &$+$34:39:35 & 9.7     &$1\times1200$  & APO       & 05/30/2007            \\    
Tyc2527-01442-1    &12:10:22.71 &$+$33:27:00 & 9.8     &$2\times500$  & KPNO      & 01/10/2008            \\    
Tyc3005-00827-1    &10:27:53.44 &$+$40:16:47 & 9.6     &$1\times700$  & KPNO      & 01/12/2008            \\    
Tyc3013-01489-1    &11:28:35.52 &$+$39:29:34 & 9.9    &$2\times800$  & APO       & 05/02/2007            \\    
Tyc3027-01042-1    &13:59:16.70 &$+$37:52:36 & 9.9      &$1\times960$  & KPNO      & 03/08/2007            \\    
Tyc3402-00280-1    &06:53:25.20 &$+$51:17:55 & 9.6     &$1\times1800$  & KPNO      & 03/06/2007            \\    
Tyc3441-00140-1    &10:15:55.88 &$+$50:46:44 & 9.5   &$2\times480$  & APO	 & 11/15/2007          \\	
Tyc3809-01017-1	   &09:14:32.77	&$+$56:12:38 & 9.6     &$2\times600$  & APO	 & 11/15/2007             \\	
Tyc5523-00830-1	   &11:51:22.28	&$-$13:50:18 & 11.0    &$2\times1300$  & KPNO	 & 01/12/2008            \\	
Tyc5868-00337-1	   &03:05:53.00	&$-$17:48:48 & 9.8     &$2\times500$  & KPNO	 & 01/11/2008            \\	
Tyc5881-01156-1	   &03:45:17.54	&$-$16:29:20 & 10.0    &$2\times500$  & KPNO	 & 01/10/2008            \\	
Tyc5981-00414-1	   &07:47:19.25	&$-$16:19:50 & 9.6     &$1\times800$  & KPNO	 & 01/10/2008            	
 \enddata
\end{deluxetable*}
\begin{deluxetable*}{lrrrrrr}
\tablewidth{\textwidth}
\tabletypesize{\scriptsize}
\tablecaption{continued\dots\label{tab:obs}}
\tablecolumns{7}
\tablenum{2}
\tablehead{
       \colhead{Star}&
       \colhead{$\alpha$}&
       \colhead{$\delta$}&
       \colhead{$V$}&
       \colhead{$n_{\rm ex}\times t_{\rm exp}$\tablenotemark{a}}&
       \colhead{Observatory}&
       \colhead{DATE-OBS} \\
       \colhead{ } &
       \colhead{(hh:mm:ss.ss)}&
       \colhead{($^\circ:':"$)}&
       \colhead{(mag)}&
       \colhead{(s)}&
       \colhead{ }&
       \colhead{(mm/dd/yyyy)}
}
\startdata
\multicolumn{7}{c}{{\bf Rapid Rotators}} \\ \hline
G0804+39.4755      &08:07:33.00 &$+$39:17:47 & 11.9    &$2\times1800$  & KPNO      & 03/06/2007           \\
G0804+39.4755      &08:07:33.00 &$+$39:17:47 & 11.9    &$4\times1500$  & KPNO      & 01/11/2008            \\
G0804+39.4755      &08:07:33.00 &$+$39:17:47 & 11.9   &$1\times700$ & APO     &  03/29/2010  \\
G0804+39.4755       &08:07:33.00 &$+$39:17:47 & 11.9  &$1\times700$ & APO     &  12/23/2010  \\
G0827-16.3424      &08:29:55.02   &$-$16:48:13  & 9.9  &$2\times500$  & KPNO      & 01/10/2008            \\
G0827-16.3424      &08:29:55.02   &$-$16:48:13  & 9.9  &$1\times130$ & APO     &  03/29/2010  \\
G0827-16.3424       &08:29:55.02   &$-$16:48:13  & 9.9  & $1\times150$ & APO     &  10/22/2010  \\
G0928+73.2600      &09:28:22.03&$+$73:09:55   & 10.3  &$1\times1800$  & KPNO      & 03/06/2007            \\
G0928+73.2600      &09:28:22.03&$+$73:09:55   & 10.3  &$2\times700$  & KPNO      & 01/11/2008            \\
G0928+73.2600      &09:28:22.03&$+$73:09:55   & 10.3  &$1\times160$ & APO     &  03/29/2010  \\
G0928+73.2600      &09:28:22.03&$+$73:09:55   & 10.3  &$1\times200$ & APO     &  10/22/2010  \\
G0946+00.48        &09:48:41.48 &$-$00:09:09   & 12.4  &$5\times1680$  & KPNO      & 03/07/2007            \\
G0946+00.48        &09:48:41.48 &$-$00:09:09   & 12.4  &$1\times1100$ & APO    &  03/29/2010  \\
G0946+00.48        &09:48:41.48 &$-$00:09:09   & 12.4  &$1\times1100$ & APO    &  12/23/2010  \\
G1213+33.15558     &12:14:23.35 &$+$33:11:45 & 11.0    &$2\times1200$  & KPNO      & 01/12/2008           \\     
G1213+33.15558     &12:14:23.35 &$+$33:11:45 & 11.0    & $1\times300$ & APO     &  03/29/2010  \\
G1213+33.15558      &12:14:23.35 &$+$33:11:45 & 11.0    &$1\times154$  & APO   &  10/22/2010  \\
G1213+33.15558      &12:14:23.35 &$+$33:11:45 & 11.0    &$1\times350$ & APO     &  08/09/2011  \\
HD112859           &12:59:03.80 &$+$47:09:05 & 8.1     &$2\times450$  & KPNO      & 01/14/2008             \\
HD112859           &12:59:03.80 &$+$47:09:05 & 8.1     &$1\times25$  & APO     &  03/29/2010  \\ 
HD112859           &12:59:03.80 &$+$47:09:05 & 8.1     &$1\times25$  & APO     &  12/23/2010  \\
HD112859            &12:59:03.80 &$+$47:09:05 & 8.1     &$1\times360 $ & APO    &  08/09/2011  \\
HD31993            &05:00:08.20 &$+$03:17:12 & 7.5     &$1\times780$  & KPNO      & 03/06/2007           \\
HD31993             &05:00:08.20 &$+$03:17:12 & 7.5     & $1\times7$  & APO      &  10/22/2010  \\
HD31993            &05:00:08.20 &$+$03:17:12 & 7.5     &$1\times15$  & APO     &  12/23/2010  \\
HD33363            &05:18:31.10 &$+$75:56:49 & 7.6     &$2\times300$  & KPNO      & 01/14/2008            \\
HD33363             &05:18:31.10 &$+$75:56:49 & 7.6     & $1\times20$  & APO     &  10/22/2010  \\
HD33363             &05:18:31.10 &$+$75:56:49 & 7.6     &$1\times20$  & APO     &  12/23/2010  \\
HD34198            &05:14:30.70 &$-$26:12:31 & 7.0     &$1\times600$  & KPNO      & 03/07/2007           \\
HD34198             &05:14:30.70 &$-$26:12:31 & 7.0    & $1\times10$  & APO     &  10/22/2010  \\
HD34198            &05:14:30.70 &$-$26:12:31 & 7.0     &$1\times15$  & APO     &  12/23/2010  \\
Tyc0347-00762-1    &15:07:42.10 &$+$05:53:01 & 9.8   &$2\times1200$  & KPNO      & 03/06/2007           \\   
Tyc0347-00762-1    &15:07:42.10 &$+$05:53:01 & 9.8   &$1\times120$  & APO    &  03/29/2010  \\
Tyc0347-00762-1    &15:07:42.10 &$+$05:53:01 & 9.8   &$1\times105$ & APO     &  08/09/2011  \\
Tyc0647-00254-1    &02:59:26.71 &$+$12:45:20 & 10.0 &$2\times600$  & KPNO      & 01/10/2008           \\  
Tyc0647-00254-1    &02:59:26.71 &$+$12:45:20 & 10.0 &$1\times150$ & APO     &  10/22/2010  \\
Tyc0647-00254-1    &02:59:26.71 &$+$12:45:20 & 10.0 &$1\times160$  & APO    &  12/23/2010  \\
Tyc0647-00254-1    &02:59:26.71 &$+$12:45:20 & 10.0 &$1\times160$ & APO     &  08/25/2011  \\
Tyc2185-00133-1    &21:07:09.65 &$+$28:21:23 & 9.8   &$2\times750$  & APO       & 07/28/2007          \\
Tyc2185-00133-1    &21:07:09.65 &$+$28:21:23 & 9.8   &$1\times120$  & APO    &  08/09/2011  \\
Tyc2185-00133-1    &21:07:09.65 &$+$28:21:23 & 9.8   &$1\times125$ & APO     &  08/25/2011  \\
Tyc3340-01195-1    &04:15:10.15 &$+$51:11:40 & 10.0 &$2\times600$  & APO       & 11/15/2007          \\      
Tyc3340-01195-1    &04:15:10.15 &$+$51:11:40 & 10.0 &$1\times120$  & APO    &  03/29/2010  \\
Tyc3340-01195-1    &04:15:10.15 &$+$51:11:40 & 10.0 &$1\times120 $ & APO    &  10/22/2010  \\
Tyc3340-01195-1    &04:15:10.15 &$+$51:11:40 & 10.0 &$1\times130$ & APO     &  08/25/2011  \\
Tyc5904-00513-1	   &04:56:02.45&$-$17:08:23 & 9.5    &$2\times1200$  & KPNO	 & 03/07/2007         	 \\
Tyc5904-00513-1	   &04:56:02.45&$-$17:08:23 & 9.5    &$2\times400$  & KPNO	 & 01/10/2008         	 \\
Tyc5904-00513-1	   &04:56:02.45&$-$17:08:23 & 9.5    &$1\times90$  & APO     &  10/22/2010  \\
Tyc5904-00513-1	   &04:56:02.45&$-$17:08:23 & 9.5    & $1\times95$  & APO     &  12/23/2010  \\
Tyc6094-01204-1	   &12:05:02.66&$-$16:23:42 & 9.5    &$1\times720$  & KPNO	 & 01/10/2008         	 \\
Tyc6094-01204-1	   &12:05:02.66&$-$16:23:42 & 9.5    &$1\times4000$  & KPNO	 & 03/05/2007         	 \\
Tyc6094-01204-1	   &12:05:02.66&$-$16:23:42 & 9.5    &$1\times90$  & APO     &  03/29/2010  \\
Tyc6094-01204-1	   &12:05:02.66&$-$16:23:42 & 9.5    &$1\times90$  & APO     &  12/23/2010  
 \enddata
\end{deluxetable*}
\begin{deluxetable*}{lrrrrrr}
\tablewidth{\textwidth}
\tabletypesize{\scriptsize}
\tablecaption{continued\dots \label{tab:obs}}
\tablecolumns{7}
\tablenum{2}
\tablehead{
       \colhead{Star}&
       \colhead{$\alpha$}&
       \colhead{$\delta$}&
       \colhead{$V$}&
       \colhead{$n_{\rm ex}\times t_{\rm exp}$\tablenotemark{a}}&
       \colhead{Observatory}&
       \colhead{DATE-OBS} \\
       \colhead{ } &
       \colhead{(hh:mm:ss.ss)}&
       \colhead{($^\circ:':"$)}&
       \colhead{(mag)}&
       \colhead{(s)}&
       \colhead{ }&
       \colhead{(mm/dd/yyyy)}
}
\startdata
\multicolumn{7}{c}{{\bf M08 Accretor Candidates}} \\ \hline
HIP35253           &07:17:03.41 &$+$26:41:22  & 6.4     &$2\times200$  & APO    &  12/02/2008   \\ 
HIP36896           &07:35:08.80 &$+$30:57:39  & 5.3     &$2\times10$  & APO      & 12/02/2008   \\
HIP81437           &16:38:00.47 &$+$56:00:56  & 5.3     &$1\times30$  & APO      & 03/29/2010   \\
\hline
\multicolumn{7}{c}{{\bf Stars with Planets}} \\ \hline
HD104985           &12:05:15.12 &$+$76:54:20 & 5.9     &$1\times480$  & KPNO      & 03/06/2007         \\     
HD122430           &14:02:22.78 &$-$27:25:47& 5.5      &$2\times300$  & KPNO      & 03/06/2007         \\     
HD13189            &02:09:40.17 &$+$32:18:59 & 7.6     &$1\times800$  & APO       & 07/28/2007        \\     
HD137759           &15:24:55.77 &$+$58:57:57 & 3.3     &$1\times(240+700)$  & KPNO      & 03/05/2007         \\     
HD177830           &19:05:20.77 &$+$25:55:16 & 7.2     &$1\times800$  & APO       & 05/02/2007         \\     
HD219449           &23:15:53.49 &$-$09:05:15 & 4.2     &$1\times35$  & APO       & 07/28/2007         \\     
HD47536            &06:37:47.62 &$-$32:20:23 & 5.3     &$2\times60$  & KPNO      & 01/11/2008         \\     
HD59686            &07:31:48.40 &$+$17:05:09  & 5.5     &$1\times600$  & KPNO      & 03/05/2007         \\     
HD73108            &08:40:12.82 &$+$64:19:40  & 4.6     &$1\times40$  & APO       & 03/30/2007                \\      
Pollux             &07:45:18.95 &$+$28:01:34 & 1.2     &$1\times(2+5)$  & APO       & 03/30/2007         \\     
\hline
\multicolumn{7}{c}{{\bf Control Stars}} \\
\hline
Aldebaran          &04:35:55.24  &$+$16:30:33  & 1.0   &$1\times(5+20)$  & KPNO      & 03/05/2007  \\
HD112127           &12:53:55.70 &$+$26:46:48 & 6.9     &$1\times700$  & KPNO      & 03/07/2007     \\
HD127665           &14:31:49.80 &$+$30:22:17 & 3.6     &$2\times60$  & APO       & 05/30/2007      \\
HD127665           &14:31:49.80 &$+$30:22:17 & 3.6     &$1\times120$  & KPNO      & 03/06/2007     \\
HD163588           &17:53:31.70 &$+$56:52:22 & 3.7     &$1\times30$  & APO       & 03/30/2007     \\
HD216228           &22:49:40.82 &$+$66:12:01 & 3.5     &$1\times20$  & APO       & 05/04/2007      \\
HD233517           &08:22:46.70 &$+$53:04:49 & 9.7     &$1\times3800$  & KPNO      & 03/05/2007      \\
HD33798            &05:15:15.50 &$+$47:10:15  & 6.9    &$1\times2400$  & KPNO      & 03/05/2007     \\
HD39853            &05:54:43.60 &$-$11:46:27 & 5.6     &$1\times300$  & KPNO      & 03/07/2007     
 \enddata
 \tablenotetext{a}{Stars with multiple exposures of unequal time are listed  as $1\times(t_1 + t_2)$. These generally occurred for the bright stars to ensure reaching very high S/N without saturating the detector. }
\end{deluxetable*}

\section{Spectral Analysis}
\label{sec:stellparams}
\subsection{Stellar Parameters}
\label{sec:teffetc}
A collection of neutral and singly ionized iron lines can be used to constrain the basic stellar parameters of effective temperature (\teff), surface gravity ($\log g$), iron abundance compared to solar ([Fe/H]), and microturbulence ($\xi$) following the prescription described, e.g.,  
in \cite{smith01}. By combining equivalent widths of these lines with stellar atmosphere models,  one can find the  model for which all the lines give the 
same iron abundance.  Generally speaking,  \ion{Fe}{1} lines spanning a wide range of excitation potentials  ($\chi$ or EP) will constrain \teff, while a large span of 
\ion{Fe}{1} line strengths will constrain $\xi$.  Transitions of \ion{Fe}{2} are much more sensitive to $\log g$ than \ion{Fe}{1}, and all of the iron lines will  constrain 
[Fe/H] in the form of $A$(Fe). 

The iron list used to derive the stellar parameters was compiled with great care to minimize  systematic errors.   We began by collecting the iron lines from \cite{smith00}, \cite{fulbright06}, and \cite{dmbiz06}. All available $\log gf$ values of these sources were tabulated together with  $\log gf$ values obtained by querying  the Vienna Atomic Line Database 
\citep[VALD;][]{kupka99}.\footnote{http://vald.astro.univie.ac.at/$\sim$vald/php/vald.php} VALD usually returns between one and three $\log gf$ values per iron line.   This initial compilation included 111 \ion{Fe}{1} and 17 \ion{Fe}{2} lines spanning 
the  wavelength range of our spectra. Equivalent widths ($W_\lambda$) of  these lines were  measured using the {\it splot} task in IRAF for both the solar and Arcturus spectra provided in the \cite{hinkle00} spectral atlas CD-ROM.   These measurements were compared to the $W_\lambda$ reported in the Smith, Fulbright, and Bizyaev papers. Where our measurements were discrepant,  the lines were remeasured to ensure accuracy of our measuring procedure (e.g., verifying the line identification, checking for overlooked line blending). Our line list was large enough that we could be  discriminating and remove lines that were severely blended with neighboring lines or that had a more uncertain continuum level.
 
\begin{figure}[tb]
\centering
\includegraphics[width=1.03\columnwidth]{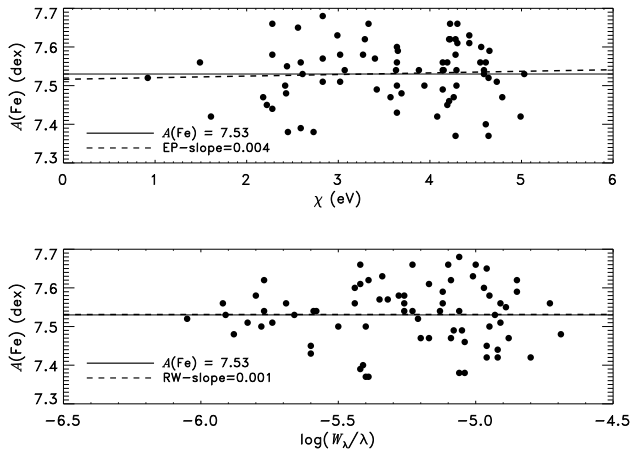} 
\caption{\label{slopes} Plot of the output $A$(Fe) (ordinate) as a function of the iron lines' excitation potential (top) and reduced equivalent width (bottom).  In each panel, the mean $A$(Fe) for all the lines is plotted with a solid line, while the trends of $A$(Fe) with either $\chi$ or $W_\lambda/\lambda$ are shown with the dotted lines. }
\end{figure}

We tested our line list by computing solar iron abundances  using the {\it abfind} driver of the MOOG stellar line analysis program\footnote{http://verdi.as.utexas.edu/moog.html} \citep{sneden73} together with the MARCS \citep{marcs08}  solar atmosphere model ($T_{\rm eff}=5777$\,K, $\log g=4.40$\,dex, and $A$(Fe)$=7.45$).  
Individual lines that suggested $A$(Fe) more than 0.2 dex from the average iron abundance of the entire list were removed.   When more than one $\log gf$ was available for a given line, we chose the $\log gf$ value that brought the iron abundance closest to the mean. 
Finally, we required that there be no trend of $A$(Fe) with either the lines' excitation potentials or reduced equivalent widths ($\log W_\lambda/\lambda$, or RW).  In other words, the ``EP-slope'' and ``RW-slope'' are both  near zero (specifically, within 0.005 dex~eV$^{-1}$, standard MOOG parlance).  Figure \ref{slopes} illustrates this zero-slope criterion;  output iron abundances are plotted on the ordinate against both the excitation potential and reduced equivalent width.  Both the mean $A$(Fe) and the trend of $A$(Fe) with the line properties are shown.  
We iteratively removed the lines that gave the most deviant $A$(Fe) until the EP-slope and RW-slope were flat and the output $A$(\ion{Fe}{ 2}) matched the output $A$(\ion{Fe}{1}), and this is the solution plotted in Figure \ref{slopes}.  The mean solar abundances derived from our list is $A$(\ion{Fe}{1})\,=\,7.53\,dex 
with a standard deviation of $\sigma_{\rm Fe\,I}=0.08$ and  $A$(\ion{Fe}{2})\,=\,7.54\,dex with $\sigma_{\rm Fe\,II}=0.06$. We derive a microturbulence of  $\xi=1.0$~\kms. 
 Ten  \ion{Fe}{1} lines were removed in this process.  The remaining lines chosen for the final list---74~\ion{Fe}{1} and 13~\ion{Fe}{2}---are given  in Table \ref{felinelist} with the lines' wavelengths,  species, excitation potential, $\log gf$, and references for the $\log gf$-values.

\begin{deluxetable*}{lrrrc|lrrrc}
\tablecaption{Iron line list \label{felinelist}}
\tablewidth{0pc}
\tabletypesize{\scriptsize}
\tablehead{
   \colhead{$\lambda$} &
   \colhead{Species} &
   \colhead{$\chi$}&
   \colhead{$\log gf$} &
   \colhead{Reference} &
   \colhead{$\lambda$} &
   \colhead{Species} &
   \colhead{$\chi$}&
   \colhead{$\log gf$} &
   \colhead{Reference} \\
   \colhead{(${\rm \AA}$)} &
   \colhead{  } &
   \colhead{ (eV)} &
   \colhead{(dex)} &
   \colhead{ } &
   \colhead{(${\rm \AA}$)} &
   \colhead{  } &
   \colhead{ (eV)} &
   \colhead{(dex)} &
   \colhead{ } }
\startdata
  5307.361 &    \ion{Fe}{1} &    1.608 &   $-$2.987  & 5, 7    &       6127.904 &    \ion{Fe}{1} &    4.143 &   $-$1.399  & 9       \\               
  5322.041 &    \ion{Fe}{1} &    2.279 &   $-$2.803  & 5       &       6151.617 &    \ion{Fe}{1} &    2.176 &   $-$3.299  & 5, 7    \\              
  5412.784 &    \ion{Fe}{1} &    4.435 &   $-$1.890  & 7       &       6165.360 &    \ion{Fe}{1} &    4.142 &   $-$1.550  & 5, 7    \\             
  5466.988 &    \ion{Fe}{1} &    3.573 &   $-$2.233  & 3, 5    &       6173.334 &    \ion{Fe}{1} &    2.223 &   $-$2.858  & 6       \\              
  5491.829 &    \ion{Fe}{1} &    4.186 &   $-$2.188  & 1, 5    &       6180.203 &    \ion{Fe}{1} &    2.728 &   $-$2.591  & 1       \\              
  5522.447 &    \ion{Fe}{1} &    4.209 &   $-$1.557  & 5, 7    &       6200.312 &    \ion{Fe}{1} &    2.609 &   $-$2.437  & 5       \\              
  5536.583 &    \ion{Fe}{1} &    2.832 &   $-$3.810  & 5       &       6232.640 &    \ion{Fe}{1} &    3.654 &   $-$1.223  & 1, 6    \\              
  5539.284 &    \ion{Fe}{1} &    3.642 &   $-$2.660  & 5, 7, 9 &       6271.282 &    \ion{Fe}{1} &    3.332 &   $-$2.950  & 7       \\              
  5560.207 &    \ion{Fe}{1} &    4.435 &   $-$1.190  & 7       &       6311.500 &    \ion{Fe}{1} &    2.832 &   $-$3.230  & 7       \\              
  5577.031 &    \ion{Fe}{1} &    5.033 &   $-$1.550  & 5, 7    &       6322.685 &    \ion{Fe}{1} &    2.588 &   $-$2.426  & 2, 5, 7 \\              
  5607.664 &    \ion{Fe}{1} &    4.154 &   $-$2.270  & 5       &       6393.601 &    \ion{Fe}{1} &    2.433 &   $-$1.620  & 7       \\              
  5611.361 &    \ion{Fe}{1} &    3.635 &   $-$2.990  & 5, 7    &       6481.870 &    \ion{Fe}{1} &    2.279 &   $-$3.113  & 8       \\              
  5618.633 &    \ion{Fe}{1} &    4.209 &   $-$1.276  & 6       &       6518.366 &    \ion{Fe}{1} &    2.832 &   $-$2.750  & 7       \\              
  5633.947 &    \ion{Fe}{1} &    4.991 &   $-$0.270  & 5, 7    &       6593.871 &    \ion{Fe}{1} &    2.437 &   $-$2.422  & 5, 7    \\              
  5635.823 &    \ion{Fe}{1} &    4.256 &   $-$1.890  & 5,7     &       6597.561 &    \ion{Fe}{1} &    4.795 &   $-$0.920  & 10      \\               
  5636.696 &    \ion{Fe}{1} &    3.640 &   $-$2.610  & 5, 7    &       6609.110 &    \ion{Fe}{1} &    2.559 &   $-$2.692  & 5, 7    \\              
  5638.262 &    \ion{Fe}{1} &    4.220 &   $-$0.870  & 5, 7    &       6699.153 &    \ion{Fe}{1} &    4.593 &   $-$2.190  & 7       \\              
  5661.363 &    \ion{Fe}{1} &    4.284 &   $-$1.736  & 5       &       6733.151 &    \ion{Fe}{1} &    4.637 &   $-$1.349  & 6       \\              
  5691.497 &    \ion{Fe}{1} &    4.301 &   $-$1.520  & 5, 7    &       6750.152 &    \ion{Fe}{1} &    2.424 &   $-$2.621  & 2, 5, 7 \\              
  5698.023 &    \ion{Fe}{1} &    3.640 &   $-$2.680  & 5, 7    &       6820.372 &    \ion{Fe}{1} &    4.638 &   $-$1.170  & 10      \\              
  5705.476 &    \ion{Fe}{1} &    4.301 &   $-$1.600  & 7       &       6837.020 &    \ion{Fe}{1} &    4.593 &   $-$1.810  & 7      \\              
  5712.134 &    \ion{Fe}{1} &    3.417 &   $-$2.060  & 7       &       6855.161 &    \ion{Fe}{1} &    4.559 &   $-$0.741  & 3      \\              
  5753.122 &    \ion{Fe}{1} &    4.260 &   $-$0.760  & 7       &       6858.150 &    \ion{Fe}{1} &    4.607 &   $-$1.046  & 8      \\              
  5760.345 &    \ion{Fe}{1} &    3.642 &   $-$2.490  & 5, 7, 9 &       6971.937 &    \ion{Fe}{1} &    3.018 &   $-$3.490  & 7      \\              
  5778.453 &    \ion{Fe}{1} &    2.588 &   $-$3.430  & 1, 6    &       7112.172 &    \ion{Fe}{1} &    2.990 &   $-$3.090  & 7       \\              
  5784.660 &    \ion{Fe}{1} &    3.397 &   $-$2.670  & 7       &       7189.155 &    \ion{Fe}{1} &    3.071 &   $-$2.771  & 1, 5    \\              
  5793.913 &    \ion{Fe}{1} &    4.220 &   $-$1.829  & 8       &       7401.658 &    \ion{Fe}{1} &    4.186 &   $-$1.690  & 7       \\              
  5807.782 &    \ion{Fe}{1} &    3.292 &   $-$3.410  & 5, 7    &       7583.790 &    \ion{Fe}{1} &    3.018 &   $-$1.990  & 7       \\              
  5809.217 &    \ion{Fe}{1} &    3.884 &   $-$1.690  & 9       &       7723.208 &    \ion{Fe}{1} &    2.279 &   $-$3.617  & 2, 5, 7 \\              
  5811.917 &    \ion{Fe}{1} &    4.143 &   $-$2.430  & 5       &       7941.085 &    \ion{Fe}{1} &    3.274 &   $-$2.580  & 7       \\              
  5814.805 &    \ion{Fe}{1} &    4.283 &   $-$1.970  & 5, 7    &       5018.440 &    \ion{Fe}{2} &   2.891 &   $-$1.345  & 4    \\                  
  5837.700 &    \ion{Fe}{1} &    4.294 &   $-$2.340  & 5, 7    &       5234.625 &    \ion{Fe}{2} &   3.221 &   $-$2.230  & 5    \\                  
  5838.370 &    \ion{Fe}{1} &    3.943 &   $-$2.337  & 9       &       5284.098 &    \ion{Fe}{2} &   2.891 &   $-$3.195  & 4    \\                  
  5849.682 &    \ion{Fe}{1} &    3.695 &   $-$2.990  & 5, 7    &       5325.559 &    \ion{Fe}{2} &   3.221 &   $-$3.324  & 4    \\                  
  5853.149 &    \ion{Fe}{1} &    1.485 &   $-$5.280  & 5       &       5414.046 &    \ion{Fe}{2} &   3.221 &   $-$3.645  & 4    \\                  
  5855.076 &    \ion{Fe}{1} &    4.608 &   $-$1.478  & 1, 5    &       5425.247 &    \ion{Fe}{2} &   3.199 &   $-$3.390  & 4    \\                  
  5856.084 &    \ion{Fe}{1} &    4.294 &   $-$1.640  & 7, 9    &       5991.368 &    \ion{Fe}{2} &   3.153 &   $-$3.560  & 10    \\                 
  5861.107 &    \ion{Fe}{1} &    4.283 &   $-$2.450  & 5, 7    &       6084.099 &    \ion{Fe}{2} &   3.199 &   $-$3.881  & 4    \\                  
  5916.246 &    \ion{Fe}{1} &    2.453 &   $-$2.832  & 6       &       6149.246 &    \ion{Fe}{2} &   3.889 &   $-$2.841  & 4    \\                  
  6024.058 &    \ion{Fe}{1} &    4.548 &   $-$0.120  & 5, 7    &       6247.577 &    \ion{Fe}{2} &   3.892 &   $-$2.310  & 5    \\                  
  6027.051 &    \ion{Fe}{1} &    4.076 &   $-$1.089  & 5       &       6416.921 &    \ion{Fe}{2} &   3.891 &   $-$2.680  & 10    \\                 
  6056.005 &    \ion{Fe}{1} &    4.733 &   $-$0.460  & 5       &       6432.682 &    \ion{Fe}{2} &   2.891 &   $-$3.687  & 4    \\                  
  6079.009 &    \ion{Fe}{1} &    4.652 &   $-$1.120  & 5, 7    &       6456.383 &    \ion{Fe}{2} &   3.903 &   $-$2.185  & 4    \\                  
  6120.250 &    \ion{Fe}{1} &    0.915 &   $-$5.950  & 5, 7    &       &&&&
\enddata
\tablerefs{(1) \citealt{bard91,bard94}; (2) \citealt{obrian91}; (3) \citealt{blackwell82a,blackwell82b,blackwell84,blackwell86,blackwell95}; (4) \citealt{raassen98}; (5) \citealt{kupka99}; (6) \citealt{kurucz94}; (7) \citealt{martin88}; (8) \citealt{barklem00,barklem05};  (9) \citealt{dmbiz06}; (10) \citealt{smith00}}
\end{deluxetable*}

Measuring $W_\lambda$  of the iron lines from the stellar spectra was accomplished both in an automated and manual way. A first pass at measuring all of the lines was made by interactively running the ``Automatic Routine for line Equivalent widths in stellar Spectra'' \citep[ARES,][]{sousa07}, which can automatically fit the continuum around each line and fit multiple Gaussians to blended lines to measure the $W_\lambda$ of the specified line.  A visual inspection of each fit is needed to ensure a proper treatment of the continuum in line-dense regions. Generally, 40\%--50\% of the ARES-measured lines are acceptable for stars with \vsini$\,<\,$11\,\kms. This fraction drops to 0\%--25\% for stars with \vsini$\,>\,$11\,\kms\ because of severe blending.  The remaining lines that are unsuitably measured by ARES are manually measured using {\it splot} in IRAF.  The high-resolution Arcturus atlas spectrum was used as a guide for identifying the continuum and  locating  individual lines in severely blended regions, particularly in the rapid rotators.  Not all of these lines are measurable in every star. Weaker lines and blended lines become increasingly difficult to measure in stars with larger \vsini\ values and lower metallicities. In addition, non-stellar spectral features  such as cosmic rays or telluric absorption lines can also contaminate some of these lines to make them unmeasurable. 

\begin{figure}[tb]
\centering
\includegraphics[width=1.03\columnwidth]{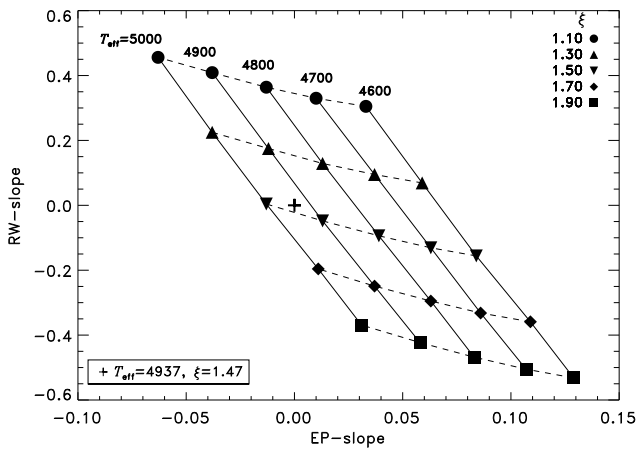} 
\caption[Illustration of how EP-slope and RW-slope vary]{Illustration of how EP-slope and RW-slope vary with temperature and microturbulence for a given $\log g$ and [Fe/H].  Solid lines connect models of constant \teff, and dashed lines connect models of constant $\xi$. The plus sign marks the best-fit $T_{\rm eff}$ and $\xi$, given in the bottom left corner,  as where both slopes are zero.}
\label{testplot}
\end{figure}

 Once the $W_\lambda$ are measured, the stellar atmosphere parameters are found with an iterative guess and update scheme.  An initial guess in temperature comes from each stars' photometric temperature, using the \cite{houd00} empirical color-temperature relations for giant stars. The initial guess of the other parameters are the same for all stars: $\log g=2.0$, $\xi=2.0$\,\kms, and  [Fe/H]\,=\,0.0.   First, abundances are found for a range of \teff\ and $\xi$  around the initial guess (holding gravity and metallicity constant) using the  grid of MARCS stellar atmosphere models. 
 These models are defined on a  grid  of effective temperature ($3700\leq T_{\rm eff} \leq 6000$, $\Delta$\teff\,$=100$ or $250$\,K), surface gravity ($0.0 \leq \log g \leq 3.5$, $\Delta \log g = 0.5$\,dex), and metallicity ($-2.5\leq$\,[Fe/H]\,$\leq +0.25$, $\Delta $[Fe/H]\,$=0.25$ or 0.5\,dex). When needed, interpolated models are created using the {\it interpol\_modeles} codes provided with the MARCS models.\footnote{http://marcs.astro.uu.se/software.php} 
 MOOG fits both the EP-slope and RW-slope for each model, and   Figure \ref{testplot} shows an example of how these two slopes vary as one runs MOOG on the same line list with different $T_{\rm eff}$ and $\xi$ but constant  $\log g$ and [Fe/H]. 
A surface is fit to the type of data shown in Figure \ref{testplot} to find the temperature and microturbulence
for which both slopes are zero.   Once these values are found, the average $A$(Fe) given by the \ion{Fe}{1} lines is compared to the metallicity of the input stellar atmosphere model.  If they disagree, it is used as the next guess of the metallicity.  The \ion{Fe}{2} lines are used to model a new guess in gravity.  If the average $A$(\ion{Fe}{2}) is less than the average $A$(\ion{Fe}{1}), then a higher surface gravity is guessed in the next iteration and vice versa for $A$(\ion{Fe}{2})\,$>$\,$A$(\ion{Fe}{1}). We found that updating the $\log g$ estimate by the same magnitude as the iron abundance discrepancy was a  successful scheme. 
In other words, $(\log g)_{\rm new}\approx(\log g)_{\rm old}+A$(\ion{Fe}{1})$ - A$(\ion{Fe}{2}).   We update  \teff\  and $\xi$ only when  the EP-slope or RW-slope is non-zero by more than 0.005,  and we update surface gravity only if  $A$(\ion{Fe}{1}) and $A$(\ion{Fe}{2}) differ by more than 0.03~dex. 
Our results are presented in the first 10 columns of Table \ref{sample:stellparam}, which lists the star name, \teff, $\log g$, the mean abundance of iron derived from the \ion{Fe}{1} lines ($A$(\ion{Fe}{1})),  the standard deviation in that mean ($\sigma_{\rm Fe\,I}$),  the number of lines contributing to the mean ($N_{\rm Fe\,I}$), $A$(\ion{Fe}{2})$, \sigma_{\rm Fe\,II}$,  $N_{\rm Fe\,II}$,  and $\xi$.  Note that a few stars appear in the table more than once because some of the stellar parameters are derived independently from more than one spectrum of that star. The year of the observation is appended to the star name in these cases.  Stellar parameters that are adopted from one spectrum but applied to another (e.g., measured only from a 2007 spectrum and listed with both the 2007 and  2008 spectra) are indicated with footnotes.
\begin{deluxetable*}{llccccrrrrrrrr} 
\tablefontsize{\scriptsize} 
\tablewidth{\textwidth}
\tabletypesize{\scriptsize}
\tablecaption{Derived stellar parameters \label{sample:stellparam}}
\setlength{\tabcolsep}{0.00in} 
\tablecolumns{14}
\tablehead{
       \colhead{Star}& \colhead{$T_{\rm eff}$}&  \colhead{$\log g$}& \colhead{$A$(\ion{Fe}{1})}& \colhead{$\sigma_{\rm Fe\,I}$}& \colhead{$N_{\rm Fe\,I}$}& \colhead{$A$(\ion{Fe}{2})}& 
       \colhead{$\sigma_{\rm Fe\,II}$}& \colhead{$N_{\rm Fe\,II}$}&\colhead{$\xi$} &\colhead{\vsini}&\colhead{Error}& \colhead{$N_{\rm l}$}& \colhead{$\zeta$} \\
       \colhead{ } & \colhead{(K)}& \colhead{(cm s$^{-2}$)}& \colhead{(dex)}& \colhead{(dex)}& \colhead{ }&
       \colhead{(dex)}& \colhead{(dex)}&  \colhead{ }& \colhead{(\kms)} & \colhead{(\kms)}&\colhead{(\kms)}& \colhead{ }& \colhead{(\kms)} \\ }
\startdata 
Arcturus\_2007  	    & 4300\tablenotemark{a}     & 1.70\tablenotemark{a}      & 6.98\tablenotemark{a}      & 0.11\tablenotemark{a}      & 74\tablenotemark{a}   &6.98\tablenotemark{a} & 0.22\tablenotemark{a}    &  12\tablenotemark{a}   &  1.64\tablenotemark{a}     &  1.69     & 0.61   &  5    &  5.00 \\ 
Arcturus\_2008         & 4300     & 1.70      & 6.98      & 0.11      & 74    & 6.98 & 0.22    &  12   &  1.64   &  2.36     & 0.98   &  4    &  5.00 \\ 
G0300+00.29           & 4590     & 2.75      & 7.52      & 0.17      & 73    & 7.52 & 0.23    &  12   &  1.42   &  2.58     & 0.84   &  5    &  4.16 \\ 
G0319+56.5830       & 5170     & 3.10      & 7.52      & 0.15      & 74    & 7.51 & 0.10    &  12   &  1.41   &  6.92     & 0.19   &  6    &  6.13 \\ 
G0319+56.6888       & 4790     & 2.90      & 7.50      & 0.15      & 74    & 7.50 & 0.16    &  12   &  1.55   &  2.78     & 0.43   &  4    &  3.12 \\ 
G0453+00.90           & 4710     & 2.80      & 7.20      & 0.12      & 74    & 7.21 & 0.11    &  12   &  1.25   &  0.66     & 0.33   &  3    &  5.23 \\ 
G0639+56.6179       & 4660     & 2.60      & 7.23      & 0.10      & 74    & 7.21 & 0.17    &  13   &  1.15   &  1.75     & 0.28   &  6    &  2.80 \\ 
G0653+16.552        & 4170     & 1.40      & 7.06      & 0.14      & 74    & 7.05 & 0.19    &  12   &  1.62   &  3.25     & 0.15   &  5    &  6.35 \\
G0827$-$16.3424       & 5090     & 3.60\tablenotemark{b}      & 7.36      & 0.21      & 54 & 7.34   & 0.15    &   7   &  2.76   &  23.85     & 0.42   &  5    &  3.84 \\ 
G0928+73.2600\_2007 &  4900 & 2.70 & 7.26 & 0.13 & 74 & 7.25 & 0.16  &12 & 1.51  & 8.36&  0.30 & 5 & 5.60 \\   
G0928+73.2600\_2008 &  4870 & 2.60 & 7.28 &  0.11 & 73 & 7.30 & 0.11 & 12 & 1.42&   8.36\tablenotemark{a}  &  0.30\tablenotemark{a}  & 5\tablenotemark{a}   & 5.60\tablenotemark{a}   \\  
HD31993	   & 4360\tablenotemark{c}   & 2.4\tablenotemark{c}\   & 7.62      & 0.31\tablenotemark{b}      & 41    & 7.43 & 0.40    &   8   &  3.0\tablenotemark{c}    &  30.4      & 0.14   &  4    &  4.59 \\ 
Tyc5904-00513-1\_2007 & 4640\tablenotemark{a}  & 2.30\tablenotemark{a} & 6.87\tablenotemark{a} & 0.17\tablenotemark{a} & 65\tablenotemark{a} & 6.87\tablenotemark{a} & 0.20\tablenotemark{a} & 12\tablenotemark{a} & 1.93\tablenotemark{a} & 14.46 & 0.32 & 3 &  5.09  \\
Tyc5904-00513-1\_2008 & 4640  & 2.30 & 6.87 & 0.17 & 65 & 6.87 & 0.20 & 12 & 1.93 & 13.95 & 0.32 & 3 & 7.12  \\
Tyc6094-01204-1\_2007 & 4320  & 1.90 & 7.06 & 0.18 & 72 & 7.06 & 0.25 & 12 & 1.53 & 13.71 & 0.46 & 4 & 4.47  \\
Tyc6094-01204-1\_2008 & 4320\tablenotemark{a}  & 1.90\tablenotemark{a} & 7.06\tablenotemark{a} & 0.18\tablenotemark{a} & 72\tablenotemark{a} & 7.06\tablenotemark{a} & 0.25\tablenotemark{a} & 12\tablenotemark{a} & 1.53\tablenotemark{a} & 12.78 & 0.39 & 3 & 6.61 
\enddata
\tablenotetext{a}{Stellar parameter  adopted from a previous measurement and not remeasured from this spectrum}
\tablenotetext{b}{High $\log g$ necessitated the use of  Kurucz plane-parallel models.}
\tablenotetext{c}{\ Literature stellar parameters from \cite{castilho00}.}
\tablecomments{This table is available in its entirety in a machine-readable form in the electronic edition of \apj. A portion is shown here for guidance regarding its form and content.}
\end{deluxetable*}

\begin{figure*}[tbh]
\centering
\includegraphics[width=0.7\textwidth]{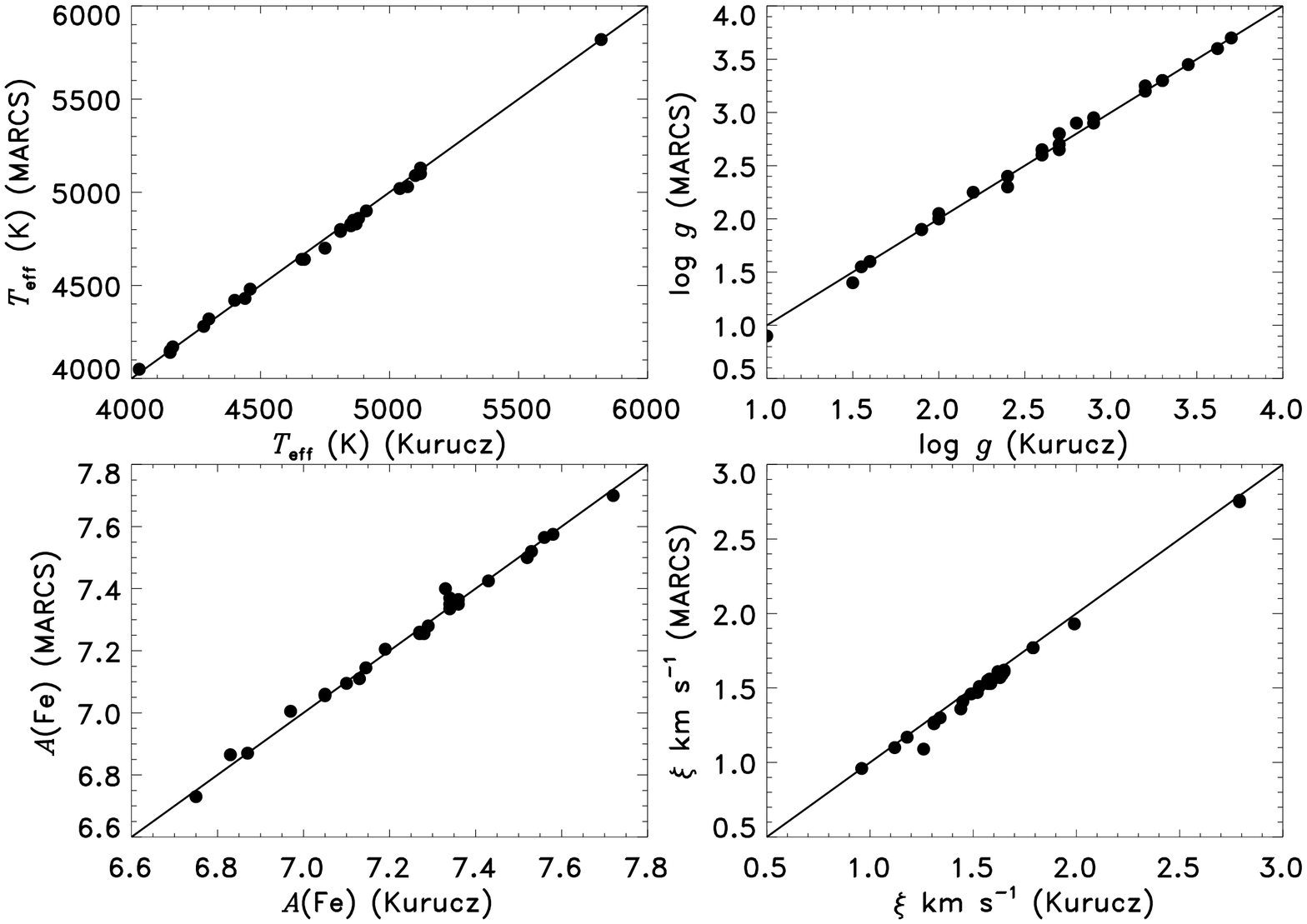} 
\caption{\label{comp_models} Comparison of the stellar parameters derived using Kurucz atmosphere models and MARCS atmosphere models.  The solid line is a unity-slope line. }
\end{figure*}

The highest surface gravity offered by the MARCS stellar atmosphere models is 3.5\,dex, and one of our stars have surface gravities exceeding this upper limit. For this stars we switched to using Kurucz plane parallel models \citep{castelli04}\footnote{ODFNEW models available at http://kurucz.harvard.edu/grids.html}; a note in Table \ref{sample:stellparam} identifies this star. 
Because we could not use spherical models for all of our stars, we are naturally concerned with systematic errors between the stellar parameters derived with the MARCS spherical models and the Kurucz  plane-parallel models.  For 27 stars in our sample, we had derived stellar parameter solutions using plane-parallel models  previous to the spherical model derivations presented here \citep{carlberg10a}.  We compare these two stellar parameters derivations in Figure \ref{comp_models}.  The one-to-one line is shown for each stellar parameter, and we generally find  good agreement between the  solutions from the two different types of models.   The only systematic difference appears for $\xi$, which is slightly lower  in the spherical models by $\sim0.04$\,\kms, on average.

  To test how well the iron list works on red giant stars,  we derived stellar parameters for Arcturus using $W_\lambda$ measured both from the \cite{hinkle00} atlas spectrum and our own data.  These solutions are tabulated along with literature stellar parameters for Arcturus in Table \ref{arcturus}.  This table shows that our line list does  well constraining $T_{\rm eff}$, $A$(Fe), and $\xi$.  We derive the same temperature from the $W_\lambda$ measured in the atlas spectrum and our observed spectrum. Our  [Fe/H] differ by only 0.03\,dex, and $\xi$ differ by only 0.04\,\kms. Comparing to the literature,  we find that our
  derived $T_{\rm eff}$  deviate from  literature values by $<20$\,K,  our [Fe/H] fall between the literature values, and  our  $\xi$ are larger by  0.03--0.08\,\kms.   The $\log g$ are less well constrained.  Our two derived values differ by 0.2\,dex.  Literature values range from 1.55 to 1.90\,dex, while we derived 1.50 and 1.70\,dex.   This larger uncertainty in $\log g$
   likely stems from the fact that $\log g$ is  primarily constrained by the stellar  \ion{Fe}{2} lines, and there are far fewer \ion{Fe}{2} lines in our list than \ion{Fe}{1} lines.

\begin{deluxetable}{llllrc} 
\tablewidth{\columnwidth}
\centering
\tabletypesize{\footnotesize}
\tablecaption{Stellar parameters of Arcturus. \label{arcturus}}
\tablecolumns{6}
\tablehead{
       \colhead{$T_{\rm eff}$}&\colhead{$\log g$}& \colhead{[Fe/H]}& \colhead{$\xi$}& \colhead{$A$(Fe)$_\sun$} &\colhead{Reference}\\
       \colhead{(K)}& \colhead{(cm s$^{-2}$ ) }& \colhead{(dex)}& \colhead{(\kms)}& \colhead{(dex)}& \colhead{ }}
\startdata
4300 & 1.70 & $-0.55$ & 1.64 & 7.53 & 1, 2 \\  
4300 & 1.50 & $-0.58$ & 1.68 & 7.53 & 1, 3 \\  
4300 & 1.7 & $-0.72$ & 1.6 & 7.50 & 4 \\ 
4283 & 1.55 & $-0.50$ & 1.61 &7.45 & 5 \\
4290 & 1.90 &$-0.68$ & \nodata& \nodata&  6
\enddata
\tablerefs{(1) Derived by us; (2) spectrum from this work; (3) \citealt{hinkle00} atlas spectrum; (4) \citealt{smith00}; (5) \citealt{fulbright06}; (6) \citealt{griffin99}, as used in \citealt{dmbiz06}.}
\end{deluxetable}
  We can also constrain our stellar parameter errors using Monte Carlo simulations. By comparing the $W_\lambda$ measured in our Arcturus spectrum to literature measurements, we find that a Gaussian distribution of $\sigma=0.1$ (i.e., $\sim10$\% errors) is a reasonable model of our expected $W_\lambda$ errors. Therefore, in each iteration of the simulation, we vary the equivalent widths of the test star by drawing random errors from a Gaussian distribution with a width of 10\% and rerun our stellar parameter finding software. Because this code is relatively time consuming, we only ran a small number of iterations.   A 10-iteration simulation for Arcturus revealed mean variations in the derived stellar parameters of 80\,K in \teff, 0.3\,dex in $\log g$, 0.03\,dex in [Fe/H], and 0.13\,\kms\ in $\xi$.

\subsection{Line Broadening: \vsini\ and $\zeta$}
\label{sec:vsini_derive}
All of the stars in our sample have \vsini\ measurements in the literature, but these measurements come from a variety of sources and from spectra of variable quality.  Therefore, we re-derive \vsini\ in our entire sample. Our high-quality spectra allows us to model the shapes of individual lines to measure the broadenings.  Six relatively isolated lines were chosen from the iron lines listed in Table \ref{felinelist}: $\lambda = $~5778.45, 6027.05, 6151.62, 6173.33, 6733.15, and 6750.15\,\AA.   For each line, we use MOOG's \emph{synth} driver to generate a synthetic line profile.  We use the iron abundance measured from that particular line, as opposed to the mean iron abundances, to more accurately fit the line depth.   The broadenings included in the profile are the instrumental broadening, macroturbulence ($\zeta$), and \vsini.  The instrumental widths were measured from each star's corresponding Th--Ar calibration lamp. The widths of five Th--Ar lines in the same echelle order as each iron line were averaged together to get the instrumental 
broadening.\footnote{In our 2008 KPNO data, the measured instrumental widths seemed too large because the broadenings of the slow rotators could be reproduced with almost no physical, atmospheric  mechanisms. 
One possible reason for this issue is that the seeing conditions were better than $1.5''$ so that the stellar lines are better resolved than the ThAr lines.
However, that explanation does not account for why the 2008 instrumental widths are $\sim 40\%$ larger than the 2007 widths when the same slit
was used for both. Within each run, the instrumental widths are relatively constant and show only some low time frequency variations that are likely due to focus drift. These variations are coherent in wavelength. 
  Fortunately, we had five stars overlapping our 2007 and 2008  KPNO observing runs.  By requiring the same \vsini\ and $\zeta$ derived from the 2007 data, we found that the 2008 KPNO instrumental widths had to be scaled by a factor of 0.83. We found no need to change the scaling factor with wavelength. }
The macroturbulence for each star was estimated from the relations of \cite{hekker07} based on the stars' temperatures and gravities.
Keeping $\zeta$ and the instrumental broadening fixed, we generated spectra with a range of \vsini\ and used a $\chi^2$-minimization to find the best \vsini.
Small adjustments to the overall continuum level and velocity shifts to the wavelength solution were
allowed to improve the fits.   As long as reasonable \vsini\ values were measured with the  initial $\zeta$ estimate (i.e., the best fits were not \vsini=0), then no adjustments to $\zeta$ were made.   Otherwise, we reduced $\zeta$ by using the \teff--$\zeta$ relationship for a dimmer luminosity class, e.g.,  changing  from class III to class IV.  We do not have the resolution or S/N to  derive \vsini\ and $\zeta$ independently.
The average derived \vsini\ value of each star is given in the eleventh column of Table \ref{sample:stellparam} followed by the standard error in the mean, the number of Fe lines comprising the average, and the  $\zeta$  of the synthetic spectra.
 
\begin{figure}[bt]
\centering
\includegraphics[width=1.03\columnwidth]{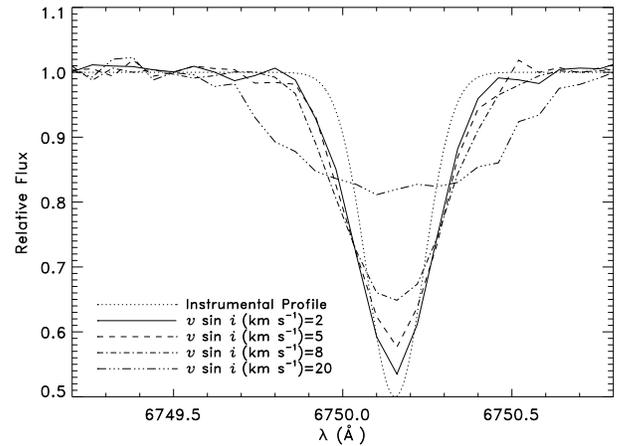} 
\caption{\label{modelstar}  Fe line at 6750.15~\AA\ for the representative star described in Section \ref{sec:vsini_derive} for  \vsini\ of 2, 5, 8, and 20~\kms\ (solid, dashed, dot-dashed, and triple-dot-dashed
 lines, respectively).  For reference, the approximate instrumental profile (dotted line, a Gaussian with FWHM=0.2~\AA) is superimposed at the central wavelength of the Fe line.  }
\end{figure}

 To address the question of degeneracy in the \vsini\ and $\zeta$ derivations, we run the following test. We compute artificial spectra at the location of the six iron lines for an artificial star with stellar parameters representative of our stellar sample, mimicking both the spectral resolution and sampling of the KPNO data. (The APO data are of higher resolution and finer sampling.) Using median stellar parameters from our sample, this representative star has $T_{\rm eff}=4670$\,K, $\log g=2.55$\,dex, [Fe/H]$~=-0.21$, $\xi=1.46$\,\kms, and $\zeta=4.74$\,\kms.
 We are interested in trying to recover the rotational broadenings for four different cases: $v \sin i < \zeta$, $v \sin i \approx \zeta$, $v \sin i > \zeta$, and $v \sin i \gg \zeta$.  These cases are modeled with \vsini\ of 2, 5, 8, and 20\,\kms, respectively, and in Figure \ref{modelstar} we plot the profiles of the 6750.15~\AA\  iron line.  For reference, we have also plotted a Gaussian with a full width at half-maximum (FWHM) of 0.2~\AA.

\begin{figure}[t!]
\centering
\includegraphics[width=1.03\columnwidth]{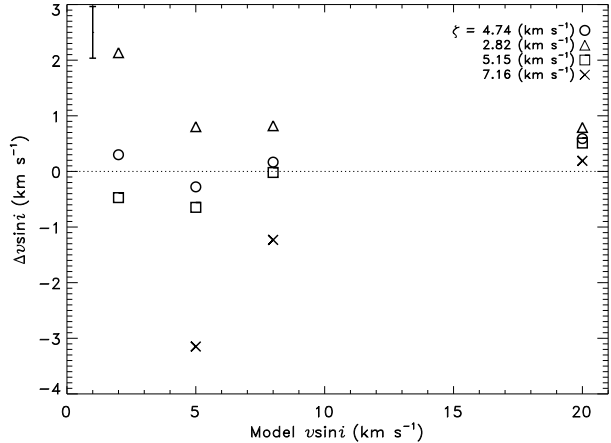} 
\caption[Difference between the measured \vsini\ and model \vsini]{\label{testbroad} Difference between the measured \vsini\ and model \vsini\ as a function of the model \vsini\ of a ``representative'' star.  Different symbols denote different values of the assumed macroturbulence, $\zeta_{\rm guess}$: the true $\zeta$ (circles), and the expected $\zeta$ from the \cite{hekker07} $T-\zeta$ relationship for luminosity class IV (triangles), III (squares), and II (crosses) stars. }
\end{figure}

 For each \vsini, we test how the assumed macroturbulence, $\zeta_{\rm guess}$, affects the measured \vsini\ using the same \vsini\ fitting program for the following values of $\zeta_{\rm guess}$: 4.74, 2.82, 5.15,  and 7.16\,\kms.  The first value is the true value of the test star, and we expect to be able to recover the correct \vsini\ in these cases.  The latter three values come from the \cite{hekker07}  $T$--$\zeta$ relationships for a \teff$=4670$\,K star in luminosity class IV, III, and II, respectively.
In Figure \ref{testbroad}, we illustrate the results of the experiment by plotting the difference between the measured and modeled \vsini\ as a function of the model \vsini.  The different points indicate the different $\zeta$'s used in the fits.  Note that there is no result for the \vsini~=~2\,\kms\ model for $\zeta_{\rm guess}=7.164$\,\kms.  Such a high $\zeta$ created synthetic lines too broad to fit the test lines even in the absence of any rotation. The  bar in the upper left shows the typical measurement error in the fit \vsini, which is 0.5\,\kms. As expected, when the assumed $\zeta$ matches the true value of 4.74\,\kms, all of the model \vsini's are recovered within the error bars.   As is also expected, the discrepancy between the true \vsini\ and the fitted \vsini\ is more dramatic for slow rotators. Both rapid rotators in our sample had their \vsini's recovered to within 1.5\,\kms\ across the full range of possible $\zeta_{\rm guess}$.   The slow rotators, on the other hand, may have large {\it fractional} errors in their derived \vsini\ values\ if the assumed $\zeta$ is incorrect. However, these larger uncertainties merely affirm their classification as ``slow rotators'' and have no other bearing on our analysis. 

\begin{figure*}[t]
\centering
\includegraphics[width=0.7\textwidth]{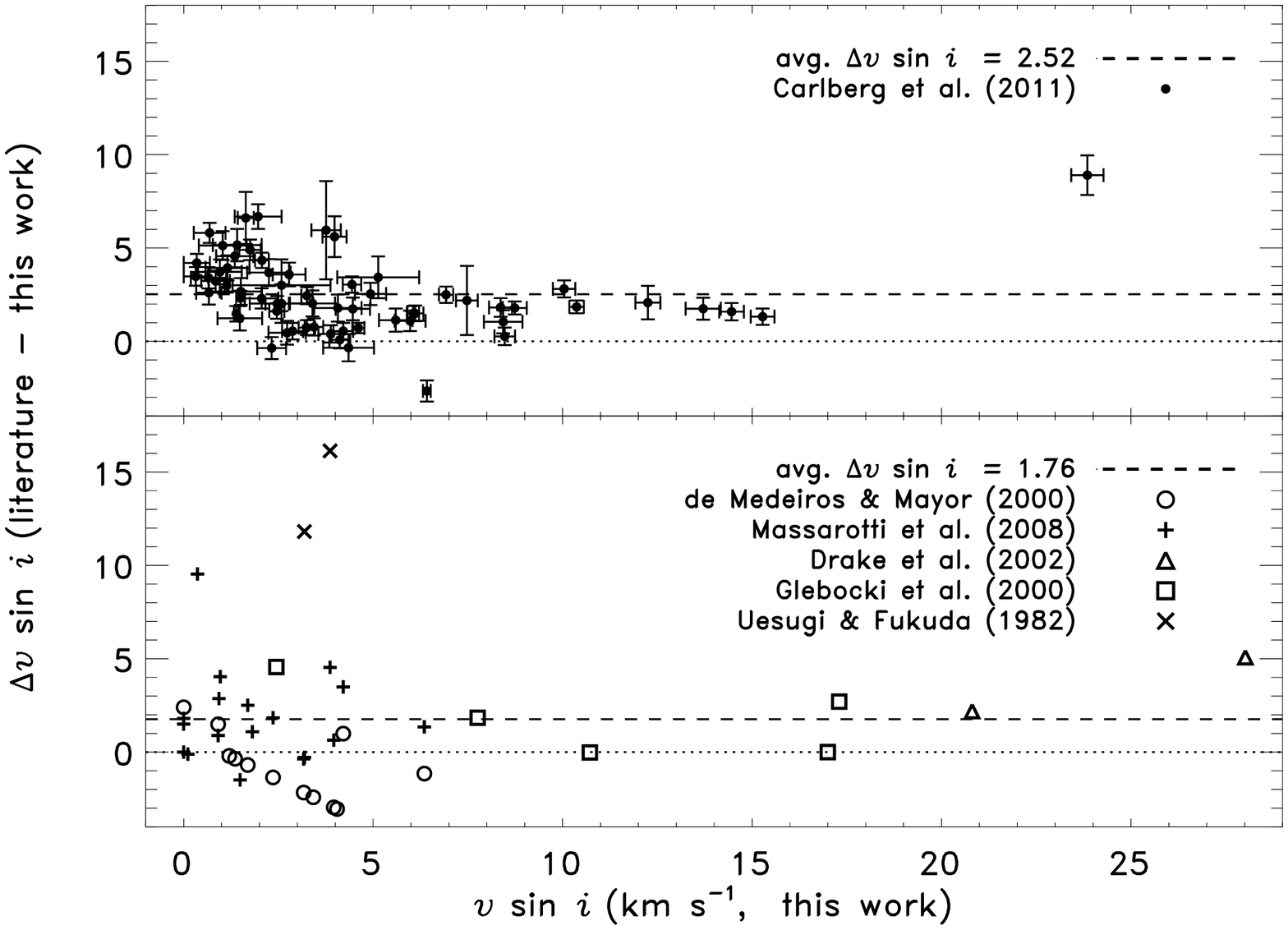} 
\caption{\label{fig:vsini}  Comparison of the \vsini\ derived in this work to those derived in the literature. The top panel shows the fainter targets selected from C11 (filled circles), while the bottom panel shows the brighter targets selected from \citet[][open circles]{deMed00},  \citet[][crosses]{massarotti08a}, \citet[][triangles]{drake02}, \citet[][squares]{glebocki00}, and \citet[][$\times$]{uesugi82}. In each panel, the lines indicate $\Delta$\vsini$=0$ (dotted) and the actual mean $\Delta$ \vsini\ of the samples (dashed).}
\end{figure*}

In Figure \ref{fig:vsini}, we compare the \vsini\ derived here to those derived in the literature sources that we used to build our sample. 
All of the comparison \vsini\ values for the GGSS and \emph{Tycho} stars (the fainter sample) come  from the low S/N spectra described in C11 and are shown in the top panel.  The brighter stars, plotted in the bottom panel, have comparison measurements from \cite{deMed00}, \cite{massarotti08a}, \cite{drake02}, \cite{glebocki00}, and \cite{uesugi82}. 
In each panel, two comparison lines are overplotted. One line represents $\Delta$\vsini$=0$, and the other indicates the mean difference between the \vsini\  in the literature and those derived here.   We tend to measure smaller \vsini\  compared to literature in both the faint and bright stellar samples. The exception is the subset of stars from \cite{deMed00}, for which we measure larger \vsini\ by $\sim0.8$~\kms. The most likely explanation for the systematically lower \vsini\ measured with our analysis is  an overestimation of either the instrumental or (more likely) the macroturbulent broadenings.  The larger systematic difference seen in the comparison to the C11 data   
may be due to blended lines inflating the measured line widths in the cross-correlation analysis of C11. In the low-\vsini\ regime ($\lesssim5$\,\kms), there is larger scatter in $\Delta$\vsini, and trends  appear for stars with \vsini\ at the lower limit of the literature surveys. The higher \vsini\ regime shows less scatter, though still a systematic offset. The literature suggests that the two most rapid rotators in our sample have even higher \vsini\ than what is measured here.
 Considering the fact that the  velocity resolution of our data is $\gtrsim$9\,\kms,    
  we consider the  discrepancy between these \vsini\ measurements and our comparison samples to be acceptably low. In particular, the \vsini\ discrepancy will only affect the classification of the stars as ``slow'' or ``rapid'' rotators.   It should have a negligible effect on our abundance measurements.  Measurements of $W_\lambda$ are relatively independent of the line shapes, and  spectral synthesis will yield correct abundance results in the presence of \vsini, $\zeta$, and instrumental broadening errors as long as the {\it total} broadening is well-represented. 

\subsubsection{Defining ``Rapid Rotation''}
We choose 8\,\kms\ as the cutoff \vsini\ between rapid and slow rotators.  As discussed previously in C11, the \vsini\ value selected to separate slow rotators from rapid rotators  is arbitrary, and values between 5 and 10\,\kms\ have been used previously in the literature.  Our choice of 8\,\kms\ is chosen partly because we are confident that we can truly distinguish stars with \vsini$\geq 8$\,\kms\  from slow rotators (which was not true for our measurements in C11) and partly for convenience (choosing 8\,\kms\ over 7, 9 or 10\,\kms\ minimizes the number of ``almost rapid rotators'').   This convention has already been
applied to the categorization of our targets in Table \ref{tab:obs}.

\subsubsection{Loss of the Independent Planet Accretor Sample}
In our observational sample, the three stars with \emph{Hipparcos} designations (HIP35253, HIP36896, and HIP81437)  were selected from the \vsini\ and RV study of M08.  These three stars have enhanced rotation (between 7 and 10\,\kms\ in the M08 study), and the authors argued that the location of these stars on the HR diagram made the accretion of planets the most likely explanation for the high rotational velocities.  With reported \vsini\ values of 9.9\,\kms (HIP35253), 8.4\,\kms\ (HIP36896), and 7.7\,\kms\ (HIP81437) the former two stars would be classified as  rapid rotators in this study, while the latter star just barely misses our cutoff. 
Therefore, we anticipated that this subsample would be an important control sample because their ``suspected planet accretor'' status was made by an independent group. 
Unfortunately, we do not reproduce the \vsini\ values reported by M08  for these three stars. We measure $<1$\,\kms\ (HIP35253), 3.8\,\kms\ (HIP36896), and 4.2\,\kms\ (HIP81437). 
Our assumed $\zeta$ was larger than that of M08 for the two latter stars, so we  re-derived \vsini\ using their values. 
Although this analysis increases the value of our \vsini\ measurements, they are still smaller than the M08  values by 2--3\,\kms\ and fall short of our rapid rotator cutoff.  Because we cannot reconcile our \vsini\ measurements with theirs, we simply add these stars to the slow rotator sample, which now numbers 61 stars.  

\subsubsection{Estimating  {\it True} Rotational Velocity}
\label{sec:truevrot}
Because \vsini\ depends on the inclination angle of the stellar rotational axis to the line of sight,  we are 
interested in estimating the fraction of slow rotators that might be rapid rotators seen nearly pole-on. To test this, we 
computed random inclinations of the rotation axis to an observer's line-of-sight for one billion test stars. We can then find the fraction of these stars that are true rapid rotators ($v_{\rm rot}\geq8$\,\kms) given the random inclination and   \vsini.  This fraction can also be thought of as the probability ($P_{\rm RR}$) that a star of a given \vsini\ is a rapid rotator. We calculate this probability for a range of \vsini\ between 1 and 5\,\kms, and find that the relationship between $P_{\rm RR}$ and the observed \vsini\ is well defined by a second-order polynomial of the form:
\begin{equation}
 P_{\rm RR}=0.0070-0.0087(v\sin i)+0.0073(v\sin i)^2.
\label{eq:inclination}
\end{equation}
The expected fraction of slow rotators that are truly rapid rotators in our entire sample is simply the sum of the probabilities for each slow rotator over the total number of slow rotators, which yields 4.5\%.
Therefore, our control sample likely has two or three unidentified rapid rotators. 

\subsection{Radial Velocity Variability}
 \begin{deluxetable*}{lrrrrr}
\tablewidth{\textwidth}
\tablecaption{Radial Velocity \label{tab:RVs}}
\tablecolumns{6}
\tablehead{
       \colhead{Star}&
       \colhead{JD$-$2450000}&
       \colhead{RV}&
       \colhead{err\_RV}&
       \colhead{$P_{\rm RV}$\tablenotemark{a}} &
       \colhead{$f_{\rm CB}$\tablenotemark{b}} \\
       \colhead{ } &
       \colhead{(days)}&
       \colhead{(\kms)}&
       \colhead{(\kms)}&
       \colhead{} &
       \colhead{}
}
\startdata
G0804+39.4755       & 4165.660    &   12.7    &   2.0  & 0\% & 92\%  \\
       & 4476.817    &   35.0    &   2.0  &   \\
       & 5284.658    &   22.3    &  0.9   &   \\
       & 5553.946    &   47.1    &  0.9   &   \\
G0827$-$16.3424       & 4475.846    &   16.1    &   2.0  & 6\% &  7\%\\
       & 5284.615    &   16.7    &   1.8  &   \\
       & 5492.033    &   21.0    &   1.4  &   \\
G0928+73.2600       & 4165.730    &   37.9    &   2.0  & 41\%  & 0.8\% \\
       & 4476.894    &   37.8    &  2.0   &   \\
       & 5284.698    &   35.9    &  0.9   &   \\
      & 5492.020    &   34.9    &   1.0  &   \\
G0946+00.48           & 4166.661     &   44.9    &    2.0  & 19\%  & 2\% \\ 
          & 5284.679     &   40.8    &    1.7  &   \\ 
           & 5553.924     &   40.7    &    1.5  &   \\ 
G1213+33.15558     & 4478.034   &    41.4    &   2.0  & 0\%  &91\% \\ 
     & 5284.713   &    39.7    &  0.8   &    \\ 
     & 5492.025   &    20.1    &   3.8  &    \\ 
    & 5782.606   &    40.1    &   1.2  &    \\ 
HD112859               & 4480.008    &   65.6     &  2.0  & 0\%  & 89\% \\ 
               & 5284.741    &  $-21.7$     &  1.8  &    \\ 
               & 5553.909    &   58.6     &  1.4  &    \\ 
               & 5782.617    &  $-14.7$     &  2.1  &    \\ 
HD31993                 & 4165.611   &    17.0   &    2.0  & 34\%  & 2\% \\ 
                 & 5491.925   &    14.2   &    2.4  &    \\ 
                 & 5553.854   &    13.2   &    1.7  &    \\ 
HD33363                 & 4479.707   &   $-57.4$   &    2.0  & 0\%  & 89\% \\ 
                 & 5491.990   &   $-39.9$   &    1.3  &    \\ 
                 & 5553.886   &   $-37.9$   &   0.8   &    \\ 
HD34198                 & 4166.596   &    5.4    &   2.0   & 89\%  & 0.7\% \\ 
                 & 5491.932   &    5.0    &   1.6   &    \\ 
                 & 5491.934   &    3.7    &   1.4   &    \\ 
                 & 5553.817   &    4.8    &   1.4   &    \\ 
Tyc0347-00762-1   & 4165.986   &   $-25.9$  &     2.0 & 0\%  &84\% \\ 
   & 5284.792   &   $-29.4$  &     1.8 &    \\ 
   & 5782.643   &   $-13.1$  &     1.8 &    \\ 
Tyc0647-00254-1   & 4475.638   &   0.1    &   2.0   & 34\% & 0.5\% \\ 
   & 5491.909   &   $-1.1$   &   0.7   &    \\ 
   & 5553.833   &   $-1.1$   &  0.8    &    \\ 
  & 5798.969   &   0.8    &  0.9    &    \\ 
Tyc2185-00133-1   & 4309.922   &   $-17.0$  &     2.0 & 71\%  & 1\% \\ 
   & 5782.692   &   $-15.2$  &    0.9  &   \\ 
   & 5798.903   &   $-15.7$  &     1.2 &    \\ 
Tyc3340-01195-1   & 4419.938   &    9.4   &    2.0  & 0\%  & 93\%  \\ 
   & 5284.600   &    1.9   &    1.3  &    \\ 
   & 5491.974   &    7.1   &   0.8   &    \\ 
   & 5798.926   &   $-22.3$  &     1.4 &    \\ 
Tyc5904-00513-1   & 4166.609   &    57.8  &     2.0 & 48\%  & 0.6\% \\ 
   & 4475.712   &    57.3  &     2.0 &    \\ 
   & 5491.916   &    55.2  &     1.5 &    \\ 
   & 5553.824   &    54.7  &     1.3 &    \\ 
Tyc6094-01204-1   & 4164.890   &   $-12.8$  &     2.0 & 0\% &  92\% \\ 
   & 4476.008   &   $-28.2$  &     2.0 &    \\ 
  & 5284.751   &   $-11.8$  &     1.1 &    \\ 
  & 5553.955   &   $-26.1$  &     1.2 &    
 \enddata
\tablenotetext{a}{Probability that the radial velocities are time independent (RV stable).}
\tablenotetext{b}{Fraction of simulated close binary systems (period $\leq 500$ days) that have a comparable likelihood of appearing RV stable---$P_{\rm RV}$ (simulated)  within 0.05 of $P_{\rm RV}$ (actual)---when ``observed'' with the same cadence and RV precision as the target star.} 
\end{deluxetable*}
 
One concern with interpreting  enhanced rotational velocity as evidence of planet accretion is the possibility that the star is instead merely rotating synchronously with a stellar companion.   RV monitoring can determine whether such a stellar companion is present.   Therefore, we collected additional high-resolution spectra of the rapid rotators 
to look for RV variability.   In Table \ref{tab:RVs}, we list the star,  date of observation, heliocentric RV, the uncertainty in the RV (err\_RV), and two measures of the likelihood that the star is a single or binary star based on the RVs.
The first of these two measures, $P_{\rm RV}$, is the probability (from the $\chi^2$ distribution) that the suite of measured RVs are consistent with the model of a single, constant RV (equal to the
weighted mean of the RVs).
The second measure, $f_{\rm CB}$, indicates the likelihood of measuring $P_{\rm RV}$ for a close-orbiting binary system, given the observing cadence and RV precision of  each of the target stars. To calculate $f_{\rm CB}$,  we generate a population of 50,000 binary systems, drawing random periods and mass ratios ($q$) from the Gaussian probability distributions described in \cite{duquennoy91}.  
We adopted a primary star mass of $2$~\msun, and the  orbital parameters of eccentricity ($e$),  longitude of periastron, and time of periastron passage were drawn from uniform probability distributions.  Inclination angles were drawn from an isotropic distribution. We chose a circularization period of 100 days (setting $e=0$ for shorter periods).  Because our simulated primary stars are giants, we discarded simulated systems that yielded binary separations smaller than $5$~\rsun.  We define ``close orbiting'' systems as those with periods less than 500 days. These are systems for which  a 100~\rsun\ primary star that is co-rotating with its companion will have $v_{\rm rot}=10$~\kms.  
 Each of the simulated binary systems is then ``observed'' using the exact same observing dates as the target star, and random noise is added to the RV measurement. These simulated RV's are then processed identically to the real data to compute $P_{\rm RV}$ for each binary simulation.  We define $f_{\rm CB}$ as the fraction of close binary stars that a have simulated $P_{\rm RV}$ within 0.05 of the measured $P_{\rm RV}$.  As an example, in Table \ref{tab:RVs} the measured $P_{\rm RV}$ for G0827$-$16.3424 is 0.06 (or 6\%) and $f_{\rm CB}=7\%$. The interpretation in this example is that   7\% of the simulated close binary systems will have simulated $P_{\rm RV}$ between 0.01 and 0.11 ($0.06 \pm 0.05$).
 
\subsubsection{ RV Variable Stars and Known Binary Systems}
\label{sec:binary}
\begin{figure}[tb]
\centering
\includegraphics[width=1.04\columnwidth]{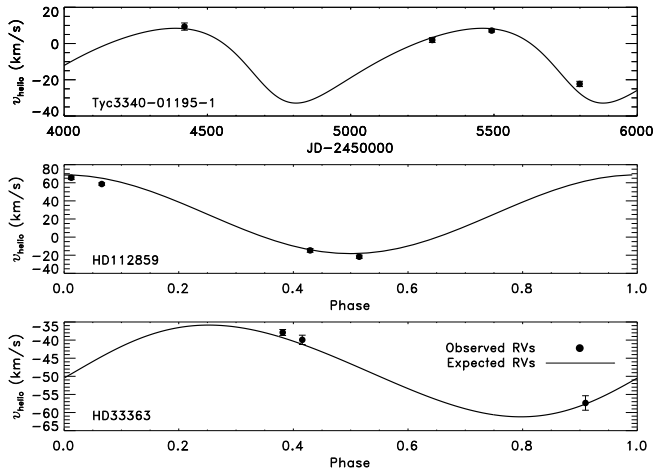} 
\caption{\label{fig:rvs} Measured heliocentric RVs (circles)  compared to the expected velocities from published orbital parameters (solid lines). The abscissa of the top panel is in days, while
the bottom two panels show orbital phase. }
\end{figure}

We find that seven of the rapid rotators have  RV variability that is inconsistent with the model of a single RV ($P_{\rm RV}=0\%$ in Table \ref{tab:RVs}).  This variability is suggestive of
orbital motion induced by a stellar companion.
One RV variable star (Tyc3340-01195-1) is a long-period binary star that was known as such before inclusion in this study. As was discussed in C11, the  separation of the stellar 
components is large enough that the enhanced rotation cannot be due to synchronous rotation.  Thus, the rapidly rotating nature of this star is just as unusual as that of the other single red giant stars in this study.  In Figure \ref{fig:rvs}, we compare our RV measurements of Tyc3340-01195-1 to the RVs predicted from the known orbital parameters provided by \cite{pourbaix04}, and we find that our measurements match the predicted curve quite well. 
Two other RV variable stars (HD112859 and HD33363) were included in this study with the expectation that they were single giant stars, but we can confirm their binary nature in the 
literature.  HD33363 appears in the ``9th Catalog of Spectroscopic Binaries'' \citep{pourbaix04}, while the binary orbit of HD112859 was recently characterized for the first time by 
\cite{griffin09}.  Our measured RVs are compared to the expected RVs calculated from the literature orbital parameters in the bottom two panels of Figure \ref{fig:rvs}. These latter 
two binary stars have periods on the order of only 20 days, and they are likely to be spinning synchronously with their stellar companions.   
The remaining four stars with variable RVs (G0804+39.4755, G1213+33.15558,  Tyc0347-00762-1, and Tyc6094-01204-1) have not been identified in the literature as binary stars 
before now.  The general good agreement between the observed and expected RVs in Figure \ref{fig:rvs} convinces us that this variability is real. At this time, however, we do not 
have enough data to constrain an orbit.  The shortest time baseline between RV epochs for these four stars is 200 days, so we also cannot distinguish whether these periods are short enough to be explained by synchronous rotation or long enough that the enhanced rotation is still unusual (as in the case of Tyc3340-01195-1). 

\subsubsection{RV Stable Stars}
\label{sec:RVgrps}
The eight remaining rapid rotators have non-zero $P_{\rm RV}$ ranging from 6\% (possibly RV stable) to 89\% (very likely RV stable).  As we just described in 
Section \ref{sec:binary}, we can use $f_{\rm CB}$ to estimate the likelihood that the star does in fact have a close-orbiting stellar companion that was missed because of the observing cadence. The largest $f_{\rm CB}$ of the apparently RV stable stars is 7\%, for G0827$-$16.3424.  
In other words, if we observed  a large sample of stars in close  binaries with the same cadence and observing precision as we have done for G0827$-$16.3424, we would only expect to measure 
$P_{\rm RV}=1$--11\% in 7\% of those systems. The total likelihood of this result is even smaller because 
the ``close binary'' designation applies to only $\sim 16\%$ of the binary systems  \citep{duquennoy91}, and only 14\% of the K giants are in binary systems \citep{famaey05}. Therefore, the total likelihood of G0827$-$16.3424 being in a close binary may be as low as $(0.07)(0.16)(0.14)= 0.2\%$.
Because G0827$-$16.3424 has the largest value of $f_{\rm CB}$, we conclude that we have observed our targets with sufficient cadence to detect close binary systems, and stars with $P_{\rm RV} > 0$ are unlikely to have binary companions with periods shorter than 500 days.  
Additional evidence of our completeness in detecting binaries is seen in the RV variable group ($P_{\rm RV}=0$).  For all of these stars, more than 84\%
 of the simulated close binaries have simulated $P_{\rm RV}<5\%$. This result can be interpreted as a $>84\%$ success rate in identifying close binaries with our observing cadence.   
Furthermore, the fact that we detect the RV variability of Tyc3340-01195-1  (a long period binary) suggests that some of the stars with $P_{\rm RV}=0$ may have companions that are too distant to cause rapid rotation.

The two classes of stars relevant to our analysis are the rapid rotators that have RV variability that allows the possibility of close binary companions ($P\leq 500$~days), which we refer to as the ``close-binary'' or CB group) and those that are likely single or have only long period companions ($P>500$~days, which we refer to as the ``single/long-period'' or SLP group).   The former group contains 
    G0804+39.4755, G1213+33.15558, HD112859, HD33363, Tyc0347-00762-1, and Tyc6094-01204-1,
 while the latter includes
  G0827$-$16.3424, G0928+73.2600, G0946+00.48, HD31993, HD34198,Tyc0647-00254-1, Tyc2185-00133-1, Tyc3340-01195-1, and Tyc5904-00513-1.

\subsection{Lithium}
\label{sec:Liabun}
\begin{figure}[t!]
\centering
\includegraphics[width=1.03\columnwidth]{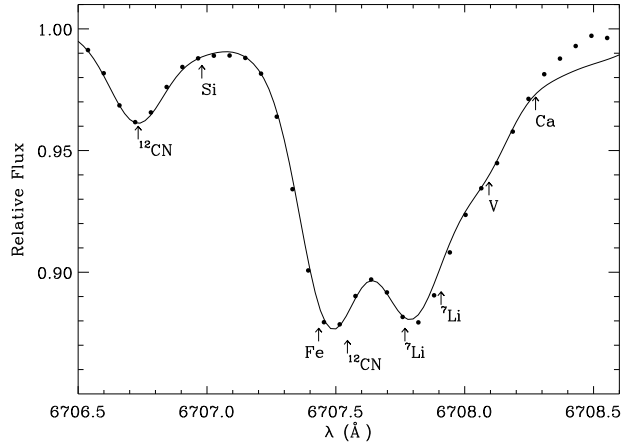} 
\caption[Example fit of a MOOG-generated synthetic spectrum]{An example fit of a MOOG-generated synthetic spectrum (solid line) to the observed spectrum of HD122430 (circles) with the location of some spectral features identified. All of the labeled atomic and molecular species are neutral.}
\label{fig:lifit2}
\end{figure}

To measure  Li abundances, we use  MOOG 
to generate synthetic stellar spectra in a $\sim5$\,\AA\ window spanning the  Li lines near  $\lambda=6708$\,\AA. The blending of the Li lines with  nearby  lines of CN, Fe, V, and Ca (see Figure \ref{fig:lifit2}) necessitates the use of spectral synthesis over 
the simple equivalent width 
abundance measurements used for the Fe lines.  
We compiled a line list  to compute the synthetic spectra  from a variety of sources including the VALD 
\citep{vald},  \cite{mandell04}, and \cite{ghezzi09}. The rest wavelengths, $\log gf$ values, and excitation potentials for the elements represented in this  spectral region  are listed in   Table \ref{tab:synli}.  

\begin{deluxetable}{rrrrr}
\tablewidth{0pc}
\tabletypesize{\scriptsize}
\tablecaption{Line list for fitting \lli \label{tab:synli}}
\tablehead{
   \colhead{$\lambda$} &
   \colhead{Species} &
   \colhead{$\chi$} &
   \colhead{$\log gf$} \\
      \colhead{(\AA)} &
   \colhead{  } &
   \colhead{(eV)} &
   \colhead{(dex)}  }
\startdata
  6706.548 &  $^{12}$CN   &  3.130 &  $-$1.359   \\ 
  6706.567 &  $^{12}$CN   &  2.190 &  $-$1.650   \\ 
  6706.657 &  $^{12}$CN   &  0.870 &  $-$3.001   \\ 
  6706.733 &  $^{12}$CN   &  0.870 &  $-$1.807   \\ 
  6706.844 &  $^{12}$CN   &  1.960 &  $-$2.775   \\ 
  6706.863 &  $^{12}$CN   &  2.070 &  $-$1.882   \\ 
  6706.880 &  \ion{Fe}{2}   &  5.956 &  $-$4.504   \\
  6706.980 &   \ion{Si}{1}   &  5.954 &  $-$2.797   \\
  6707.205 & $^{12}$CN    &  1.970 &  $-$1.222   \\  
  6707.282 & $^{12}$CN    &  2.040 &  $-$1.333   \\  
  6707.371 & $^{12}$CN    &  3.050 &  $-$0.522   \\  
  6707.433 &  \ion{Fe}{1}    &  4.608 &  $-$2.288   \\
  6707.457 & $^{12}$CN    &  0.790 &  $-$3.055   \\  
  6707.470 & $^{12}$CN    &  1.880 &  $-$1.451   \\  
  6707.473 &  \ion{Sm}{2}    &  0.933 &  $-$1.477   \\
  6707.518 &  \ion{V}{1}   &  2.743 &  $-$1.995   \\
  6707.545 & $^{12}$CN    &  0.960 &  $-$1.548   \\  
  6707.595 & $^{12}$CN    &  1.890 &  $-$1.851   \\  
  6707.596 &  \ion{Cr}{1}   &  4.208 &  $-$2.767   \\
  6707.645 & $^{12}$CN    &  0.960 &  $-$2.460   \\  
  6707.740 &  \ion{Ce}{2}   &  0.500 &  $-$3.810   \\
  6707.752 &  \ion{Sc}{1}   &  4.049 &  $-$2.672   \\
  6707.756 &  $^7$\ion{Li}{1}    &  0.000 &  $-$0.428   \\
  6707.768 &  $^7$\ion{Li}{1}    &  0.000 &  $-$0.206   \\
  6707.771 &  \ion{Ca}{1}   &  5.796 &  $-$4.015   \\
  6707.807 & $^{12}$CN    &  1.210 &  $-$1.853   \\  
  6707.848 & $^{12}$CN    &  3.600 &  $-$2.417   \\  
  6707.899 & $^{12}$CN    &  3.360 &  $-$3.110   \\  
  6707.907 &  $^7$\ion{Li}{1}    &  0.000 &  $-$1.509   \\
  6707.908 &  $^7$\ion{Li}{1}    &  0.000 &  $-$0.807   \\
  6707.919 &  $^7$\ion{Li}{1}    &  0.000 &  $-$0.807   \\
  6707.920 &  $^7$\ion{Li}{1}    &  0.000 &  $-$0.807   \\
  6707.930 & $^{12}$CN    &  1.980 &  $-$1.651   \\  
  6707.964 &  \ion{Ti}{1}   &  1.879 &  $-$6.903   \\
  6707.980 & $^{12}$CN    &  2.390 &  $-$2.027   \\  
  6708.023 &  \ion{Si}{1}   &  6.000 &  $-$2.910   \\
  6708.026 & $^{12}$CN    &  1.980 &  $-$2.031   \\  
  6708.094 &  \ion{V}{1}   &  1.218 &  $-$3.113   \\
  6708.147 & $^{12}$CN    &  1.870 &  $-$1.434   \\  
  6708.275 &  \ion{Ca}{1}   &  2.710 &  $-$3.377   \\
  6708.375 & $^{12}$CN    &  1.979 &  $-$1.097   \\  
  6708.499 & $^{12}$CN    &  1.868 &  $-$1.423   \\  
  6708.577 &  \ion{Fe}{1}    &  5.446 &  $-$2.728   \\
  6708.635 & $^{12}$CN    &  1.870 &  $-$1.584   
  \enddata
\end{deluxetable}

In Figure \ref{fig:lifit2}, we show an example synthetic spectral fit to the spectrum of HD122430 with the positions of major spectral features marked. 
When fitting, \lli\ was varied in intervals of 0.01 dex, the smallest abundance change specifiable in MOOG. 
Additionally, we varied the abundances of C and N (keeping the ratio C/N constant) as well as Fe, Ca, Si, and V.  Small adjustments were also allowed in the RV, overall continuum scaling, and line broadenings to get the best fits. The quality of the fits was estimated using MOOG's plot of the difference between the observed and computed spectra. Although the total number of free parameters could be numerous, we introduced a new free parameter (such as a change in the Si abundance or a broadening adjustment) only if the spectra could not be well fit without it.
For 32 stars in our sample, only one of which is a rapid rotator, we can only derive upper limits for the Li abundance. 

There are two major sources of uncertainty in the \lli\ measurements. The first comes from the somewhat subjective decision of what constitutes the best fit to the data. This uncertainty was estimated by seeing how much  \lli\ could vary before the model was no longer a good fit to the data, and we refer to this uncertainty as $\delta$\lli.  The second source of uncertainty comes from propagating the errors in the stellar parameters to errors in \lli.  This propagation could not be computed analytically because of the complex relationship between the stellar structure (as specified in the model stellar atmospheres)  and the resulting line profile, and it would be excessively time-consuming to remeasure \lli\ for each star on a grid of different stellar atmosphere models in the same manner that we made the initial \lli\ measurements. Instead, we used  
MOOG's {\it abfind } driver to find the  $W_\lambda$ of the Li lines  that correspond to our measured \lli\ from spectral fitting.   We then rerun
MOOG keeping  $W_\lambda$ fixed while changing the stellar atmosphere models within the errors of the stellar parameters to measure the variations of \lli. 
The $T_{\rm eff}$ and $\xi$  errors were estimated to be 100\,K and 0.5\,\kms\ for all stars. The errors in $A\rm{(Fe)}$ and $\log g$ are represented by $\sigma_{A{\rm (Fe\,I)}}$  and $\sigma_{A{\rm (Fe\,II)}}$\footnote{Recall from Section \ref{sec:teffetc} that we found that changing $\log g$ of the stellar atmosphere models by $\Delta \log g$ effected a change in $A$(\ion{Fe}{2}) that was $\sim -\Delta \log g$.}
 from Table  
\ref{sample:stellparam}, respectively. For most stars, only the variation in $T_{ \rm eff}$ yielded variations in \lli\  greater than 0.01\,dex (the output abundance precision of MOOG).
The total formal error for each star, $\epsilon_{A({\rm Li})}$, 
is the quadrature sum of the propagated  stellar parameter error  and $\delta$\lli. 

Finally, these  Li abundances, which were derived in local thermodynamic equilibrium (LTE),  were corrected for non-LTE (NLTE) departures using the grid of corrections supplied in \cite{lind09}.   The grid points include $4000 \leq T_{\rm eff} \leq 8000$\,K with $\Delta T_{\rm eff}=500$\,K, $1 \leq \log g \leq 5$ with $\Delta \log g =1$, 
$-3.0\leq$~[Fe/H]~$\leq0.0$ with $\Delta$[Fe/H]$=1.0$, $-0.3\leq$~\lli~$\leq 4.2$\,dex with $\Delta$\lli~$=0.3$, and $\xi=$ 1, 2, and 5\,\kms. We used simple linear 
interpolation to estimate corrections between the Lind grid points to match the stellar properties of $T_{\rm eff}$, $\log g$, [Fe/H], and \lli$_{\rm LTE}$, and
we  chose to use the grid that most closely matched each star's $\xi$ (instead of interpolating between $\xi$).  
A number of stars had stellar parameters outside of the grid, most commonly for having \lli$_{\rm LTE}<-0.3$\,dex.  However, we noticed that when all other stellar parameters were held constant, the NLTE corrections tended toward a constant value for small \lli$_{\rm LTE}$.  Therefore, we used the corrections supplied for  \lli$_{\rm LTE}=-0.3$\,dex for these low lithium stars. 
Similarly, we used the corrections computed at [Fe/H]$=0.0$ for the 13 stars with super-solar metallicity and the corrections computed for $\log g=1$ for Tyc0913-01248-1, which has $\log g=0.9$. 

However, we found that the  \cite{lind09} NLTE correction grids were not completely filled over the stellar parameter ranges described above, which led to a problem for 44 stars using the $\xi=1$\,\kms\ grid.   The interpolations would have included   ``empty'' NLTE corrections  in multiple dimensions of the grid. The problem was rectified
for 43 of these problem stars by using  the $\xi=2$\,\kms\ grid.   Over the stellar parameter range covered by our target stars, the NLTE corrections in the $\xi=1$\,\kms\  and $\xi=2$\,\kms\  grids (when corrections existed for both) differed by only 0.008\,dex.  For the remaining problem star, HD177830, we simply report the LTE abundance. 
Our final Li measurements are given in the first five columns of Table \ref{tab:li_carbon}, which lists the star name, the measured LTE abundance, $\delta$\lli,  $\epsilon_{A({\rm Li})}$, and the NLTE-corrected \lli\  for each of our program stars.  Upper limits  to \lli\ are indicated with ``$<$'' in the $\delta$\lli\ column. 
\begin{deluxetable}{lrrrrr}
\tablewidth{\columnwidth}
\tablecaption{LTE and NLTE \lli\ and \cratio\ of the program stars \label{tab:li_carbon}}
\tablehead{
  \colhead{Star}&\colhead{\lli$_{\rm L}$}&\colhead{$\delta$\lli}&\colhead{$\epsilon_{A({\rm Li})}$}&\colhead{\lli$_{\rm N}$}&\colhead{\cratio} \\
\colhead{ }&\colhead{(dex)}&\colhead{(dex)}&\colhead{(dex)}&\colhead{(dex)}&\colhead{ }    }
\startdata
Arcturus\_2007	& $-1.23$ &  $<$   & 0.17  & $-0.93$  &   7.5$\pm$0.3  \\ 
Arcturus\_2008  & $-0.73$ &  $<$   & 0.17  & $-0.43$  &    6.9 $\pm$0.5  \\ 
G0300+00.29     & $-0.19$ &   0.2  & 0.25  & $+0.05$  &   40.3 $\pm$  5  \\ 
G0319+56.5830   & $+0.01$ &  $<$   & 0.11  & $+0.14$  &   17.7 $\pm$  3.6  \\ 
G0319+56.6888   & $-0.61$ &  $<$   & 0.13  & $-0.41$  &   14.0 $\pm$  3  \\ 
G0453+00.90     & $+0.36$ &  0.02  & 0.13  & $+0.55$  &   21.5 $\pm$  5    \\ 
G0639+56.6179   & $-0.99$ &  $<$   & 0.14  & $-0.78$  &   19.0 $\pm$  1.2  \\ 
G0653+16.552    & $+0.35$ &  0.01  & 0.16  & $+0.70$  &   11   $\pm$  3  \\ 
G0654+16.235    & $+0.13$ &  0.01  & 0.16  & $+0.43$  &   11.9 $\pm$  1.5  \\ 
G0804+39.4755   & $+1.24$ &  0.01  & 0.13  & $+1.41$  &        $>30$       \\ 
G0928+73.2600\tablenotemark{a}   & $+3.62$ &  0.05  & $0.20$ & $3.30$ &    28  $\pm$  8    \\
HD31993\tablenotemark{b}	& $+0.98$ &  0.1   & 0.19  & $+1.03$  &   13.6 $\pm$  2    	
\enddata
\tablenotetext{a}{See \cite{carlberg10b} for details of the \lli\ and \cratio\ measurement.}
\tablenotetext{b}{Using literature stellar parameters from \cite{castilho00}.}
\tablecomments{This table is available in its entirety in a machine-readable form in the electronic edition of \apj. A portion is shown here for guidance regarding its form and content. }
\end{deluxetable}

 As a check of our Li measurements, we compare our measurements to published values for eight stars, as listed in Table \ref{compLi}. 
 Because the literature \lli\ were derived under the assumption of LTE, we use our LTE-derived values for comparison. The literature sources also
 each provide \vsini\ so we include a \vsini\ comparison in Table \ref{compLi}  as well. 
 Only four  stars have literature \lli\ measurements that are not upper limits.  Of these, all of the Li measurements agree within the
 quoted uncertainties.
\tabletypesize{\scriptsize}
\begin{deluxetable}{lrrrr} 
\tablewidth{\columnwidth}f
\tablecaption{Comparison of \lli$_{\rm LTE}$ measurements. \label{compLi}}
\tablehead{ & \multicolumn{2}{c}{This Work} & \multicolumn{2}{c}{Literature} \\
\colhead{Name} &\colhead{\lli} & \colhead{\vsini}& \colhead{\lli} & \colhead{\vsini}}
\startdata
HD34198  & 0.08$\pm$0.2 & 17.3 & 0.4$\pm$0.2  & 18.7\tablenotemark{a} \\
HD33798  & 1.66$\pm$0.1 & 28.0 & 1.5$\pm$0.2 & 30\tablenotemark{b} \\
HD31993  & 0.98\tablenotemark{c},1.80$\pm$0.2 & 30.4 & 1.4$\pm$0.2 & 31.1\tablenotemark{a} \\
HD108255 &$<-0.49$ &0.1 & $<+0.6$\tablenotemark{d} & 1.4\tablenotemark{b} \\
HD115478 &$-0.36\pm0.2$ &0.0 & $-0.4$\tablenotemark{d} & 1.7\tablenotemark{b} \\
HD116010 &$<-1.5$ &0.5 & $<-0.6$\tablenotemark{d} & 1.4\tablenotemark{b} \\
Arcturus &$<-0.73$ &2.0 & $<-0.8$\tablenotemark{d} & $<1.0$\tablenotemark{b} \\
HIP81437 &$-0.03\pm0.2$ &4.2 & $<+0.2$\tablenotemark{d} & 8\tablenotemark{e}
\enddata
\tablenotetext{a}{\cite{bohm04}.}
\tablenotetext{b}{\cite{demed99}.}
\tablenotetext{c}{Our derivation using stellar parameters from \cite{castilho00}.}
\tablenotetext{d}{\cite{brown89}.}
\tablenotetext{e}{\cite{massarotti08a}.}
\end{deluxetable}

\subsection{Carbon Ratio}
\label{sec:cratio}
The measurement of \cratio\ comes from fitting a small group of lines in the spectral region between 8001 and 8005\,\AA, where synthetic spectra are again generated with MOOG.   
The line list used for spectral synthesis is presented in Table  \ref{carbonlines}, which lists each line's wavelength, species, excitation potential, $\log gf$, and the primary references.  
 The CN wavelengths are taken from the laboratory measurements
by \cite{davis63} for $^{12}$CN and \cite{wyller66} for $^{13}$CN.
The $gf$ values were calculated based on the absolute $f$ values from the
analysis of \cite{sneden82} and the adopted dissociation energy
is 7.65 eV.
The atomic line list was complied using VALD \citep{vald}.  The $\log gf$ value of the \ion{Fe}{1} lines at 8002.567\,\AA\ and 8003.227\,\AA\ are  reduced by $\sim0.6$\,dex from the values found in VALD.  Without this reduction, the models predict significantly stronger absorption at these wavelengths than what is present in the observed spectra---both in the atlas solar spectrum and our red giant spectra. 
\begin{deluxetable*}{lrrrr|lrrrr}
\tablecaption{Line list for fitting \cratio \label{carbonlines}} 
\tablewidth{0pc}
\tabletypesize{\scriptsize}
\tablehead{
   \colhead{$\lambda$} &
   \colhead{Species} &
   \colhead{$\chi$} &
   \colhead{$\log gf$} &
   \colhead{Reference} &
      \colhead{$\lambda$} &
   \colhead{Species} &
   \colhead{$\chi$} &
   \colhead{$\log gf$} &
   \colhead{Reference} \\
    \colhead{(\AA)} &
   \colhead{  } &
   \colhead{(eV)} &
   \colhead{(dex)} &
   \colhead{  }   &
   \colhead{(\AA)} &
   \colhead{  } &
   \colhead{(eV)} &
   \colhead{(dex)} &
   \colhead{  }   }
\startdata
  7990.388  & $^{12}$CN     &   1.38  & $-$2.0585 &  7, 8 &       8002.367  & $^{13}$CN     &   1.49  & $-$1.8327  & 7, 9 \\    
  7990.790  & $^{12}$CN     &   1.45  & $-$1.6234 &  7, 8 &       8002.412  & $^{12}$CN     &   0.18  & $-$1.4962  & 7, 8 \\    
  7990.957  & $^{13}$CN     &   1.37  & $-$1.6861 &  7, 9 &       8002.571  & $^{13}$CN     &   0.21  & $-$1.7212  & 7, 9 \\    
  7991.128  & $^{13}$CN     &   0.05  & $-$2.1512 &  7, 9 &       8002.576  &  \ion{Fe}{1}   &  4.580  & $-$2.2400  &  2 \\         
  7991.128  & $^{13}$CN     &   0.18  & $-$1.7496 & 7, 9 &        8003.185  &  \ion{Al}{1}   &  4.087  & $-$1.8791  & 4  \\         
  7991.583  & $^{12}$CN     &   1.47  & $-$1.6216 &  7, 8 &       8003.213  & $^{12}$CN     &   0.12  & $-$1.9431  & 7, 8 \\    
  7991.711  & $^{13}$CN     &   0.04  & $-$2.2700 & 7, 9 &        8003.227  & \ion{Fe}{1}    & 5.539  & $-$2.3889   &  2 \\         
  7992.126  & $^{13}$CN     &   0.10  & $-$1.6364 & 7, 9 &        8003.311  & $^{13}$CN     &   1.34  & $-$2.0883  & 7, 9 \\    
  7992.297  & $^{12}$CN     &   0.09  & $-$2.0114 & 7, 8 &        8003.485  &  \ion{Ti}{1}   &  3.724  & $-$0.2000  & 1  \\         
  7992.297  & $^{12}$CN     &   0.18  & $-$1.5143 & 7, 8&        8003.553  & $^{12}$CN     &   0.31  & $-$1.6440  & 7, 8 \\    
  7993.144  & $^{13}$CN     &   0.21  & $-$1.7423 &  7, 9 &       8003.910  & $^{12}$CN     &   0.33  & $-$1.6478  & 7, 8 \\    
  7993.600  &  \ion{Ti}{1}   &  1.873  & $-$2.4970 & 1  &             8004.036  & $^{12}$CN     &   0.06  & $-$2.9245  & 7, 8 \\        
  7993.869  & $^{13}$CN     &   1.49  & $-$1.8508 & 7, 9 &        8004.550  & $^{13}$CN     &   0.12  & $-$1.5918  & 7, 9 \\    
  7994.018  & $^{12}$CN     &   1.42  & $-$1.6326 & 7, 8 &        8004.715  & $^{13}$CN     &   0.07  & $-$2.0814  & 7, 9 \\    
  7994.312  & $^{13}$CN     &   0.09  & $-$1.6615 & 7, 9 &        8004.801  & $^{13}$CN     &   0.10  & $-$1.6144  &  7, 9 \\   
  7994.688  & $^{13}$CN     &   1.34  & $-$1.6536 & 7, 9 &        8005.248  &  Zr{1}  &  0.623  & $-$2.1901   & 4  \\         
  7994.694  & $^{12}$CN     &   0.11  & $-$1.9666 & 7, 8 &        8006.065  & $^{13}$CN     &   1.41  & $-$1.6517  &  7, 9\\   
  7995.015  & $^{12}$CN     &   0.16  & $-$1.5143 & 7, 8 &        8006.126  & $^{13}$CN     &   0.24  & $-$1.7122  &  7, 9 \\   
  7995.158  & $^{13}$CN     &   0.02  & $-$2.8996 &  7, 9 &       8006.459  &  \ion{Si}{1}  &  6.261  & $-$1.7231   & 5  \\         
  7995.640  & $^{12}$CN     &   0.06  & $-$2.9172 & 7, 8 &        8006.703  &  \ion{Fe}{1}  &  5.067  & $-$2.1280   &  2 \\         
  7995.640  & $^{12}$CN     &   0.29  & $-$1.6556 & 7, 8 &        8006.925  & $^{12}$CN     &   1.60  & $-$1.7878  & 7, 8 \\    
  7995.640  & $^{12}$CN     &   0.31  & $-$1.6615 & 7, 8 &        8007.211  & $^{13}$CN     &   1.48  & $-$1.2652  & 7, 9 \\    
  7996.435  &  \ion{Ti}{1}   &  3.337  &  0.2660 & 1  &             8007.242  &   \ion{Co}{1} &   4.146 &   0.1159  &  6 \\             
  7996.706  & $^{13}$CN     &   0.19  & $-$1.7352 & 7, 9 &        8007.582  & $^{12}$CN     &   0.11  & $-$1.9586  & 7, 8\\    
  7996.761  & $^{12}$CN     &   1.58  & $-$1.7986 &  7, 8 &       8007.882  & $^{13}$CN     &   0.03  & $-$2.8962  &  7, 9 \\   
  7996.807  & $^{13}$CN     &   0.04  & $-$2.2233 &  7, 9 &       8007.904  & $^{13}$CN     &   0.05  & $-$2.1415  &  7, 9 \\   
  7996.816  & \ion{Fe}{1}    & 4.584  & $-$2.4860  & 2  &             8008.387  &  \ion{Si}{1}  &  6.079  & $-$1.8289   &  5 \\            
  7997.334  & $^{13}$CN     &   1.39  & $-$1.6696 & 7, 9 &        8008.493  & $^{12}$CN     &   0.20  & $-$1.4802  &  7, 8\\   
  7997.800  & $^{13}$CN     &   0.06  & $-$2.1152 & 7, 9 &        8008.652  & $^{13}$CN     &   1.36  & $-$1.6421  &  7, 9 \\   
  7998.216  & $^{13}$CN     &   0.11  & $-$1.6144 & 7, 9 &        8008.737  & $^{13}$CN     &   0.22  & $-$1.7077  & 7, 9 \\    
  7998.312  & $^{12}$CN     &   1.54  & $-$1.2480 & 7, 8 &        8009.278  & $^{12}$CN     &   1.44  & $-$1.6216  & 7, 8 \\    
  7998.876  & $^{13}$CN     &   1.38  & $-$1.6696 & 7, 9 &        8009.703  & $^{12}$CN     &   1.62  & $-$1.7932  &  7, 8 \\   
  7998.944  &  \ion{Fe}{1}   &   4.371 &   0.1489 &  2 &            8010.084  & $^{12}$CN     &   0.19  & $-$1.4802  &  7, 8 \\        
  7999.214  & $^{12}$CN     &   1.40  & $-$2.0287  & 7, 8 &       8010.084  & $^{12}$CN     &   1.40  & $-$2.0200  &  7, 8 \\   
  7999.214  & $^{12}$CN     &   1.60  & $-$1.8041  & 7, 8 &       8010.468  & $^{13}$CN     &   0.11  & $-$1.5918  &  7, 9 \\   
  7999.408  & $^{13}$CN     &   0.09  & $-$1.6383  & 7, 9 &       8011.101  & $^{13}$CN     &   1.51  & $-$1.8210  &  7, 9 \\   
  7999.460  & $^{13}$CN     &   1.46  & $-$1.2644  &  7, 9 &      8011.159  & $^{13}$CN     &   0.14  & $-$1.5702  & 7, 9 \\    
  7999.465  & $^{13}$CN     &   0.22  & $-$1.7282  &  7, 9 &      8011.484  & $^{13}$CN     &   1.63  & $-$1.4828  &  7, 9 \\   
  7999.846  & $^{12}$CN     &   0.10  & $-$1.9830  &  7, 8 &      8011.732  & $^{12}$CN     &   0.33  & $-$1.6326  & 7, 8 \\    
  8000.261  & $^{12}$CN     &   0.19  & $-$1.4962  &  7, 8 &      8011.836  & $^{13}$CN     &   1.52  & $-$1.8268  & 7, 9 \\    
  8000.316  & $^{12}$CN     &   1.47  & $-$1.6091  & 7, 8 &       8011.899  & $^{13}$CN     &   0.08  & $-$2.0506  & 7, 9 \\    
  8000.757  &  \ion{Nd}{2}   &  1.091  & $-$1.2220  &  3 &           8011.950  & $^{12}$CN     &   0.13  & $-$1.9208  & 7, 8 \\         
  8001.369  & $^{13}$CN     &   0.03  & $-$2.8962  & 7, 9 &       8011.950  & $^{12}$CN     &   1.50  & $-$1.5952  & 7, 8 \\    
  8001.524  & $^{12}$CN     &   1.42  & $-$1.6253  &  7, 8 &      8012.546  & $^{12}$CN     &   0.35  & $-$1.6364  & 7, 8 \\    
  8001.652  & $^{12}$CN     &   1.48  & $-$1.6091  & 7, 8 &       8012.620  & $^{13}$CN     &   1.35  & $-$2.0675  & 7, 9 \\    
  8002.214  & $^{13}$CN     &   0.05  & $-$2.1805  & 7, 9 & & & & &
\enddata
 \tablerefs{(1) \citealt{kuruczCD20}; 
 (2) \citealt{kurucz94}; 
 (3) \citealt{blaise84};   
 (4) \citealt{kuruczCD18}; 
 (5) Kurucz 2007 from http://cfaku5.cfa.harvard.edu/atoms/1400/; 
 (6) \citealt{kurucz94b};  
 (7) \citealt{sneden82}
 (8) \citealt{davis63}; (9) \citealt{wyller66}.}
\end{deluxetable*}

The strengths of the molecular features we are fitting depend on the abundances of both C and N, and we allow both of these abundances to change. However, we fix 
the ratio of C to N to 1.5 \citep[following ][]{marcs08}, which is  appropriate for moderate CN processing 
in giant stars. This method allows us to add only one free parameter to describe C and N abundances that deviate from scaled-solar.
Fitting this red region of the spectrum is also complicated by the presence of numerous telluric features, mainly H$_2$O.  No telluric calibration stars (such as a rapidly rotating [\vsini~$>200$\,\kms]
O or B stars, which  have  nearly featureless continua) were observed with the target stars; however, the high quality atlas telluric spectrum from \cite{hinkle00} provides a good model.  Although it is common to  ``divide out'' the telluric features that are superimposed on  stellar spectra, we instead chose to add 
a telluric spectrum that is scaled, broadened (to maintain the instrument response), and RV shifted (to place it in the stellar rest frame)  to the synthetic spectrum models.
To account for the exponentially decreasing transmission of the stellar spectrum through the atmosphere at wavelengths of telluric absorption, i.e., applying Beer's Law, the scaled telluric spectrum $T(\lambda)$ is given as  
$T(\lambda)=T_0(\lambda)^f$, where $T_0(\lambda)$ is the broadened but  unscaled telluric spectrum  and $f$ is the scaling factor, a free parameter.

Other free parameters in the overall fit include small adjustments in the stellar broadening (generally within 2\,\kms\ each for the \vsini\ and $\zeta$), RV (usually less than 2\,\kms),  iron abundance (up to $\pm~1~\sigma_{A\rm(Fe\,I)}$), and continuum adjustments.
The continuum fitting is the most subjective and, in some cases, the most difficult part of the fitting procedure because of the high density of stellar lines.    Unlike the \lli\ fitting, where a global scaling factor was sufficient to get good fits to the continuum, we allow linear adjustments to be made. Higher order fits were considered, but ultimately it was decided that higher order adjustments were more likely to introduce errors rather than reduce them.   For stars with extreme rotation (\vsini~$\gtrsim15$\,\kms), there is no continuum  in the spectral region being fit. In these cases, we relied heavily on the models for guidance.

\begin{figure}[t]
\centering
\includegraphics[width=1.03\columnwidth]{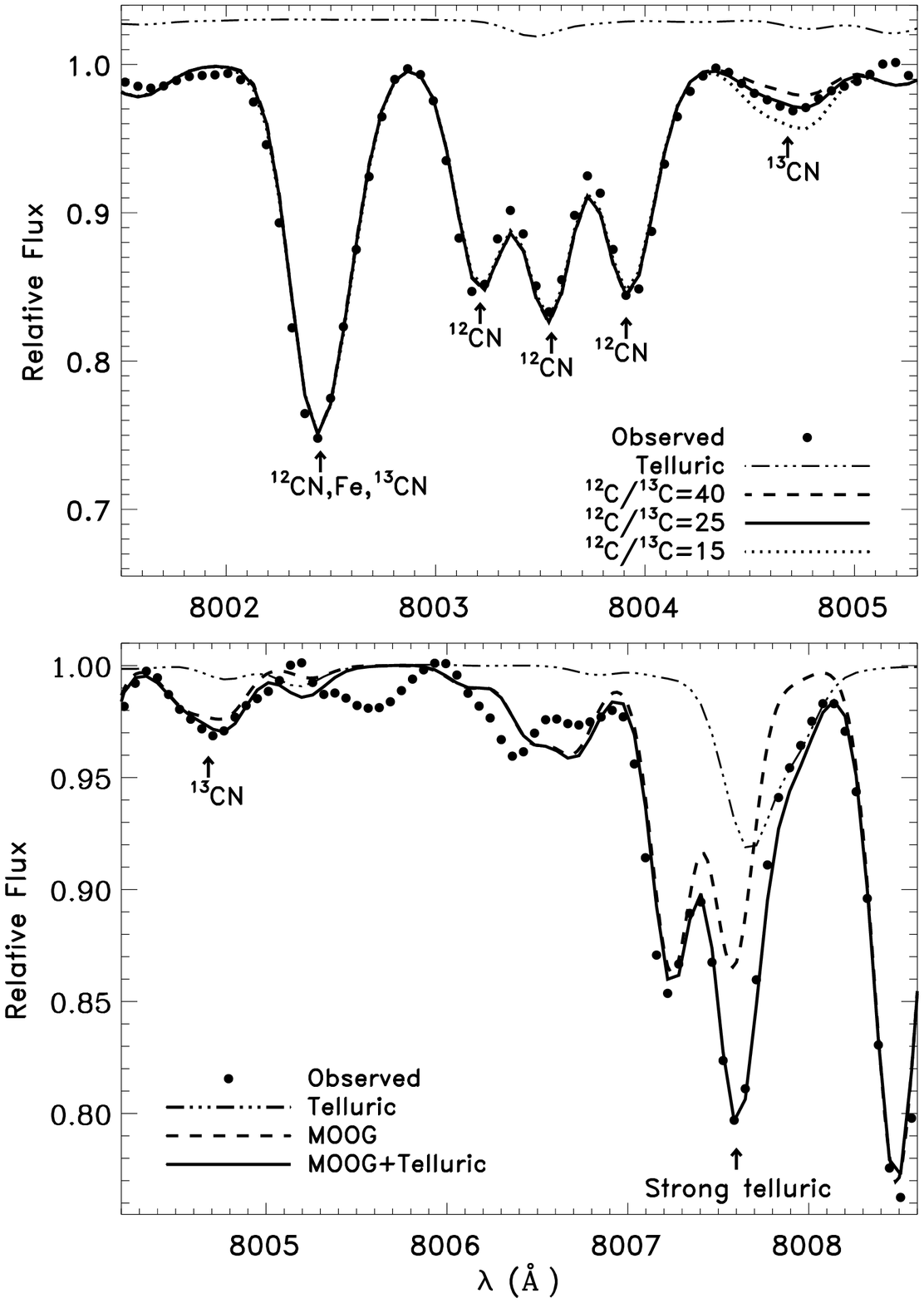} 
\caption{Top: sample fits to measure \cratio\ from the observed spectrum of G1130+37.9414 (circles).  The three synthetic spectra have \cratio\ 
of 40 (dashed), 25 (solid, best fit), and 15 (dotted). Major spectral features are labeled with the contributing atomic or molecular species. The weak $^{13}$CN lines near $\lambda=8004.7$\,\AA\ provide the leverage for measuring \cratio. 
The \cite{hinkle00} atlas telluric spectrum, broadened to match our instrumental resolution, is shown for reference (triple-dot-dashed line, offset vertically for clarity).  Bottom: redward of the  $^{13}$CN feature is a telluric line in the observed spectrum (circles) that is suitable for scaling the atlas telluric spectrum.  The lines show the best-fit \cratio$=25$ model (solid) and the contributions from the MOOG-generated stellar spectrum (dashed)   and the atlas telluric spectrum (triple-dot-dashed).}
\label{excn}
\end{figure}

A best-case example of measuring \cratio\ in our stellar spectra is shown in Figure \ref{excn}  for G1130+37.9414. In the top panel, the three synthetic spectra demonstrate  the best fit \cratio\ of 25 and \cratio\ that are both higher and lower than the best fit. Major spectral features are identified. The synthetic spectra include a telluric contribution, which is plotted separately for reference.  In the bottom panel of Figure \ref{excn}, we show the best fit decomposed  into the pure stellar spectrum that is generated with MOOG and the atlas telluric spectrum.  We used the strong telluric line indicated in that panel to find  the scaling parameter, $f$, for the  telluric spectrum.

In Table \ref{compCratio}, we compare our measured \cratio\ (from the technique described above) to published values for  five of our control stars and three of our program stars.
One control star, HD39853, had a strong telluric absorption feature fall precisely  at 8004.7\,\AA\ in the stellar rest frame, rendering \cratio\ unmeasurable. For the remaining stars, Table \ref{compCratio} demonstrates that we do tend to measure similar \cratio\ to those found in the literature. For example, with the exception
of Aldebaran, the ranking of stars from low to high \cratio\ would be the same using our measured \cratio\  or using the literature \cratio. 
However, the measurements often disagree by more than the quoted uncertainties, indicating that either we or the  literature sources are underestimating 
the uncertainty.   

We add the derived  \cratio\ and uncertainty  for our stellar sample in the final column of Table \ref{tab:li_carbon}.   The uncertainties for some of the stars come from
averaging several different attempts at fitting the line that all gave equally good results.  In these situations, the listed uncertainty is the standard deviation in the individual measurements.  In the absence of multiple measurements we assign  an uncertainty of $\pm$3 for \cratio~$<20$ and $\pm$5 for \cratio~$\geq 20$.   These choices come from the median uncertainties in the stars with multiple measurements. 
\begin{deluxetable}{lrr} 
\tablewidth{\columnwidth}
\tablecaption{Comparison of \cratio\ measurements \label{compCratio}}
\tablehead{ &  \multicolumn{2}{c}{\cratio} \\
\cline{2-3}
\colhead{Name} &\colhead{This Work} &\colhead{Literature } }
\startdata
Aldebaran\tablenotemark{a} & $19.7\pm3$ & 10\tablenotemark{b} \\ 
Arcturus & $7.2\pm0.2$ & 7\tablenotemark{c} \\ 
HD112127 & $34.4\pm2$ & 22\tablenotemark{d} \\
HD127665 &  $11.3\pm0.5$ &$15\pm2$\tablenotemark{e} \\
HD163588 & $25.8\pm2$  & $20\pm2$\tablenotemark{e} \\
HD216228\tablenotemark{a} & $11.4\pm0.4$  & $16\pm2$\tablenotemark{e} \\
HD39853   & Telluric contamination& 7\tablenotemark{d} \\
Pollux & $15.9\pm3$ & 16\tablenotemark{c} 
\enddata
\tablenotetext{a}{Used literature values for model atmosphere.}
\tablenotetext{b}{\cite{smith85}.}
\tablenotetext{c}{\cite{gilroy89}.}
\tablenotetext{d}{\cite{fekel93}.}
\tablenotetext{e}{\cite{dearborn75}.}
\end{deluxetable}

\section{Results: Evolution of Abundances and Rotation} 
\label{sec:placc}
In this paper, we are investigating the hypothesis that rapid rotators can be spun up by the accretion of planets by searching for the chemical signatures of   \lli\ replenishment and increased \cratio\ that should accompany the enhanced rotation under the planet accretion scenario. 
 If the enhanced rotation is {\it not} caused
by planet accretion, then we expect to find that either (1) \lli\ and \cratio\ will  not differ between the rapid rotator and slow rotators samples, or (2) if  rapid rotation  drives additional
mixing  conducive to lithium production, any  \lli\ enhancement will be accompanied by {\it lower} \cratio. 
We explore these scenarios  by comparing both the group abundance properties of  the slow and rapid rotators (Section \ref{sec:meanlicarbon}) and dividing the samples into common stellar evolution groups to disentangle the
effects of normal evolution abundance changes    (Section \ref{sec:stellev}).

\subsection{Mean Abundances of the Stellar Samples} 
\label{sec:meanlicarbon}
\begin{figure}[tb]
\centering
\includegraphics[width=1.03\columnwidth]{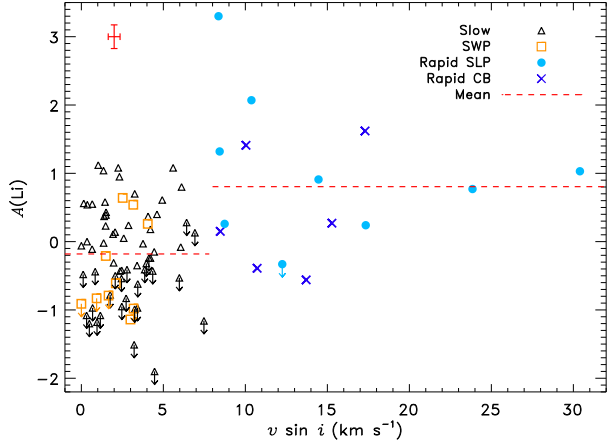} 
\caption{Plot of \lli\ as a function of \vsini\ for the slower rotators with and without known planets (triangles and squares, respectively),  rapid rotators that are single or in long-period binaries (SLP group, circles), and rapid rotators suspected to be in close binaries (CB group, $\times$).
 The horizontal dashed lines show the average abundances of the slow (\vsini$< 8$\,\kms) and rapid (\vsini$\geq 8$\,\kms) rotators. Upper limits are indicated with downward facing arrows. The sizes of the typical errors are shown in the upper left corner.}
\label{liplot}
\end{figure}

The first question we wish to address is whether the abundance distribution of light elements differs between the slow and rapid rotator samples.  
We plot the Li abundance as a function of projected rotational velocity in Figure \ref{liplot}.  This plot gives us the first evidence from our own data that there is a relationship between rotational velocity and \lli, as \cite{drake02} previously suggested.   
The average  \lli\ and standard errors in the mean for the  entire slow rotator and rapid rotators samples are $-0.18 \pm 0.08$ and $+0.81\pm0.27$, respectively, and these mean values (without the errors) are indicated on the plot. The average abundances include the upper limit measurements. If the SWPs are excluded from the slow rotator sample, the mean abundance increases to $-0.14\pm\ 0.09$.
Five of the fifteen rapid rotators have \lli\ exceeding the {\it maximum} \lli\ of all of the slow rotators. A two-sided Kolmogorov Smirnov (K-S) test reveals that the probability that the lithium abundances of the slow and  rapid rotators are from the same parent population is only 0.6\%. 
We can subdivide the rapid rotators into the two RV groups defined in Section \ref{sec:RVgrps}: the CB group, which contains the stars that may be synchronously rotating with binary companions,  and the SLP group, which require some other explanation (such as planet accretion) for the rapid rotation.   The CB group has a mean \lli\ of $+0.41\pm0.37$~dex, which is much closer to the mean \lli\  of the slow rotators than the \lli\ of the entire ensemble of rapid rotators is to the slow rotators. The CB group is small enough that it has a comparable probability of being drawn from either the slow rotator sample (48\%) or the RV stable, rapid rotator sample (53\%). 
The SLP group, in contrast, has a mean \lli\ of  $+1.06\pm0.36$~dex, which is even more Li enhanced than the entire rapid rotator ensemble. The K-S probability is 0.5\% when comparing the distribution of \lli\ of the SLP group to the slow rotators.

\begin{figure}[tb]
\includegraphics[width=1.03\columnwidth]{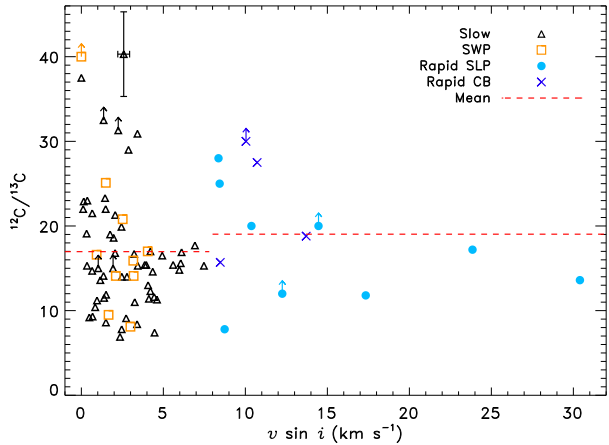} 
\caption{Plot of \cratio\ as a function of \vsini\ for the slow and rapid rotators, using the same symbols as in Figure \ref{liplot}.  The dashed line  indicates the mean \cratio\ of the slow and rapid rotators. Lower limits are indicated with upward facing arrows. A typical error bar is shown on one of the slow rotators. }
\label{cratiovrot}
\end{figure}

We make a similar comparison of \cratio\ in our stellar samples by plotting \cratio\ against \vsini\ in Figure \ref{cratiovrot}. 
The average \cratio\ of the slow and rapid rotators subsamples are overplotted. The rapid rotators again seem to show the  enhancement
expected from planet accretion; they have   $\overline{^{12}{\rm C}/^{13}{\rm C}}=19.0\pm1.9$  
compared to the slow rotators that have   $\overline{^{12}{\rm C}/^{13}{\rm C}}=17.0\pm0.9$. 
By again subdividing the rapid rotators, we find that the nine stars in the SLP group have  $\overline{^{12}{\rm C}/^{13}{\rm C}}=17.3\pm2.2$. Only four of the  CB 
stars have measurable \cratio, and the average of those stars are $23.0 \pm 3.4$.    However, because of the larger relatively uncertainties in the \cratio\ measurements,  K-S tests of the \cratio\ distributions show no statistically significant differences
between the slow rotators, SWPs, RRs, and subsamples of RRs. 

\subsection{Common Evolution Groups}
\label{sec:stellev}
Both the strength and weakness in the above comparison is that it averages over the normal variations in the abundances  to probe how rotational velocity is potentially correlated to the abundances. 
However, there are well-known  changes that are expected in these elements as the stars evolve through the red giant phase. Our slow and rapid rotators are not distributed equally across the various stages of RGB evolution; therefore,
it is worth repeating the comparison of \lli\ and \cratio\ with \vsini\ in subgroups of similarly evolved stars.

\begin{figure*}[th] 
\centering
\includegraphics[width=0.7\textwidth]{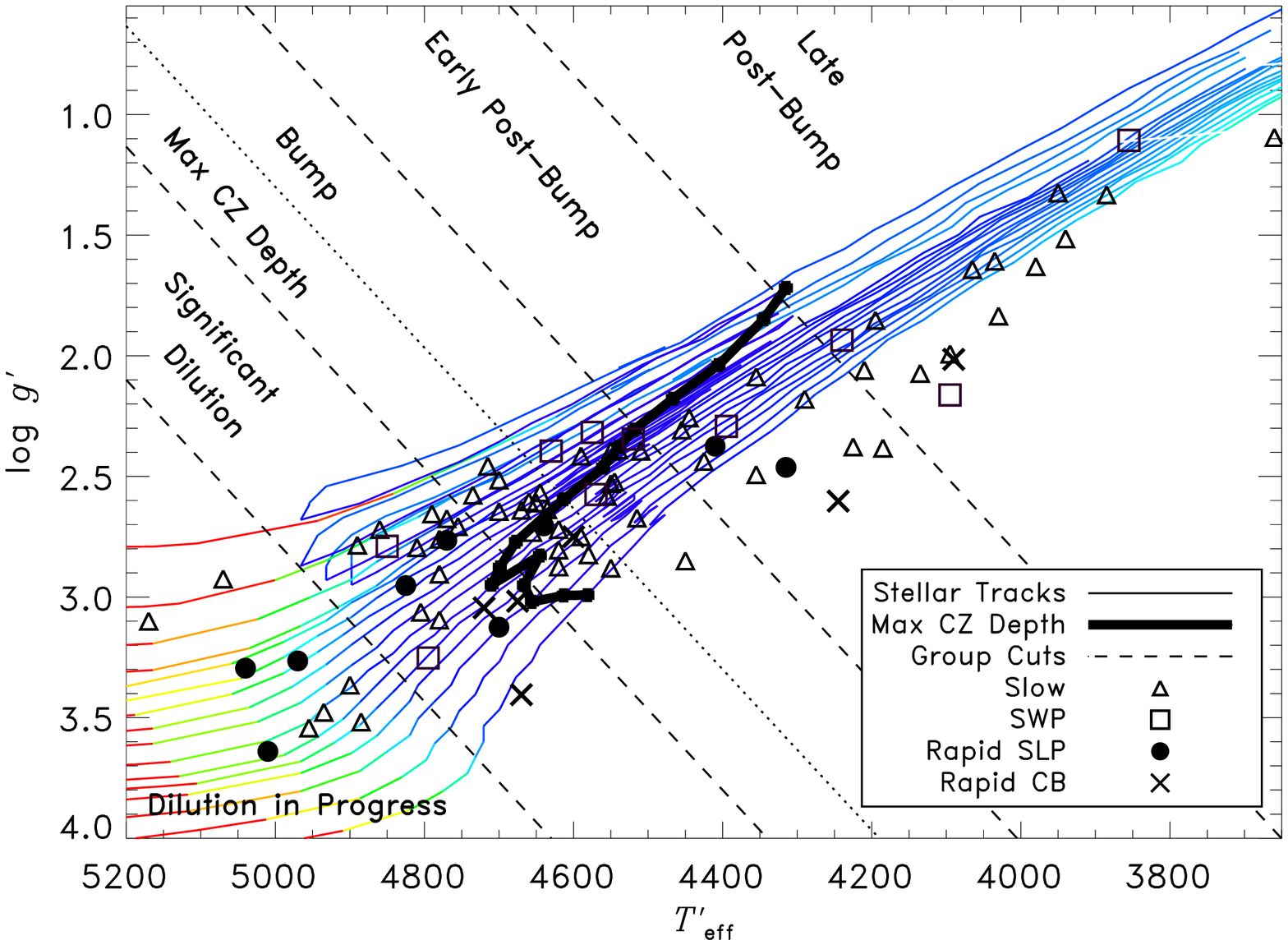}   
\caption{Illustration of our selection of stellar evolution groups. The ``primed''   \teff\ and \logg\ are the observed values plus a metallicity-dependent shift applied to align them to  solar metallicity evolution tracks (see the text for details). The thin solid lines are \cite{giard00} solar metallicity  evolution tracks for  masses between
0.8~\msun\ (bottom track) and 3.0~\msun\ (top track). Evolution progress along each track from  lower left to upper right.  The color scale of the models illustrates the relative depth of the convection zone (CZ) from shallow (light gray or red) to deep (dark gray or blue).  
The heavy solid line marks the stage at which  the maximum CZ depth is achieved.  Our stellar sample is plotted with the same symbols identified in Figure \ref{liplot}. The dashed  lines were chosen by eye to group stars of similar evolutionary stages.  The dotted line is used to isolate the group of stars that fall on the luminosity bump.  The bump can be identified as where lower mass tracks briefly double back in the direction of evolution. \label{model_dilution}}
\end{figure*}

 We define evolution groups by comparing our data to model evolution tracks.  In Figure \ref{model_dilution}, we plot  \cite{giard00}  solar metallicity stellar evolution tracks  on  a modified HR diagram (replacing luminosity with $\log g$) for the mass range of 0.8--3.5\,\msun.  The color scale of the evolution tracks indicate the relative depth of the convection zone ranging from the shallower
 depth the convection zones have on the subgiant branch to the  full FDU depth. 
Major stages of stellar evolution---the first ascent RGB phase,  the luminosity bump, and the post RGB-tip evolution---shift to cooler temperatures and larger surface gravities at super-solar metallicities (and vice versa for sub-solar metallicities).  
The rapid rotator sample is small enough that grouping stars first by similar metallicity and then by evolution stages would generate subsamples too small to make useful comparisons. 
Instead, to compare our stars to these solar-metallicity tracks, we  offset the observed \teff\ and $\log g$ of our program stars by metallicity-dependent factors to create $T_{\rm eff}' = T_{\rm eff}+ {\rm [Fe/H]}\times500\,{\rm K\,dex}^{-1}$ and $\log g'=\log g-0.25\times {\rm [Fe/H]}$.  These ``primed'' stellar parameters represent
the temperatures and gravities the target stars would have if they were solar-metallicity stars. The offsets were found by comparing \cite{giard00}  evolution tracks of $Z=0.004$, 0.008, 0.019, and 0.030.  
Stellar evolution groups are defined by making cuts perpendicular to the general direction that stars evolve through the $T_{\rm eff}'$--$\log g'$ plane, preferably in locations where there were gaps in the distribution of our program stars.  These cuts were done by eye, and are indicated by the dashed lines in Figure \ref{model_dilution}.  One additional cut, indicated by the dotted line, was made to separate stars that are at the luminosity bump from those stars  have not yet reached this important stage. None of the rapid  rotators fall into the luminosity bump class. 

The groups are labeled  according to the normal stellar evolutionary processes affecting the \lli\ and \cratio\  of solar-mass stars at that stage. These groups (from bottom left to upper right in Figure \ref{model_dilution}) are as follows.
	\begin{description}
	\item[Dilution in Progress.] Stars are still undergoing Li dilution. These stars should have the largest \lli\ and \cratio\ in normal stellar evolution.
	\item[Significant Dilution.] The convection zone is near its maximum depth; therefore,  Li dilution is essentially complete. $^{13}$C is being dredged-up, and \cratio\ is falling.
	\item[Max CZ Depth.] The convection zone (CZ) has reached its maximum depth and is receding. \lli\  and \cratio\  are at standard dilution values.  The H-burning shell is still interior to the maximum CZ depth. 
	\item[Bump.]  The H-burning shell reaches the maximum CZ depth, and non-convective mixing can occur. $^7$Be created by the Cameron Fowler chain may be carried into the convection zone where it decays to $^7$Li, which is {\it not} destroyed. There is a potential for large \lli\ at this stage, perhaps even exceeding that seen in the ``Dilution in Progress'' group.  \cratio\ remains constant at the standard value. 
	\item[Early Post-Bump.] Stars have evolved just beyond bump. Non-convective mixing may be reducing both \lli\ and \cratio. Li can be either rather high (if Li was briefly regenerated) or very low (if no Li was regenerated). \cratio\ may be dropping.
	\item[Late Post-Bump.] Low to very low values of \lli\ and \cratio\ depending on the amount of extra mixing.
	\end{description}

\begin{figure*}[th] 
\centering
\includegraphics[width=0.7\textwidth]{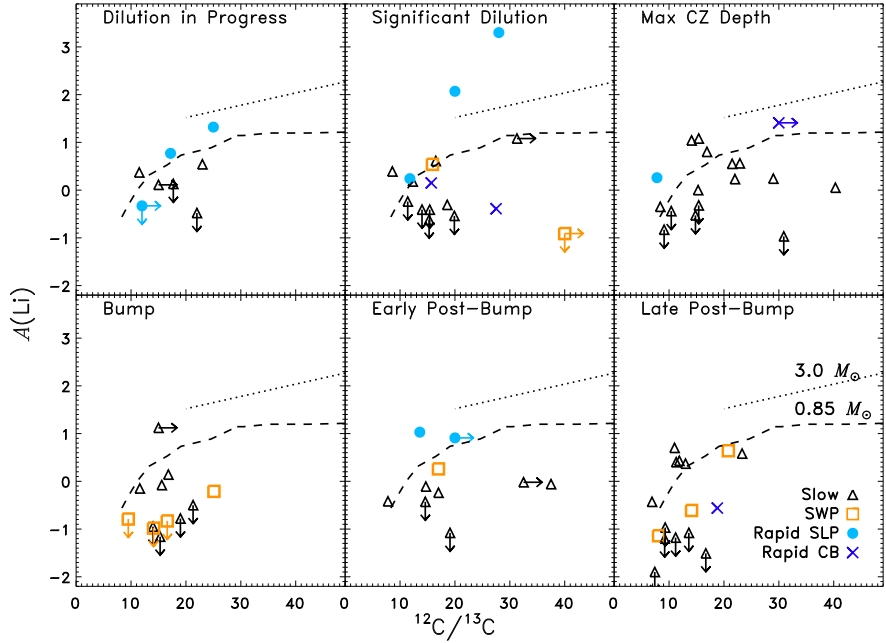}
\caption{\lli\ plotted against \cratio\  for the slower rotators (triangles), giants with known planets (squares), the SLP group of rapid rotators (circles), and the CB group ($\times$). Each panel shows a stellar evolution group, progressing from less evolved to more evolved (left to right, then top to bottom). Upper limits in \lli\ and lower limits in \cratio\ are indicated with downward and rightward facing arrows, respectively.  
The lines show the expected relationship of these two properties for a 3\,\msun\ star (dotted), for  which  only standard dilution is expected \citep[e.g., ][]{iben67,lambert80}, and for a low-mass star (0.85\,\msun, dashed line), for which  both standard dilution and  canonical extra-mixing are expected \citep{denissenkov03}. }
\label{ali_cratio_new}
\end{figure*}

In Figure \ref{ali_cratio_new}, we plot   \lli\ versus \cratio\ for these six  stellar groups along  with two models  of how these properties should change in stars of different masses.  
The first model star is 3\,\msun, for which only standard dilution is  expected to 
affect the abundances \citep{iben67,lambert80}. In this model, the \lli\ at the end of the  MS  evolution is $\sim 3$~dex.  The second model is for a 0.85\,\msun\ star for which standard dilution and canonical extra-mixing are expected to 
occur \citep{denissenkov03}. The \lli\ at the end of the MS for this model is $\sim 2.4$~dex. 
We  expect the stars in the first three evolution groups to  fall  between the two model lines in the upper right portion of each panel.  For the remaining three evolution groups, we expect any higher mass stars in our sample to remain in the upper
right region of the plot (near the dotted line), while lower mass stars should move down and left along the low mass model. 
Our data, however, do not follow these expectations.  In all of the subgroups, there are no clear correlations between \lli\ and \cratio\ 
 (though, one should remember that the errors in \cratio\ are often large: typically $\pm 5$ for \cratio$>20$).  Almost all of the stars appear to show substantial Li depletion (their Li abundances fall under the 3~\msun\ model), and at the coolest temperatures (``Late post-bump'' stage), some of the stars have Li below the extra-mixing model. 
More than half of the stars at each stellar evolution stage have \cratio\ that is lower than the standard model (i.e, lie to the left of the 3\,\msun\ model), but none of the stars have a \cratio\ (within the 
measurement errors) that is lower than that predicted by the extra-mixing model. If real, the unexpectedly low \cratio\ at early RGB stages could indicate a variation in the initial \cratio\ of the stars.  Similarly, because  the variation in \lli\ at each evolutionary stage exceeds the measurement 
errors, we are likely  seeing intrinsic scatter in \lli\ that makes a simple comparison to only two models difficult. \lli\ that is {\it lower} than model predictions can be 
explained by the  variations of slow Li-depletion processes on the MS.  Both models use MS \lli\ abundances that are at least an order of magnitude larger than the present-day solar values. 
On the other hand, the two Li-rich rapid rotators in the ``Significant Dilution'' group
exceed the  predicted values even if no MS Li-depletion has occurred. 

\begin{figure*}[th]
\centering
\includegraphics[width=0.7\textwidth]{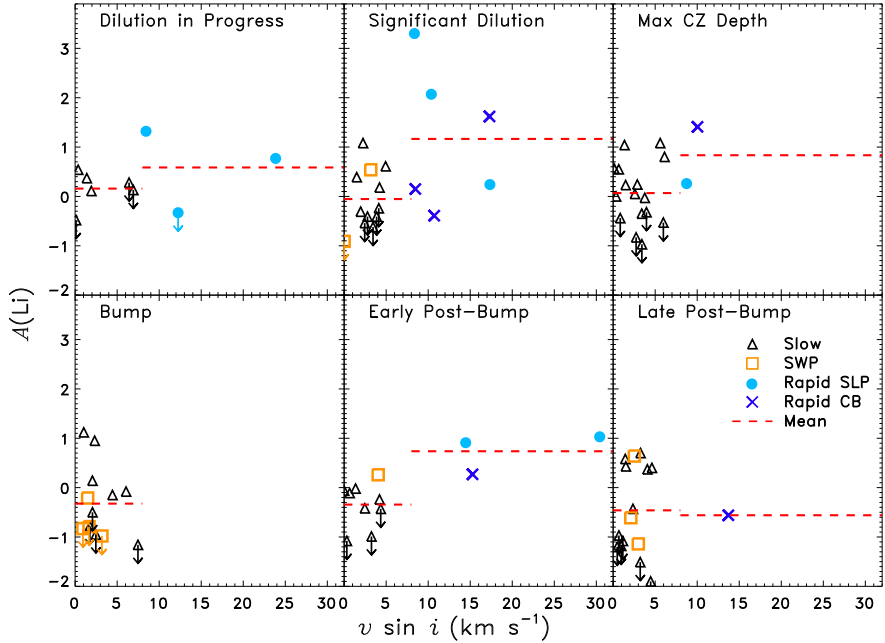} 
\caption{Same data as Figure \ref{liplot} but plotted for the stellar evolution groups progressing from less evolved to more evolved (left to right, then top to bottom).  The symbols have the same meaning as in Figure \ref{ali_cratio_new}.  The dashed lines in each panel show the mean \lli\ for the slow rotators (left line) and rapid rotators (right line).  }
\label{ali_by_group} 
\end{figure*}

To compare the effect of rotation  on \lli\ more explicitly, we plot \lli\ against \vsini\ for each of the stellar evolution bins in Figure \ref{ali_by_group}. The average \lli\ of the slow and rapid rotators in each bin are indicated with lines. 
 This figure is meant to address the question of whether the larger mean \lli\ of the rapid rotators (discovered in Figure \ref{liplot}) was merely result of the rapid rotators tending to be less evolved on average compared to the slow rotators.  
It is noteworthy, therefore, that in all of the evolution bins that contain more than one rapid rotator, the mean \lli\ of the rapid rotators exceeds that of the slow rotators.
Furthermore,  the most Li-rich star in the bin is a rapid rotator. 
Curiously, in both evolution bins that contain fewer than two rapid rotators, there are a distinct group of slow rotators with \lli\ larger than the rest.  

\section{Discussion}
\label{sec:discuss}
 In the previous section, we found two important  results. First, the rapid rotators tend to be more Li-rich than the slow rotators regardless of evolutionary stage. 
 Second,  there are at least two stars with \lli\ exceeding even the most conservative estimates of Li dilution (the standard dilution models) as well as a handful of stars that appear to be Li-rich when comparing their abundances to the other stars of similar evolutionary stage, but whose abundances compared to the models in Figure \ref{ali_cratio_new}  are not atypical.  
Here, we consider various scenarios for understanding these two results. 

\subsection{Post-bump Lithium Regeneration}
The prevailing explanation for Li-rich stars is  internal Li regeneration via the Cameron-Fowler chain, which is thought to begin  at the luminosity bump when the convective envelope can be connected to the 
H-burning shell via non-convective (generally diffusive) mixing processes.   At the luminosity bump, the hydrogen burning shell erases the chemical discontinuity, or ``$\mu$-barrier,'' that inhibits mixing.
Considerable work has been put into trying to understand the physical causes of this ``cool-bottom processing'' (CBP) mechanism---whether the mixing  results from, e.g.,  a thermohaline instability or buoyant magnetic flux tubes.
Regardless of the underlying physics, one observational  consequence of this explanation is that the star must have evolved
to the luminosity bump, and indeed, \cite{charbonnel00} found  most Li-rich giants to be associated with this evolutionary stage. 
Another observational  signature of this processing is the further reduction of \cratio\ that occurs after the Li regeneration ceases.  Stars with the largest \lli\ may appear to have normal \cratio, while stars with more modest Li enrichments will show abnormally low \cratio.   On average, therefore, \cratio\ should be lower in stars that have undergone Li regeneration. 
Considering these two observational signatures of Li regeneration, we can conclude that our Li-rich rapid rotators do not fit this picture for the following reasons. First, on average the rapid rotators do not exhibit
lower \cratio\ than the slow rotators. Second, if our stellar evolution groups are defined correctly, then the two most Li-rich stars are at pre-bump stages and should not have  yet experienced  Li regeneration.

\subsection{Rotation-induced Mixing}
An alternative Li-regeneration solution is one in which the enhanced rotation of the rapid rotators drives the non-convective mixing that connects the envelope to the Li-burning interior.
The assumption is that the rapid rotation creates favorable mixing currents that can deposit the newly regenerated Li (or the parent
isotope $^{7}$Be) into the stellar envelope.  In this scenario, the star's evolutionary proximity to the luminosity bump is irrelevant because the enhanced rotation is itself responsible for the non-canonical mixing. 
Rotation-induced mixing has been explored by both \cite{sack99} and \cite{denissenkov04} to explain Li-rich giants, and both models are also capable of lowering \cratio\ from the standard model predictions.
However, both groups were also exploring rotation-induced mixing as a possible CBP mechanism. In other words, their models required the erasure of the chemical discontinuity at the luminosity bump, whereas
a successful model for explaining our pre-bump Li-rich rapid rotators is one which does not require the erasure of the mean molecular weight discontinuity to function. 
\cite{chaname05} were the first to relax the assumption that the $\mu$-barrier inhibits all extra mixing  to see whether rotational mixing would occur at other evolutionary phases. 
They explored the evolution of  different rotational profiles and found that rotational mixing does indeed become active beyond the luminosity bump but only if the CZ was  differentially rotating.
In other words, in the case of solid body rotation in the stellar envelope, a star {\it would not} experience extra mixing even after the $\mu$-barrier was erased. 
Additionally, they found that the FDU mixing of \cratio\ occurred earlier and more gradually in their rotating models compared to standard models (though both models resulted in the same post-FDU value). 
On the other hand, to reproduce the abundance observations, the initial rotation rates  of their models must exceed the values actually observed.
The \cite{chaname05} paper did not explore the explicit effect of rotation on the Li abundance, in particular whether their models allow the special mixing cases capable of creating at least short-lived periods of enhance Li (as opposed to 
further destruction of Li that generally occurs with extra mixing).  Further work is needed to see whether such conditions are possible and whether those conditions require the removal of the ``$\mu$-barrier'' or not.

\subsection{Helium Flash}
Recently,  \citet[][hereafter, K11]{kumar11} conducted a large survey for Li-rich stars in  a sample of 2,000 red giants  and found 15 previously unidentified Li-rich stars. Like our study, they find  some  Li-rich giants that are too warm to be associated with the luminosity bump.  K11 suggest that these stars are  likely   
associated with the red clump---the core helium-burning stage of relatively metal-rich stars.  They suggest two scenarios for explaining the presence of large \lli\ in post-RGB stars. The 
first explanation  is simply the survival of Li that was regenerated at the luminosity bump; however, this scenario is  contradicted by the relatively few Li-rich stars between 
the bump and RGB tip.  K11 also hypothesize the possibility that Li could be regenerated  during the He flash, although this suggestion is based simply on the observed coincidence  of 
their  Li-rich giants near the red clump more so than on any physical mechanism.
The idea is that some fraction of the $^3$He remaining in the  RGB tip star could be converted to Li through the Cameron-Fowler process, whereby $^3$He($^4$H,$\gamma)^7$Be$(e^-,\nu)^7$Li. 
They suggest that this model would only operate in stars within the narrow mass range of $1.5M_\sun\leq M_\star \leq 2.25 M_\sun$,  where the star has a low enough mass to experience a He-flash but has a high enough mass for some $^3$He to survive. 
As a point of interest, we believe that K11 put a conservative lower limit on the mass range.  \cite{eggleton08} modeled the destruction of $^3$He 
and found that as much as $\sim 95$\% of the $^3$He created during the MS evolution will be destroyed by the time the star reaches the tip of the RGB. For the 1.5 \msun\ model, the destruction fraction is only 75--83\%. Nevertheless, the amount of $^3$He created in low mass stars is substantial. \cite{iben67} estimates $\sim10^{-3}$~\msun, which corresponds to $A(^3{\rm He})\sim 8.6$. Assuming 95\% of the original $^3$He was destroyed, only $10^{-4}$ of the remaining $^3$He must end up as $^7$Li to yield \lli$=3.3$~dex (the abundance of our most Li-rich star). In the 1.5 \msun\ case, the fraction of $^3$He remaining is $\sim20$~\%, but the 
original $^3$He budget is smaller, so that the reservoir of $^3$He at the RGB tip only about 2.5 times larger in the 1.5~\msun\ case than in the 1~\msun\ case. 
 K11 do not report \vsini, so we cannot comment on any relationship of rapid rotation to \lli\ in their results.  However, K11 do provide \cratio, and in contrast to  our stars, their warm Li-rich giants tend to have \cratio$\leq16$, indicating that some sort
 of extra mixing process has occurred in their Li-rich stars that does not appear to have occurred in the Li-rich stars studied here.

\subsection{Planet Accretion}
One of the main drawbacks to the Li-regeneration models is that they fail to explain the excess angular momentum of the rapid rotators.
The accretion of a planet is one means by which a red giant star can acquire sufficient angular momentum to become a rapid rotator.  We can test whether our abundance results are consistent with the planet accretion paradigm by estimating the masses and chemical compositions  that  accreted planets would 
have had to account for  the mean abundance differences between the slow and rapid rotators. The expected  stellar abundances of Li after planet accretion---$A$(Li)$_{\rm new}$---are  given by
\begin{equation}
\label{eq:liplanet}
 A({\rm Li})_{\rm new}= \log(q_{\rm e}10^{A({\rm Li})_{\rm p}}+10^{A({\rm Li})\star}) - \log(1+q_{\rm e}) , 
 \end{equation} 
where $q_{\rm e}$ is the ratio of the planet mass ($M_{\rm p}$) to the mass in the stellar convective envelope ($M_{\rm env}$) and $A({\rm Li})_{\rm p}$ and $A({\rm Li})_\star$ 
are the initial abundances of Li in the planet and star, respectively.  (See the Appendix for the derivation of this equation.)
For the present argument, we assume that the slow rotators are representative of the initial stellar abundances 
(i.e., $A({\rm Li})_\star =\overline{A({\rm Li})}_{\rm slow}= -0.18$\,dex) while the RV stable rapid 
rotators represent the post-planet-accretion  abundances (i.e., $A({\rm Li})_{\rm new} = 1.06$\,dex).  If we assume that the meteoritic 
Li abundance  of our solar system  \citep[\lli~$\sim$~3.3\,dex,][]{lodders98} is a good representation of the Li abundances of hypothetically accreted planets (i.e, $A({\rm Li})_{\rm p}=3.3$),  then we find from Equation \eqref{eq:liplanet} that  $q_{\rm e}=5.4\times10^{-3}$. This  $q_{\rm e}$ corresponds to  $\sim 6 M_{\rm Jup}$  assuming $M_{\rm env}=1$\,\msun, which is a reasonably-sized planet accreted into a reasonably-sized red giant convection zone. 

Similarly, we can test our non-detection of a \cratio\ increase using our estimated  $q_{\rm e}$.
The carbon ratio expected after the accretion of a planet, $(^{12}{\rm C}/^{13}{\rm C})_{\rm new}$, is given  by (see the Appendix)
\begin{equation}
\label{eq:cratio}
(^{12}{\rm C}/^{13}{\rm C})_{\rm new} = \frac{10^{A({\rm C})_{\rm p}}\frac{r_{\rm p} q_{\rm e}}{1+r_{\rm p}}  + 10^{A({\rm C})_\star }\frac{r_\star }{1+r_\star}}{10^{A({\rm C})_{\rm p}}\frac{ q_{\rm e}}{1+r_{\rm p}} + 10^{A({\rm C})_\star }\frac{1}{1+r_\star}},
\end{equation}
where  $r_{\rm p}=(^{12}{\rm C}/^{13}{\rm C})_{\rm p}$ and  $r_\star=(^{12}{\rm C}/^{13}{\rm C})_\star$.
 For  a solar metallicity red giant star, we take $A({\rm C})_\star=8.26$, which is the solar-metallicity \citep[$A$(C)$_\sun=8.39$\,dex; ][]{grevesse07} adjusted for some post-MS  processing of C and N such that [C/Fe]~$=-0.13$ \citep{marcs08}. As with the lithium example, we use the average slow rotator \cratio\ to represent pre-planet accretion so that $r_\star = 17.0$. 
 Using Jupiter as an analog, we expect the assimilated planet to have had  $3\times$  the  solar carbon abundance \citep{wong04}. Thus, 
$A({\rm C})_{\rm p}=A({\rm C})_\sun+\log(3)=8.87$\,dex.  We adopt $r_{\rm p} = 89$, which is a standard value in the solar system \citep{lodders98}.  Under these assumptions and adopting  the $q_{e}$ derived from the Li enhancement,  we  calculate that  a sample of planet accreting stars (such as the rapid rotators) is expected to have $\overline{^{12}{\rm C}/^{13}{\rm C}}=17.3$. 
Saturn is even more carbon-enriched \citep{mousis09} with up to 10$\times$ the solar value.  Using Saturn's composition as the analog for accreted planets, the expected carbon abundance increases to 
$\overline{^{12}{\rm C}/^{13}{\rm C}}=18.0$.   
The average \cratio\ of the RV stable rapid rotators (17.3)  is similar to the value expected from accreting a Jupiter analog, supporting the plausibility of planet accretion by these stars.
However, given the relative sizes of the signal (the \cratio\ increase) compared to our uncertainties, we caution that we cannot conclusively assert that we are measuring a true \cratio\ difference.

We can further test the plausibility of the planet accretion hypothesis by assessing whether both the enhanced angular momentum and  Li enrichment  are consistent. 
 \cite{carlberg09} introduced an equation (their Equation (1)) to relate an observed stellar \vsini\ to the properties of the star and the planet it accreted, i.e., 
$v\sin i = 8[ M_{\rm p}\sin i \sqrt{GM_\star a_{\rm p}(1-e^{2})} ]/M_{\rm env}R_\star$, where $M_{\rm p}$ is the planet mass, $a_{\rm p}$ is the planet's initial orbital 
separation, and $M_{\rm env}$ is the mass in the stellar convection envelope. The terms in  brackets describe the initial angular momentum of a planet orbiting the star. 
To estimate the angular momentum gained by the star, both the approximate stellar envelope mass ($M_{\rm env}$) and radius must be known  ($R_\star$).   Because the latter can change by two orders of magnitude during RGB evolution,
it is helpful to perform the analysis on individual stars as opposed to using the average stellar properties we have used thus far.
To this end, we select a small group of Li-rich rapid rotators that represent the best candidates in our sample of stars that have accreted planets. In addition to the enhanced rotation and enriched Li, these are stars that have not yet evolved to the luminosity bump and are not suspected to be in close binary systems. Three stars meet all of these requirements and are listed in Table \ref{tab:accretors}.  
The first two stars in that table are  the two most Li-rich stars in the ``Significant Dilution'' group, as seen in Figure \ref{ali_by_group}. The third star, Tyc3340-01195-1, is the most Li-rich star  in the ``Dilution in Progress'' group and has a long-period binary companion. 
\begin{deluxetable*}{lrrrrrr}
\tablewidth{\textwidth}
\tablecaption{Best Case Candidates for Planet Accretion \label{tab:accretors}}
\tablehead{
  \colhead{Star}&\colhead{$T_{\rm eff}'$}&\colhead{\vsini}&\colhead{\lli$_{\rm LTE}$}&\colhead{\lli$_{\rm NLTE}$}&\colhead{\cratio} &\colhead{[Fe/H]}  \\
\colhead{ }&\colhead{(K)}&\colhead{(\kms)}&\colhead{(dex)}&\colhead{(dex)}&\colhead{}&\colhead{ (dex)} }
\startdata
G0928+73.2600   & 4770   & 8.4  &    +3.62  &  +3.30 &     $28\pm 8$ &$-0.26$ \\
Tyc0647-00254-1 & 4825   & 10.4  &   +1.92  &   +2.06 &    $20 \pm 3$  &$-0.01$  \\
 Tyc3340-01195-1 & 5040   & 8.4  &    +1.21  &   +1.32 &    $25 \pm 5$ &$-0.18$  \\
\enddata
\end{deluxetable*}

\begin{figure}[tb]  
\centering
\includegraphics[width=1.03\columnwidth]{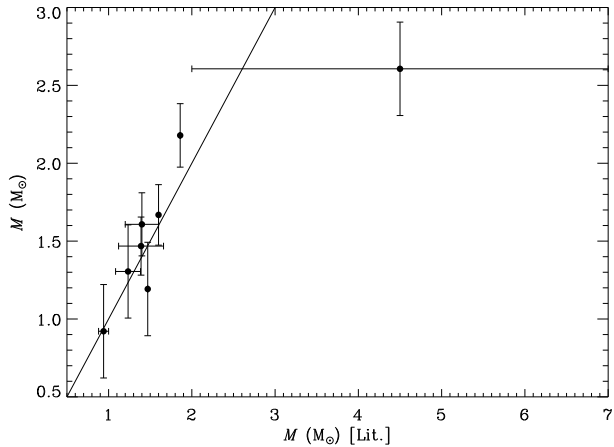}   
\caption{Comparison of the isochrone-derived masses of the giant SWPs ($y$-axis) with the mass measurements reported in the Exoplanet Encyclopedia ($x$-axis). A unity-slope line is plotted for reference. 
\label{comp:massage}  }
\end{figure}

Matching the stars' effective temperatures, surface gravities, and metallicities to  grids of either stellar evolution tracks or isochrones can be used to estimate the stellar masses and radii. We opt to use isochrones in this analysis because the isochrones of \cite{marigo08} can be interpolated onto a finer grid in $Z$ than what is available for the evolutionary tracks.   
We downloaded\footnote{http://stev.oapd.inaf.it/cmd accessed on 2011 May 17.} isochrones with $Z=0.0002$, 0.001, 0.004, 0.008, 0.012, 0.016, 0.020, 0.024,  and 0.03 (Fe/H~= $-2.05$,  $-1.33$, $-0.72$,  
$ -0.41$, $-0.23$, $-0.10$, $0.00$, $+0.08$, and $+0.18$), and $\log t$ ranging from 8.5 to 10.1 in increments of 0.1\,dex (where $t$ is the age of the stellar population in years). 
The stellar masses and radii of the stars are estimated in the following manner.   First, we identify the isochrone metallicity that most closely matches the observed stellar metallicity.  For each age at constant $Z$, we find where the isochrone intersects the stellar \teff.  
If there are multiple intersections, which is common for red giant evolutionary stages, we use the spectroscopically derived $\log g$ to select the best intersection. 
The isochrones are computed on finite grids of \teff\  points, so we identify the isochrone points that straddle the stellar \teff\  and interpolate the masses and $\log g$ associated with those two points to the \teff\ we are interested in. 
In other words,  for each star we reduce the isochrones to a grid of age, mass, and $\log g$ for the stellar temperature. 
We then find adjacent grid points of isochrone-derived $\log g$ that straddle the stellar $\log g$,  and we interpolate the ages and masses associated with those points to estimate the age and mass of the stellar $\log g$.   This last step generally finds between one and  three unique age/mass solutions, and the final mass and age estimates of our program stars  average over these unique solutions.   
Once we estimate the stellar mass, it is trivial to calculate an estimate in radius.  Using the measured surface gravity, the radius is given by $R_\star=\sqrt{GM_\star/g}$, where $G$ is the gravitational constant. 
Although we acknowledge the large uncertainty inherent in this analysis, especially with the overlap of the RGB and horizontal branch at the stars' \teff\ and $\log g$, we find that our isochrone fitting works rather well.  The sample of  giant SWPs has independent measurements  of mass  with which we can compare our results. In Figure \ref{comp:massage}, we plot the comparison of our mass measurements to literature mass measurements and find that our masses agree with literature values within the quoted uncertainties, with the exception of Pollux, which has $M_{\rm Lit.}=1.9$\,\msun\ and $M_{\rm isoch}=2.2\pm0.2$\,\msun. 

The isochrones do not provide information on the internal structure of the stars. Therefore, we use the \cite{giard00} stellar evolution models that most closely match the mass and metallicity of our 
stars to estimate the fraction of the mass in the convective envelope.  The least evolved star in Table \ref{tab:accretors} is Tyc3340-01195-1,  which has $\sim32$\% of the total mass  in the stellar envelope.  The other two stars have envelope
mass fractions of $\sim 80$\%.
\begin{figure}[tb]
\centering
\includegraphics[width=1.03\columnwidth]{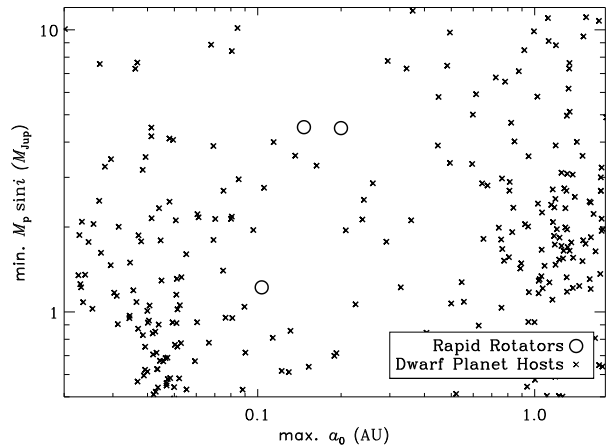}   
\caption{Minimum ``projected mass'' ($M_{\rm p} \sin i$) needed to account for the observed \vsini\ of the best case planet accretion candidates as a function of the expected maximum initial semimajor axis  ($a_0$) of the accreted planet. The true mass of any accreted planets will be larger both for smaller inclination angles (more pole-on angles) and for $a_0<{\rm max.}(a_0)$. For reference, the planetary masses and orbital separations of known extrasolar planets orbiting MS stars  are shown with crosses.
\label{fig:pacc}  }
\end{figure}

Obviously, the unknown inclination angle ($i$) limits us to estimating the minimum planet mass ($M_{\rm p}\sin i$).  
There is also a degeneracy between the planetary mass and orbital separation.  A massive planet at a small separation can create the same degree of stellar rotation
as a less massive planet at an initially larger separation. 
Fortunately, the tidal modeling of \cite{carlberg09} provides a means of breaking this degeneracy in the limiting case of the maximum initial orbital separations of planets that could have been accreted ($a_{\rm max}$)---planets that are too distant from their stars will not be accreted.  In the  exoplanetary systems  modeled in \citet[][99 systems with a total of 115 planets]{carlberg09}, we find that the 
ratio of the planets' initial semimajor axis ($a_{\rm 0}$) to the stellar radius at the time of planet accretion, $R_\star(t_{\rm acc})$,  is  typically 4.2. 
Therefore,  we can  estimate $a_{\rm max}\sim 4.2\, R_\star$.   Substituting this maximum orbital separation into the angular momentum equation yields the 
minimum planet mass capable of producing the observed rotational velocities in the rapid rotators.   This derived mass is both a minimum $M_{\rm p}$ because  of the unknown inclination {\it and} a minimum $M_{\rm p}\sin i$ because more massive planets with smaller initial $a_{\rm p}$ are also capable of producing the observed rotation. 
The results of this calculation for our selected Li-rich rapid rotators   are presented in Figure \ref{fig:pacc}, which shows the minimum $M_{\rm p}\sin i$ required to spin up the  star as a function of $a_{\rm max}\sim 4.2 R_\star$.    The least evolved star could have accreted planets that originally orbited within 0.1~AU and requires a Jupiter-mass planet to account for the angular momentum.   The other two red giants could have accreted planets within 0.2~AU, and the planets must have been at least 4~$M_{\rm Jup}$ to explain their enhanced rotation. This minimum planet mass is  slightly less than the average of 6~$M_{\rm Jup}$ planet needed to account for the Li-enrichment in Section \ref{sec:meanlicarbon}.   For comparison, we also plot in Figure \ref{fig:pacc} the planetary masses and orbital separations of known exoplanets.\footnote{Data come from the Exoplanet Orbit Database \cite{wright11}, accessed on  2011 December 18.} Our angular momentum analysis implies that our rapid rotators can be explained by accreting planets to the left and above the  rapid rotators' positions on the plot, and
we find such planet do exist among the known
exoplanetary systems.    However, it is also worth noting that there is a dearth of Jupiter mass planets (and larger) between $\sim 0.08$ and $\sim 0.7$~AU, the region where stars like those listed in Table \ref{tab:accretors}  would be actively  clearing out their planets by tidal decay.

\subsection{A Few Noteworthy Stars}
{\it G0928+73.2600.}  
This star was analyzed separately  by \cite{carlberg10b}, with attention brought to its apparent pre-bump evolutionary stage but high enough \lli\ to require accreting a Li-enriched object. 
We highlight this star again to contrast it to another star in the study that has almost identical stellar parameters, but very different \lli\ and \cratio.
Using the stellar parameters from the 2007 observations, G0928+43.2600 has \teff$=4900$~K, \logg$=2.70$, and [Fe/H]$=-0.26$, and the slow rotator G1200+67.3882 has 
\teff$=4900$~K, \logg$=2.70$, and [Fe/H]$=-0.24$.   Aside from rotation speeds (8.4~\kms\ and 1.5~\kms), these stars differ in their abundances. The rapid rotator has \lli$=3.30$ and \cratio$=28$, while the
slow rotator has \lli$=+0.39$ and \cratio$=8.6$.  Such a low \cratio\ is suggestive of more than the usual mixing, which implies that this star could have regenerated Li internally and is now in the process of destroying that
regenerated Li and reducing  \cratio\ below the normal values. 
If this scenario is indeed true,  it implies that G0928+73.2600 is at the slightly earlier stage where newly regenerated Li has not yet been destroyed and so the \cratio\ is still at the standard value.

The problem with using Li regeneration to explain the abundance differences of these otherwise similar stars is that the stars are both too hot to have evolved through the bump phase  ($T_{\rm bump}\sim $~4600--4700~K for [Fe/H]=$-0.25$). 
This difficulty could be resolved if our choice of stellar evolution tracks predicts too cool temperatures for the luminosity bump or if our derived spectroscopic temperatures are systematically too hot.
To test the first scenario, we compare stellar evolution tracks from two independent sources.
\cite{charbonnel00} plot solar metallicity stellar evolution tracks in their Figure 1. A visual inspection of that plot shows the luminosity bump at temperatures between 4450 and 4500~K, which is $\sim 100$~K cooler
than the luminosity bump of the solar-metallicity models we plot in Figure \ref{model_dilution}. As a second test, we computed the evolution of a 2~\msun, $Z=0.011$ star using the MESA code \citep{paxton11}\footnote{MESA version 3661, using the ``inlist'' file provided in the  ``1M\_pre\_ms\_to\_wd''  test suite as a template. We changed only the mass and metallicity of that file.}.  The luminosity bump of that model extends to a temperature of 4700~K, comparable to the models shown in Figure \ref{model_dilution}.
Next, we compile two different photometric temperatures to test for a systematic offset in our spectroscopic derivation.
The first set of temperatures comes from \cite{carlberg11} and were derived  using the stars' Washington $M-T_2$ colors \citep[which are converted to Cousin $V-I$, ][]{majewski00} and \cite{houd00} color-temperature relations.  This calculation yields $T_{\rm phot1}= 4773$~K
for the rapid rotator and $T_{\rm phot1}= 4875$~K for the slow rotator.
Both stars also have {\it Tycho} designations.   G0928+73.2600 is Tyc4382-00780-1, and G1200+67.3882 is Tyc4160-00999-1.  We converted their observed {\it Tycho} $B-V$ magnitudes \citep{hoeg00} to Johnson $B-V=0.85(B-V)_{\rm Tycho}$, dereddened
the colors using \cite{schlegel98} maps, and again used \cite{houd00} color-temperature relations to find  $T_{\rm phot2}= 4755$~K
for the rapid rotator and $T_{\rm phot2}= 5002$~K for the slow rotator.
Thus, the rapid rotator may be slightly cooler and more evolved than our spectroscopic analysis suggested, while the slow rotator may be slightly warmer (and less evolved).  However, the photometric temperatures are still warmer than the luminosity bump temperatures. 
Therefore, these two stars may represent  examples of the He-flash lithium regeneration hypothesized by K11. 

{\it Tyc3340-01195-1.}
This star is a long period binary star, and C11 argued that the  relatively large separation of the stellar components in the former system  ($a_\star\sin i\sim 425 R_\sun\sim 2$~AU)    made
tidal synchronization an unlikely  explanation for the primary star's enhanced rotational velocity  (\vsini=8.4\,\kms).  The presence of a stellar companion raises  questions about the stability of planets in the system. However,  planets
can have stable orbits around the primary star  interior to the  stellar companion  if the orbits are small enough.  
 In a study of such ``S-type'' planetary orbits, \cite{rabl88} found a quadratic relationship
  relating the largest Lowest Critical Orbit (LCO---orbits larger than this may be unstable, while smaller orbits are stable) to the binary system's separation ($a_\star$) and eccentricity ($e$) to be 
${\rm LCO}=0.262a_\star-0.254a_\star e-0.060a_\star e^2$.  For the Tyc3340-01195-1 system, a planet orbiting with $a_{\rm p}<0.4$~AU would be stable around the primary. In Figure \ref{fig:pacc} we find that
the maximum initial separation of planets that could have been accreted by Tyc3340-01195-1 is only $\sim0.1$~AU; therefore, the orbits of any accreted planets would have been dynamically
stable in the binary star system during the primary star's MS lifetime.

\section{Summary}
\label{sec:summary}

We have studied the global abundance patterns of \lli\ and \cratio\ in a homogeneously selected sample of slow and rapid rotators to see whether  the rotation is correlated with the replenishment of elements destroyed during the stellar evolution, as expected from planet accretion.   
Our final sample contains 71 slow rotator stars (58 chosen purely for their slow rotation, 10 selected because they are known to host planets, and the three M08 stars) and 15 rapid rotators.
From an analysis of line-of-sight considerations, we expect that 4.5\% of our slow rotators have true rotational velocities qualifying as rapid rotation.

Overall, the rapid rotators show an enhancement of \lli\ over the slow rotators by 0.99\,dex. When selecting the subset of rapid rotators that are RV stable, the average enhancement of \lli\   increases to 1.24\,dex over the slow rotators.  These Li enhancements are  consistent with the accretion of a $\sim 6$ Jupiter masses of planetary material with a Li abundance similar to meteoritic abundances (thought to be relic of the solar nebular abundance).  Consistent with this explanation and our relatively large \cratio\ uncertainties, we find no statistically significant difference between the \cratio\  measured in the rapid and slow rotator samples. A more massive or more carbon-enriched object would have to be accreted to be measurable at the level of our \cratio\ uncertainties. However, we do not measure smaller \cratio\ in our Li-rich stars or rapid rotators, a signature that  would  need to be seen to prove  the replenishment of Li through nuclear processing.
 We also compared our stellar sample to evolutionary tracks to ascertain how the relative Li and carbon abundances varied within subsets of similar evolutionary stage.  This comparison was necessary because the rapid rotators  and slow rotators are not distributed evenly across the RGB.   We found that in all groups that contained more than one rapid rotator, the most Li-rich star
 in the group was a rapid rotator.  In other words, the result that  rapid rotators  are more Li-rich than the slow rotators persists even within groups of similarly-evolved stars. 

These main conclusions were drawn by comparing the global properties of our two main samples, where it is safe to assume 
that the peculiar properties of individual stars are  likely to average out.  
 However,  to determine whether both the Li enrichment and enhanced angular momentum were consistent with the planet accretion hypothesis, we needed estimates of both the stellar mass and radius. The latter can vary by two orders of magnitude during RGB evolution.  Instead of averaging these properties over our stellar sample, we 
selected three Li-rich rapid rotators  that are the best examples in our sample of stars that may have engulfed planets. 
  Stellar isochrones were used to estimate stellar masses and radii.  Combining these stellar properties with the stars' measured \vsini\ and an estimate of the maximum orbital separations of accreted planets, we find that minimum planet masses of $\sim4.5$~$M_{\rm Jup}$ can account for the rotation of the two most Li-rich, rapidly rotating stars.  
  This planetary mass estimate is comparable to the $\sim 6$~$M_{\rm Jup}$ needed to explain  the  global Li enrichment seen in the rapid rotators.
The rotation of the third (and least Li-rich) of the selected stars can be explained with a minimum planet mass of only $1$~$M_{\rm Jup}$.

 However, we also found in the stellar evolution analysis that our stellar sample did not reproduce the detailed \lli\ and \cratio\ abundance patterns we expected. First, there is significant scatter in the \lli\ abundances in every evolution group, and we found stars with low \cratio\ at stages earlier than the completion of FDU. This scatter may be due to the wide range of stellar metallicities and masses represented in our sample combined with both a temperature-dependent \lli\ sensitivity and a difficulty in measuring precise values for \cratio~$\gtrsim 20$.  
Second, we did not find any classical Li-rich stars (\lli\ $\gtrsim 1.5$~dex) at the luminosity bump. 
The most Li-rich slow rotator had \lli$=1.12$ (and may be associated with the luminosity bump, the fourth of our six evolutionary classes). Furthermore, all stars that are more Li rich than this  are in the second of our defined classes---likely at pre-bump 
evolutionary stages.   This finding is reminiscent of the \cite{kumar11} study that also found Li-rich giants at pre-bump evolutionary stages.   Based only on this concentration in \teff, they suggested that the stars in their sample are red clump 
stars and that Li regeneration may occur during the He flash.  However, most of the warm (\teff$\leq 4600$\,K), pre-bump Li-rich stars in the K11 study had \cratio$\leq 16$, whereas our 
Li-rich pre-bump stars generally have \cratio$\sim 20$.
On the other hand, we noted that there is a slow rotator in our sample with nearly identical \teff, \logg, and [Fe/H] as the most Li-rich star in our sample. That slow rotator has a low \cratio\ suggestive of the enhanced extra mixing that should only occur
at post-luminosity bump stages. Together, these two stars appear to be in adjacent stages of the Li-regeneration phenomena that is thought to occur at they luminosity bump except that they are both too hot to be luminosity bump stars.

In conclusion, the \lli\ and \cratio\ of our sample showed far greater complexity than we anticipated.  The tendency for the rapid rotators  to show Li-enrichment implies that either planet accretion
or some sort of rotational mixing has taken place in these stars.   If planet accretion is not responsible for the Li enriched stars, then the fact that these stars are hotter than the luminosity bump presents
a problem for the Li-regeneration models that generally require the removal of the mean molecular weight barrier. 
  Many of the slow rotators in our sample showed lower than expected \lli, suggesting
that a variety in either the initial Li abundances or amount of Li destruction on the MS may exist. 

\acknowledgments
We  thank the anonymous referee for a  thorough reading of this manuscript and helpful suggestions for improving the presentation of our results. 
This work has been supported by NASA/JPL through the Space Interferometry Mission Preparatory Science Grants 1201670 and 1222563, which supported the generation of the
original giant star catalog and some follow-up observations used here, as well as NASA/JPL 
grant NRA-99-04-OSS-058.  JKC acknowledges financial  support by NASA Headquarters under the NASA Earth and Space Science Fellowship Program for proposal 
NNX08AX03H and the VSGC Graduate Research Fellowship. Travel support was provided by the NOAO Thesis Student  Travel Fund to obtain the observations taken at Kitt Peak National Observatory.
This work  made use of data products from the following online resources: the Vienna Atomic Line Database, the Exoplanet Orbit Database
and the Exoplanet Data Explorer at exoplanets.org,
the NASA ADS database, and the SIMBAD database.

\appendix 
\label{appendix:eqs}
 \renewcommand{\theequation}{A\arabic{equation}}
 
The derivation of Equations \eqref{eq:liplanet} and  \eqref{eq:cratio} is conceptually straightforward. Both \lli\ and \cratio\ are ratios of the number of atoms. In the former,
it is the ratio of Li atoms to H atoms. In the latter, it is the ratio of the isotope $^{12}$C to the isotope $^{13}$C.
\subsection*{Lithium Enhancement}
After a planet is accreted,  the new \lli---defined in Equation \eqref{eq:li1}---can be calculated by summing the contributions of Li and H from the planet and from the stellar envelope, as shown in Equation \eqref{eq:li1b}.
\begin{equation}
\label{eq:li1}
 A({\rm Li})_{\rm new}\equiv  \log(N({\rm Li})_{\rm new} / N({\rm H})_{\rm new})+12.00  \end{equation}
\begin{equation}
\label{eq:li1b}
 A({\rm Li})_{\rm new}   =     \log((N({\rm Li})_\star + N({\rm Li})_{\rm p}) / (N({\rm H})_\star + N({\rm H})_{\rm p})) +12.00,
\end{equation}
where $N({\rm Li})$ and $N({\rm H})$ are the number of Li  and H atoms, respectively, and the subscripts ``new,'' ``$\star$,'' and ``p'' refer to the star after planet accretion, the star before planet accretion, and the original planet, respectively. 
Rewriting the definition of \lli\ gives an equation for the number of Li atoms as
\begin{equation}
N({\rm Li}) = N({\rm H}) 10^{A({\rm Li})}10^{-12}.
\end{equation}
Therefore, Equation \eqref{eq:li1b} becomes
\begin{equation}
\label{eq:li2}
 A({\rm Li})_{\rm new}  = \log( ( N({\rm H})_\star 10^{A({\rm Li})_\star} 10^{-12} +  N({\rm H})_{\rm p} 10^{A({\rm Li})_{\rm p}}10^{-12})  / (N({\rm H})_\star + N({\rm H})_{\rm p})) +12.00,
\end{equation}
Rewriting  Equation \eqref{eq:li2} as a sum of logarithms (after factoring out the $10^{-12}$) yields
\begin{equation}
\label{eq:li3}
 A({\rm Li})_{\rm new}  = \log(N({\rm H})_\star 10^{A({\rm Li})_\star} +  N({\rm H})_{\rm p} 10^{A({\rm Li})_{\rm p}}) +\log 10^{-12} -\log (N({\rm H})_\star + N({\rm H})_{\rm p}) +12.00,
\end{equation}
 which simplifies to
\begin{equation}
\label{eq:li3b}
 A({\rm Li})_{\rm new}  = \log(N({\rm H})_\star 10^{A({\rm Li})_\star} +  N({\rm H})_{\rm p} 10^{A({\rm Li})_{\rm p}})-\log (N({\rm H})_\star + N({\rm H})_{\rm p}).
\end{equation}
 
 The actual number of H atoms comes from the hydrogen mass fraction of the object, $X$, which is generally known (or assumed).  That is, 
$ N({\rm H}) = XM/m_{\rm H}$, where $M$ is the mass of the object, and $m_{\rm H}$ is the mass of a single H atom.  
Assuming that the stellar envelope and planet have the same mass fraction of H (i.e., $X_\star = X_{\rm p}$), then $N({\rm H})_{\rm p}$ can be written in terms of $N({\rm H})_\star$ as  
\begin{equation}
\label{eq:Np}
N({\rm H})_{\rm p} = M_{\rm p}/M_{\rm env} N({\rm H})_\star = q_{\rm e}  N({\rm H})_\star,
\end{equation}
where $q_{\rm e} \equiv M_{\rm p}/M_{\rm env}$.  Substituting Equation\ \eqref{eq:Np} into Equation\ \eqref{eq:li3b},  factoring out $N({\rm H})_\star$, and rewriting as  a sum of logarithms yields 
\begin{equation}
\label{eq:li4}
 A({\rm Li})_{\rm new}  =\log N({\rm H})_\star + \log(10^{A({\rm Li})_\star} +  q_{\rm e} 10^{A({\rm Li})_{\rm p}}) - \log N({\rm H})_\star - \log (1+q_{\rm{e}}),
\end{equation}
which simplifies to Equation\ \eqref{eq:liplanet} in the paper
\begin{equation}
 A({\rm Li})_{\rm new}= \log(q_{\rm e}10^{A({\rm Li})_{\rm p}}+10^{A({\rm Li})_\star}) - \log(1+q_{\rm e})
\end{equation}

\subsection*{Carbon Ratio Enhancement}
In the following equations, we simplify the notation by dropping the $N()$ notation when referring to the isotopes. In other words, instead of $N(^{12}$C$)$, we simply write $^{12}$C.
After planet accretion, the new \cratio\ comes from the sum of the contributions of each isotope from the stellar envelope and the planet. Thus,

 \begin{equation}
\label{eq:cr1}
 (^{12}{\rm C}/^{13}{\rm C})_{\rm new} = \frac{^{12}{\rm C}_{\rm p}+^{12}{\rm C}_\star}{^{13}{\rm C}_{\rm p}+^{13}{\rm C}_\star}
\end{equation}

For both the stellar envelope and the bulk planet composition, the following relationships are true.   We assume that all of the carbon is in the two most abundant isotopes. Therefore,
\begin{equation}
\label{eq:cr2}
^{12}{\rm C}+^{13}{\rm C} = N({\rm C}).
\end{equation}
If we use $r$ to represent \cratio\ then Eq.\ \eqref{eq:cr2} can be expressed  in order to solve for $^{12}$C as

\begin{equation}
\label{eq:cr3a}
^{12}{\rm C}=  rN({\rm C}) /(1+r)
\end{equation}
Similarly,  Eq.\ \eqref{eq:cr2} can be expressed as a solution for $^{13}$C  as 
\begin{equation}
\label{eq:cr3b}
^{13}{\rm C}= N({\rm C} ) /(1+r)
\end{equation}

The total number of carbon atoms,  $N$(C), in the object comes from the abundance \ac\ (the number ratio of C to H atoms) and the
total number of H atoms from $N({\rm H}) = XM/m_{\rm H}$:
\begin{equation}
\label{eq:totC}
N({\rm C})=10^{A({\rm C})-12}X M/m_{\rm H}
\end{equation}
Using Equations \eqref{eq:cr3a}--\eqref{eq:totC} with subscripts for the planet and star, Equation \eqref{eq:cr1} expands to become
\begin{equation}
\label{eq:cbig}
(^{12}{\rm C}/^{13}{\rm C})_{\rm new} = \frac{r_{\rm p} 10^{A({\rm C})_{\rm p}}10^{-12}\frac{X_{\rm p} M_{\rm p}}{m_{\rm H}(1+r_{\rm p})}  + r_\star 10^{A({\rm C})_\star }10^{-12}\frac{X_\star M_{\rm env}}{m_{\rm H}(1+r_\star)}}{10^{A({\rm C})_{\rm p}}10^{-12}\frac{X_{\rm p} M_{\rm p}}{m_{\rm H}(1+r_{\rm p}) } + 10^{A({\rm C})_\star }10^{-12}\frac{X_\star M_{\rm env}}{m_{\rm H}(1+r_\star)}}.
\end{equation}
Assuming $X_{\rm p} \approx X_\star$, substituting $q_{\rm e} M_{\rm env}$ for $M_{\rm p}$,  and simplifying yields  the following expression for the post-planet accretion \cratio:
\begin{equation}
\label{eq:cfinal}
(^{12}{\rm C}/^{13}{\rm C})_{\rm new} = \frac{10^{A({\rm C})_{\rm p}}\frac{r_{\rm p} q_{\rm e}}{1+r_{\rm p}}  + 10^{A({\rm C})_\star }\frac{r_\star }{1+r_\star}}{10^{A({\rm C})_{\rm p}}\frac{ q_{\rm e}}{1+r_{\rm p}} + 10^{A({\rm C})_\star }\frac{1}{1+r_\star}},
\end{equation}


\begin{thebibliography}
\expandafter\ifx\csname natexlab\endcsname\relax\def\natexlab#1{#1}\fi

\bibitem[{{Alexander}(1967)}]{alexander67}{Alexander}, J.~B. 1967, The Observatory, 87, 238

\bibitem[Anders \& Grevesse(1989)]{anders89} Anders, E., \& Grevesse, N.\ 1989, \gca, 53, 197 

\bibitem[{{Balachandran} {et~al.}(2000){Balachandran}, {Fekel}, {Henry}, \&  {Uitenbroek}}]{balach00}
{Balachandran}, S.~C., {Fekel}, F.~C., {Henry}, G.~W., \& {Uitenbroek}, H.  2000, \apj, 542, 978

\bibitem[{{Bard} {et~al.}(1991){Bard}, {Kock}, \& {Kock}}]{bard91}
{Bard}, A., {Kock}, A., \& {Kock}, M. 1991, \aap, 248, 315

\bibitem[{{Bard} \& {Kock}(1994)}]{bard94}
{Bard}, A., \& {Kock}, M. 1994, \aap, 282, 1014

\bibitem[{{Barklem} \& {Aspelund-Johansson}(2005)}]{barklem05}
{Barklem}, P.~S., \& {Aspelund-Johansson}, J. 2005, \aap, 435, 373

\bibitem[{{Barklem} {et~al.}(2000){Barklem}, {Piskunov}, \&
  {O'Mara}}]{barklem00}
{Barklem}, P.~S., {Piskunov}, N., \& {O'Mara}, B.~J. 2000, \aaps, 142, 467

\bibitem[{{Bizyaev} {et~al.}(2006){Bizyaev}, {Smith}, {Arenas}, {Geisler}, {Majewski}, {Patterson}, {Cunha}, {Del Pardo}, {Suntzeff}, \&
  {Gieren}}]{dmbiz06}{Bizyaev}, D., {et~al.} 2006, \aj, 131, 1784

\bibitem[{{Blackwell} {et~al.}(1986){Blackwell}, {Booth}, {Haddock}, {Petford},  \& {Leggett}}]{blackwell86}
{Blackwell}, D.~E., {Booth}, A.~J., {Haddock}, D.~J., {Petford}, A.~D., \&  {Leggett}, S.~K. 1986, \mnras, 220, 549

\bibitem[{{Blackwell} {et~al.}(1984){Blackwell}, {Booth}, \&  {Petford}}]{blackwell84}
{Blackwell}, D.~E., {Booth}, A.~J., \& {Petford}, A.~D. 1984, \aap, 132, 236

\bibitem[{{Blackwell} {et~al.}(1995){Blackwell}, {Lynas-Gray}, \&  {Smith}}]{blackwell95}
{Blackwell}, D.~E., {Lynas-Gray}, A.~E., \& {Smith}, G. 1995, \aap, 296, 217

\bibitem[{{Blackwell} {et~al.}(1982{\natexlab{a}}){Blackwell}, {Petford},  {Shallis}, \& {Simmons}}]{blackwell82a}
{Blackwell}, D.~E., {Petford}, A.~D., {Shallis}, M.~J., \& {Simmons}, G.~J.  1982{\natexlab{a}}, \mnras, 199, 43

\bibitem[{{Blackwell} {et~al.}(1982{\natexlab{b}}){Blackwell}, {Petford}, \&  {Simmons}}]{blackwell82b}
{Blackwell}, D.~E., {Petford}, A.~D., \& {Simmons}, G.~J. 1982{\natexlab{b}},  \mnras, 201, 595

\bibitem[{{Blaise} {et~al.}(1984){Blaise}, {Wyart}, {Tahar Djerad}, \& {Ben
  Ahmed}}]{blaise84}
{Blaise}, J., {Wyart}, J.-F., {Tahar Djerad}, M., \& {Ben Ahmed}, Z. 1984,
  \physscr, 29, 119

\bibitem[{{B{\"o}hm-Vitense}(2004)}]{bohm04}
{B{\"o}hm-Vitense}, E. 2004, \aj, 128, 2435

\bibitem[{{Boothroyd} \& {Sackmann}(1999)}]{boothroyd99}
{Boothroyd}, A.~I., \& {Sackmann}, I. 1999, \apj, 510, 232

\bibitem[{Boesgaard} {et~al.}(1988)]{bosegaard88} Boesgaard, A.~M., Budge, K.~G., \& Ramsay, M.~E.\ 1988, \apj, 327, 389 

\bibitem[{{Brown} {et~al.}(1989){Brown}, {Sneden}, {Lambert}, \&  {Dutchover}}]{brown89}
{Brown}, J.~A., {Sneden}, C., {Lambert}, D.~L., \& {Dutchover}, E.~J. 1989,  \apjs, 71, 293
  
  \bibitem[Busso et al.(2007)]{busso07} Busso, M., Wasserburg, 
G.~J., Nollett, K.~M., \& Calandra, A.\ 2007, \apj, 671, 802 

\bibitem[{{Butler} {et~al.}(2006){Butler}, {Wright}, {Marcy}, {Fischer},  {Vogt}, {Tinney}, {Jones}, {Carter}, {Johnson}, {McCarthy}, \&
  {Penny}}]{butler06}{Butler}, R.~P., {et~al.} 2006, \apj, 646, 505

\bibitem[{{Cameron} \& {Fowler}(1971)}]{cameron71}
{Cameron}, A.~G.~W., \& {Fowler}, W.~A. 1971, \apj, 164, 111

\bibitem[{{Carlberg} {et~al.}(2009){Carlberg}, {Majewski}, \&  {Arras}}]{carlberg09}
{Carlberg}, J.~K., {Majewski}, S.~R., \& {Arras}, P. 2009, \apj, 700, 832

\bibitem[{{Carlberg} {et~al.}(2011){Carlberg}, {Majewski}, {Patterson},
  {Bizyaev}, {Smith}, \& {Cunha}}]{carlberg11}
{Carlberg}, J.~K., {Majewski}, S.~R., {Patterson}, R.~J., {Bizyaev}, D.,
  {Smith}, V.~V., \& {Cunha}, K. 2011, \apj, 732, 39

\bibitem[{{Carlberg} {et~al.}(2010{\natexlab{a}}){Carlberg}, {Majewski},
  {Smith}, {Cunha}, {Patterson}, {Bizyaev}, {Arras}, \& {Rood}}]{carlberg10a}
{Carlberg}, J.~K., {Majewski}, S.~R., {Smith}, V.~V., {Cunha}, K., {Patterson},
  R.~J., {Bizyaev}, D., {Arras}, P., \& {Rood}, R.~T. 2010{\natexlab{a}}, in
  IAU Symp.\ 265, Chemical Abundances in the Universe: Connecting First Stars to Planets, ed. {K.~Cunha, M.~Spite, \&
  B.~Barbuy}(Cambridge: Cambridge Univ. Press), 408

\bibitem[{{Carlberg} {et~al.}(2010{\natexlab{b}}){Carlberg}, {Smith}, {Cunha},
  {Majewski}, \& {Rood}}]{carlberg10b}
{Carlberg}, J.~K., {Smith}, V.~V., {Cunha}, K., {Majewski}, S.~R., \& {Rood},
  R.~T. 2010{\natexlab{b}}, \apjl, 723, L103

\bibitem[{{Carney} {et~al.}(2003){Carney}, {Latham}, {Stefanik}, {Laird}, \&  {Morse}}]{carney03}
{Carney}, B.~W., {Latham}, D.~W., {Stefanik}, R.~P., {Laird}, J.~B., \&  {Morse}, J.~A. 2003, \aj, 125, 293

\bibitem[{{Castelli} \& {Kurucz}(2004)}]{castelli04}
{Castelli}, F., \& {Kurucz}, R..~L.\  2004, arXiv:astro-ph/0405087 


\bibitem[{{Castilho} {et~al.}(2000){Castilho}, {Gregorio-Hetem}, {Spite},
  {Barbuy}, \& {Spite}}]{castilho00}
{Castilho}, B.~V., {Gregorio-Hetem}, J., {Spite}, F., {Barbuy}, B., \& {Spite},
  M. 2000, \aap, 364, 674

\bibitem[Chanam{\'e} et al.(2005)]{chaname05} Chanam{\'e}, J., 
Pinsonneault, M., \& Terndrup, D.~M.\ 2005, \apj, 631, 540 

\bibitem[{{Charbonnel} \& {Balachandran}(2000)}]{charbonnel00}
{Charbonnel}, C., \& {Balachandran}, S.~C. 2000, \aap, 359, 563

\bibitem[{{Charbonnel} {et~al.}(1998){Charbonnel}, {Brown}, \&
  {Wallerstein}}]{charbonnel98}
{Charbonnel}, C., {Brown}, J.~A., \& {Wallerstein}, G. 1998, \aap, 332, 204

\bibitem[{{Charbonnel} \& {Zahn}(2007)}]{charbonnel07}{Charbonnel}, C., \& {Zahn}, J.-P. 2007, \aap, 467, L15

\bibitem[{{da Silva} {et~al.}(1995){da Silva}, {de La Reza}, \&
  {Barbuy}}]{dasilva95} {da Silva}, L., {de La Reza}, R., \& {Barbuy}, B. 1995, \apjl, 448, L41

\bibitem[{{Davis} \& {Phillips}(1963)}]{davis63} {Davis}, S.~P., \& {Phillips}, J.~G. 1963, The Red System (A$^{2}$$\Pi$ --
X$^{2}$$\Sigma$) of the CN Molecule (Berkeley, CA: Univ.\  California Press)

\bibitem[{{Dearborn} {et~al.}(1975){Dearborn}, {Lambert}, \&  {Tomkin}}]{dearborn75}
{Dearborn}, D.~S.~P., {Lambert}, D.~L., \& {Tomkin}, J. 1975, \apj, 200, 675

\bibitem[{{de Medeiros} {et~al.}(1996){de Medeiros}, {Da Rocha}, \&{Mayor}}]{deMed96b}
{de Medeiros}, J.~R., {Da Rocha}, C., \& {Mayor}, M. 1996, \aap, 314, 499

\bibitem[{{de Medeiros} {et~al.}(2000){de Medeiros}, {do Nascimento},
  {Sankarankutty}, {Costa}, \& {Maia}}]{deMed00}
{de Medeiros}, J.~R., {do Nascimento}, J.~D., Jr., {Sankarankutty}, S.,
  {Costa}, J.~M., \& {Maia}, M.~R.~G. 2000, \aap, 363, 239

\bibitem[{{de Medeiros} \& {Mayor}(1999)}]{demed99}{de Medeiros}, J.~R., \& {Mayor}, M. 1999, \aaps, 139, 433

\bibitem[{{Denissenkov} \& {Herwig}(2004)}]{denissenkov04}{Denissenkov}, P.~A., \& {Herwig}, F. 2004, \apj, 612, 1081

\bibitem[{{Denissenkov} \& {VandenBerg}(2003)}]{denissenkov03}{Denissenkov}, P.~A., \& {VandenBerg}, D.~A. 2003, \apj, 593, 509

\bibitem[{{D{\"o}llinger} {et~al.}(2007){D{\"o}llinger}, {Hatzes}, {Pasquini},  {Guenther}, {Hartmann}, {Girardi}, \& {Esposito}}]{dollinger07}
{D{\"o}llinger}, M.~P., {Hatzes}, A.~P., {Pasquini}, L., {Guenther}, E.~W.,  {Hartmann}, M., {Girardi}, L., \& {Esposito}, M. 2007, \aap, 472, 649

\bibitem[{{Drake} {et~al.}(2002){Drake}, {de la Reza}, {da Silva}, \&  {Lambert}}]{drake02}
{Drake}, N.~A., {de la Reza}, R., {da Silva}, L., \& {Lambert}, D.~L. 2002,  \aj, 123, 2703

\bibitem[Duquennoy \& Mayor(1991)]{duquennoy91} Duquennoy, A., \& Mayor, M.\ 1991, \aap, 248, 485 

\bibitem[{{Eggleton} {et~al.}(2006){Eggleton}, {Dearborn}, \&  {Lattanzio}}]{eggleton06}
{Eggleton}, P.~P., {Dearborn}, D.~S.~P., \& {Lattanzio}, J.~C. 2006, Science,  314, 1580

\bibitem[{{Eggleton} {et~al.}(2008){Eggleton}, {Dearborn}, \&  {Lattanzio}}]{eggleton08} {Eggleton}, P.~P., {Dearborn}, D.~S.~P., \& {Lattanzio}, J.~C. 2008, \apj, 677, 581

\bibitem[{{Famaey} {et~al.}(2005){Famaey}, {Jorissen}, {Luri}, {Mayor}, {Udry},  {Dejonghe}, \& {Turon}}]{famaey05}{Famaey}, B., {Jorissen}, A., {Luri}, X., {Mayor}, M., {Udry}, S., {Dejonghe},  H., \& {Turon}, C. 2005, \aap, 430, 165

\bibitem[{{Fekel}(1997)}]{fekel97}{Fekel}, F.~C. 1997, \pasp, 109, 514

\bibitem[{{Fekel} \& {Balachandran}(1993)}]{fekel93}{Fekel}, F.~C., \& {Balachandran}, S. 1993, \apj, 403, 708

\bibitem[{{Frink} {et~al.}(2002){Frink}, {Mitchell}, {Quirrenbach}, {Fischer},  {Marcy}, \& {Butler}}]{frink02}
{Frink}, S., {Mitchell}, D.~S., {Quirrenbach}, A., {Fischer}, D.~A., {Marcy},  G.~W., \& {Butler}, R.~P. 2002, \apj, 576, 478

\bibitem[{{Fuhr} {et~al.}(1988){Fuhr}, {Martin}, \& {Wiese}}]{martin88}{Fuhr}, J.~R., {Martin}, G.~A., \& {Wiese}, W.~L. 1988, J.\ Phy.\ 
  Chem.\ Ref.\ Data,  17, Suppl.~4

\bibitem[{{Fulbright} {et~al.}(2006){Fulbright}, {McWilliam}, \&  {Rich}}]{fulbright06}
{Fulbright}, J.~P., {McWilliam}, A., \& {Rich}, R.~M. 2006, \apj, 636, 821

\bibitem[{{Ghezzi} {et~al.}(2009){Ghezzi}, {Cunha}, {Smith}, {Margheim},  {Schuler}, {de Ara{\'u}jo}, \& {de la Reza}}]{ghezzi09}
{Ghezzi}, L., {Cunha}, K., {Smith}, V.~V., {Margheim}, S., {Schuler}, S., {de  Ara{\'u}jo}, F.~X., \& {de la Reza}, R. 2009, \apj, 698, 451

\bibitem[{{Gilroy}(1989)}]{gilroy89}{Gilroy}, K.~K. 1989, \apj, 347, 835

\bibitem[{{Gilroy} \& {Brown}(1991)}]{gilroy91}{Gilroy}, K.~K., \& {Brown}, J.~A. 1991, \apj, 371, 578

\bibitem[{{Girardi} {et~al.}(2000){Girardi}, {Bressan}, {Bertelli}, \&  {Chiosi}}]{giard00}
{Girardi}, L., {Bressan}, A., {Bertelli}, G., \& {Chiosi}, C. 2000, \aaps, 141,  371

\bibitem[{{Glebocki} \& {Stawikowski}(2000)}]{glebocki00}{Glebocki}, R., \& {Stawikowski}, A. 2000, Acta Astronomica, 50, 509

\bibitem[Gonzalez et~al.(2009)]{gonzalez09Li} {Gonzalez}, O.~A., {Zoccali}, M., {Monaco}, L., {Hill}, V., {Cassisi}, S.,  {Minniti}, D.,  {Renzini}, A.,  {Barbuy}, B.,  
	{Ortolani}, S. \& {Gomez}, A.\ 2009, \aap, 508, 289 

\bibitem[{{Gratton} {et~al.}(2000){Gratton}, {Sneden}, {Carretta}, \&  {Bragaglia}}]{gratton00}
{Gratton}, R.~G., {Sneden}, C., {Carretta}, E., \& {Bragaglia}, A. 2000, \aap,  354, 169

\bibitem[{{Gray}(1981)}]{gray81}{Gray}, D.~F. 1981, \apj, 251, 155

\bibitem[{{Gray}(1982)}]{gray82}{Gray}, D.~F. 1982, \apj, 262, 682

\bibitem[{{Grevesse} {et~al.}(2007){Grevesse}, {Asplund}, \&  {Sauval}}]{grevesse07}{Grevesse}, N., {Asplund}, M., \& {Sauval}, A.~J. 2007, \ssr, 130, 105

\bibitem[{{Griffin} \& {Lynas-Gray}(1999)}]{griffin99}{Griffin}, R.~E.~M., \& {Lynas-Gray}, A.~E. 1999, \aj, 117, 2998

\bibitem[{{Griffin}(2009)}]{griffin09}{Griffin}, R.~F. 2009, The Observatory, 129, 317

\bibitem[{{Guandalini} {et~al.}(2009){Guandalini}, {Palmerini}, {Busso}, \&  {Uttenthaler}}]{guandalini09}
{Guandalini}, R., {Palmerini}, S., {Busso}, M., \& {Uttenthaler}, S. 2009,  PASA, 26, 168

\bibitem[{{Gustafsson} {et~al.}(2008){Gustafsson}, {Edvardsson}, {Eriksson},  {J{\o}rgensen}, {Nordlund}, \& {Plez}}]{marcs08}
{Gustafsson}, B., {Edvardsson}, B., {Eriksson}, K., {J{\o}rgensen}, U.~G.,  {Nordlund}, {\AA}., \& {Plez}, B. 2008, \aap, 486, 951

\bibitem[{{Hatzes} \& {Cochran}(1993)}]{hatzes93}{Hatzes}, A.~P., \& {Cochran}, W.~D. 1993, \apj, 413, 339

\bibitem[{{Hatzes} {et~al.}(2006){Hatzes}, {Cochran}, {Endl}, {Guenther},  {Saar}, {Walker}, {Yang}, {Hartmann}, {Esposito}, {Paulson}, \&
  {D{\"o}llinger}}]{hatzes06}{Hatzes}, A.~P., {et~al.} 2006, \aap, 457, 335

\bibitem[{{Hatzes} {et~al.}(2005){Hatzes}, {Guenther}, {Endl}, {Cochran},  {D{\"o}llinger}, \& {Bedalov}}]{hatzes05}
{Hatzes}, A.~P., {Guenther}, E.~W., {Endl}, M., {Cochran}, W.~D.,  {D{\"o}llinger}, M.~P., \& {Bedalov}, A. 2005, \aap, 437, 743

\bibitem[{{Hekker} \& {Mel{\'e}ndez}(2007)}]{hekker07}{Hekker}, S., \& {Mel{\'e}ndez}, J. 2007, \aap, 475, 1003

\bibitem[{{Henry} {et~al.}(1995){Henry}, {Fekel}, \& {Hall}}]{henry95}{Henry}, G.~W., {Fekel}, F.~C., \& {Hall}, D.~S. 1995, \aj, 110, 2926

\bibitem[{{Hinkle} {et~al.}(2000){Hinkle}, {Wallace}, {Valenti}, \&  {Harmer}}]{hinkle00}{Hinkle}, K., {Wallace}, L., {Valenti}, J., \& {Harmer}, D. 2000, {Visible and
  Near Infrared Atlas of the Arcturus Spectrum 3727--9300~\AA} (San Francisco, CA:  ASP)
  
  \bibitem[H{\o}g et~al.(2000)]{hoeg00} H{\o}g, E., Fabricius, C., Makarov, V.~V., et al.\ 2000, \aap, 355, L27 

\bibitem[{{Houdashelt} {et~al.}(2000){Houdashelt}, {Bell}, \&  {Sweigart}}]{houd00}{Houdashelt}, M.~L., {Bell}, R.~A., \& {Sweigart}, A.~V. 2000, \aj, 119, 1448

\bibitem[{{Iben}(1967)}]{iben67}{Iben},  I., Jr. 1967, \apj, 147, 624

\bibitem[{{Kumar} \& {Reddy}(2009)}]{kumar09}{Kumar}, Y.~B., \& {Reddy}, B.~E. 2009, \apjl, 703, L46

\bibitem[{{Kumar} {et~al.}(2011){Kumar}, {Reddy}, \& {Lambert}}]{kumar11}{Kumar}, Y.~B., {Reddy}, B.~E., \& {Lambert}, D.~L. 2011, \apjl, 730, L12

\bibitem[{{Kupka} {et~al.}(1999){Kupka}, {Piskunov}, {Ryabchikova}, {Stempels},  \& {Weiss}}]{kupka99}
{Kupka}, F., {Piskunov}, N., {Ryabchikova}, T.~A., {Stempels}, H.~C., \&  {Weiss}, W.~W. 1999, \aaps, 138, 119

\bibitem[{{Kurucz}(1993)}]{kuruczCD18}
{Kurucz}, R. 1993, SYNTHE Spectrum Synthesis Programs and Line Data:~Kurucz
  CD-ROM No.~18 (Cambridge, MA.: Smithsonian Astrophysical Observatory)

\bibitem[{{Kurucz}(1994{\natexlab{a}})}]{kuruczCD20}
{Kurucz}, R. 1994{\natexlab{a}}, Atomic Data for Ca, Sc, Ti, V, and Cr:~ Kurucz CD-ROM
  No.~20 (Cambridge, MA.: Smithsonian Astrophysical Observatory)

\bibitem[{{Kurucz}(1994{\natexlab{b}})}]{kurucz94}
{Kurucz}, R. 1994{\natexlab{b}}, Atomic Data for Fe and Ni:~ Kurucz CD-ROM
  No.~22 (Cambridge, MA.: Smithsonian Astrophysical Observatory)

\bibitem[{{Kurucz}(1994{\natexlab{c}})}]{kurucz94b}
{Kurucz}, R. 1994{\natexlab{c}}, Atomic Data for Mn and Co:~ Kurucz CD-ROM
  No.~21 (Cambridge, MA.: Smithsonian Astrophysical Observatory)

\bibitem[{{Lambert} {et~al.}(1980){Lambert}, {Dominy}, \&
  {Sivertsen}}]{lambert80}
{Lambert}, D.~L., {Dominy}, J.~F., \& {Sivertsen}, S. 1980, \apj, 235, 114

\bibitem[{{Lambert} \& {Ries}(1981)}]{lambert81}{Lambert}, D.~L., \& {Ries}, L.~M. 1981, \apj, 248, 228

\bibitem[Lebzelter et~al.(2012)]{lebzelter12} Lebzelter, T., Uttenthaler, S., Busso, M., Schultheis, M., \& Aringer, B.\ 2012, \aap, 538, A36 


\bibitem[{{Lind} {et~al.}(2009){Lind}, {Asplund}, \& {Barklem}}]{lind09}
{Lind}, K., {Asplund}, M., \& {Barklem}, P.~S. 2009, \aap, 503, 541

\bibitem[Lodders \& Fegley(1998)]{lodders98} Lodders, K., \& Fegley, B.\ 1998, {\it The Planetary Scientist's Companion} (New York : Oxford Univ.\ Press)

\bibitem[{{Luck}(1994)}]{luck94}{Luck}, R.~E. 1994, \apjs, 91, 309

\bibitem[{{Majewski} {et~al.}(2000){Majewski}, {Ostheimer}, {Kunkel}, \& {Patterson}}]{majewski00}{Majewski}, S.~R., {Ostheimer}, J.~C., {Kunkel}, W.~E., \& {Patterson}, R.~J.  2000, \aj, 120, 2550

\bibitem[{{Mandell} {et~al.}(2004){Mandell}, {Ge}, \& {Murray}}]{mandell04}{Mandell}, A.~M., {Ge}, J., \& {Murray}, N. 2004, \aj, 127, 1147

\bibitem[{{Marigo} {et~al.}(2008){Marigo}, {Girardi}, {Bressan}, {Groenewegen},  {Silva}, \& {Granato}}]{marigo08}
{Marigo}, P., {Girardi}, L., {Bressan}, A., {Groenewegen}, M.~A.~T., {Silva},  L., \& {Granato}, G.~L. 2008, \aap, 482, 883

\bibitem[{{Massarotti} {et~al.}(2008){Massarotti}, {Latham}, {Stefanik}, \&  {Fogel}}]{massarotti08a}
{Massarotti}, A., {Latham}, D.~W., {Stefanik}, R.~P., \& {Fogel}, J. 2008, \aj,  135, 209

\bibitem[{{Mitchell} {et~al.}(2003){Mitchell}, {Frink}, {Quirrenbach},  {Fischer}, {Marcy}, \& {Butler}}]{mitchell03}
{Mitchell}, D.~S., {Frink}, S., {Quirrenbach}, A., {Fischer}, D.~A., {Marcy},  G.~W., \& {Butler}, R.~P. 2003,  BAAS, 35, 1234

\bibitem[Monaco et~al.(2011)]{monaco11} Monaco, L., Villanova, S., Moni Bidin, C., Carraro, G., Geisler, D., Bonifacio, P., Gonzalez, O.~A., Zoccali, M., Jilkova, L. 2011, \aap, 529, A90 

\bibitem[Mousis et al.(2009)]{mousis09} {Mousis}, O., {Marboeuf}, U., {Lunine}, J.~I., {Alibert}, Y., 
	{Fletcher}, L.~N., {Orton}, G.~S., {Pauzat}, F. \& {Ellinger}, Y. 2009, \apj, 696, 1348 

\bibitem[Nordhaus et al.(2008)]{nordhaus08} Nordhaus, J., Busso, M., Wasserburg, G.~J., Blackman, E.~G., \& Palmerini, S.\ 2008, \apjl, 684, L29 

\bibitem[{{O'Brian} {et~al.}(1991){O'Brian}, {Wickliffe}, {Lawler}, {Whaling},  \& {Brault}}]{obrian91}
{O'Brian}, T.~R., {Wickliffe}, M.~E., {Lawler}, J.~E., {Whaling}, W., \&  {Brault}, J.~W. 1991, J.\ Opt.\ Soc.\ Am.\ B: Opt.\ Phys., 8, 1185

\bibitem[{{Palacios} {et~al.}(2006){Palacios}, {Charbonnel}, {Talon}, \&  {Siess}}]{palacios06}
{Palacios}, A., {Charbonnel}, C., {Talon}, S., \& {Siess}, L. 2006, \aap, 453,  261

\bibitem[{{Palmerini} \& {Maiorca}(2010)}]{palmerini10}{Palmerini}, S., \& {Maiorca}, E. 2010, Journal of Physics Conference Series, 202, 012030

\bibitem[Paxton et al.(2011)]{paxton11} {Paxton}, B., {Bildsten}, L., {Dotter}, A., {Herwig}, F., {Lesaffre}, P. \& {Timmes}, F., \ 2011, \apjs, 192, 3 

\bibitem[{{Peterson} {et~al.}(1983){Peterson}, {Tarbell}, \& {Carney}}]{pete83}{Peterson}, R.~C., {Tarbell}, T.~D., \& {Carney}, B.~W. 1983, \apj, 265, 972

\bibitem[{{Pilachowski} {et~al.}(2003){Pilachowski}, {Sneden}, {Freeland}, \&  {Casperson}}]{pilachowski03}
{Pilachowski}, C., {Sneden}, C., {Freeland}, E., \& {Casperson}, J. 2003, \aj,  125, 794

\bibitem[{{Piskunov} {et~al.}(1995){Piskunov}, {Kupka}, {Ryabchikova}, {Weiss}, \& {Jeffery}}]{vald}
{Piskunov}, N.~E., {Kupka}, F., {Ryabchikova}, T.~A., {Weiss}, W.~W., \&  {Jeffery}, C.~S. 1995, \aaps, 112, 525

\bibitem[{{Pourbaix} {et~al.}(2004){Pourbaix}, {Tokovinin}, {Batten}, {Fekel}, {Hartkopf}, {Levato}, {Morrell}, {Torres}, \& {Udry}}]{pourbaix04}
{Pourbaix}, D., {Tokovinin}, A.~A., {Batten}, A.~H., {Fekel}, F.~C., {Hartkopf}, W.~I., {Levato}, H., {Morrell}, N.~I., {Torres}, G. \& {Udry}, S. 2004, \aap, 424, 727

\bibitem[{{Raassen} \& {Uylings}(1998)}]{raassen98}{Raassen}, A.~J.~J., \& {Uylings}, P.~H.~M. 1998, \aap, 340, 300

\bibitem[{{Rabl} \& {Dvorak}(1988)}]{rabl88}{Rabl}, G., \& {Dvorak}, R. 1988, \aap, 191, 385

\bibitem[{{Raghavan} {et~al.}(2006){Raghavan}, {Henry}, {Mason}, {Subasavage},  {Jao}, {Beaulieu}, \& {Hambly}}]{raghavan06}
{Raghavan}, D., {Henry}, T.~J., {Mason}, B.~D., {Subasavage}, J.~P., {Jao}, W.-C., {Beaulieu}, T.~D., \& {Hambly}, N.~C. 2006, \apj, 646, 523

\bibitem[{{Recio-Blanco} \& {de Laverny}(2007)}]{reicoblanco07}{Recio-Blanco}, A., \& {de Laverny}, P. 2007, \aap, 461, L13

\bibitem[{{Reddy} \& {Lambert}(2005)}]{reddy05}{Reddy}, B.~E., \& {Lambert}, D.~L. 2005, \aj, 129, 2831

\bibitem[{{Reddy} {et~al.}(2002){Reddy}, {Lambert}, {Hrivnak}, \&  {Bakker}}]{reddy02a}{Reddy}, B.~E., {Lambert}, D.~L., {Hrivnak}, B.~J., \& {Bakker}, E.~J. 2002, \aj, 123, 1993

\bibitem[Ruchti et~al.(2011)]{ruchti11} {Ruchti}, G.~R., {Fulbright}, J.~P., {Wyse}, R.~F.~G., et al.\ 2011, \apj, 743, 107 

\bibitem[{{Sackmann} \& {Boothroyd}(1999)}]{sack99}{Sackmann}, I.-J., \& {Boothroyd}, A.~I. 1999, \apj, 510, 217

\bibitem[{{Sato} {et~al.}(2003){Sato}, {Ando}, {Kambe}, {Takeda}, {Izumiura},  {Masuda}, {Watanabe}, {Noguchi}, {Wada}, {Okada}, {Koyano}, {Maehara},
  {Norimoto}, {Okada}, {Shimizu}, {Uraguchi}, {Yanagisawa}, \&  {Yoshida}}]{sato03}
{Sato}, B., {Ando}, H., {Kambe}, E.\ {et~al.} 2003, \apjl, 597, L157

\bibitem[{{Scalo} {et~al.}(1975){Scalo}, {Despain}, \& {Ulrich}}]{scalo75}{Scalo}, J.~M., {Despain}, K.~H., \& {Ulrich}, R.~K. 1975, \apj, 196, 805

\bibitem[Schlegel et al.(1998)]{schlegel98} Schlegel, D.~J., Finkbeiner, D.~P., \& Davis, M.\ 1998, \apj, 500, 525 

\bibitem[{{Setiawan} {et~al.}(2003){Setiawan}, {Hatzes}, {von der L{\"u}he}, {Pasquini}, {Naef}, {da Silva}, {Udry}, {Queloz}, \& {Girardi}}]{setiawan03}
{Setiawan}, J.,  {Hatzes}, A.~P., {von der L{\"u}he}, O.,  {Pasquini}, L., {Naef}, D., {da Silva}, L.,  {Udry}, S., {Queloz}, D. \& {Girardi}, L. 2003, \aap, 398, L19

\bibitem[{{Setiawan} {et~al.}(2004){Setiawan}, {Pasquini}, {da Silva},
  {Hatzes}, {von der L{\"u}he}, {Girardi}, {de Medeiros}, \&  {Guenther}}]{setiawan04}
{Setiawan}, J., {Pasquini}, L., {da Silva}, L., {Hatzes}, A.~P., {von der
  L{\"u}he}, O., {Girardi}, L., {de Medeiros}, J.~R., \& {Guenther}, E. 2004,  \aap, 421, 241

\bibitem[{{Shetrone}(2003)}]{shetrone03}{Shetrone}, M.~D. 2003, \apjl, 585, L45

\bibitem[{{Siess} \& {Livio}(1999)}]{siess99}{Siess}, L., \& {Livio}, M. 1999, \mnras, 308, 1133

\bibitem[{{Smiljanic} {et~al.}(2009){Smiljanic}, {Gauderon}, {North}, {Barbuy},  {Charbonnel}, \& {Mowlavi}}]{smiljanic09}
{Smiljanic}, R., {Gauderon}, R., {North}, P., {Barbuy}, B., {Charbonnel}, C.,  \& {Mowlavi}, N. 2009, \aap, 502, 267

\bibitem[{{Smith} {et~al.}(2001){Smith}, {Cunha}, \& {Lazzaro}}]{smith01}{Smith}, V.~V., {Cunha}, K., \& {Lazzaro}, D. 2001, \aj, 121, 3207

\bibitem[{{Smith} \& {Lambert}(1985)}]{smith85}{Smith}, V.~V., \& {Lambert}, D.~L. 1985, \apj, 294, 326

\bibitem[{{Smith} {et~al.}(2000){Smith}, {Suntzeff}, {Cunha}, {Gallino},  {Busso}, {Lambert}, \& {Straniero}}]{smith00}
{Smith}, V.~V., {Suntzeff}, N.~B., {Cunha}, K., {Gallino}, R., {Busso}, M.,  {Lambert}, D.~L., \& {Straniero}, O. 2000, \aj, 119, 1239

\bibitem[Sneden \& Lambert(1982)]{sneden82} Sneden, C., \& Lambert, D.~L.\ 1982, \apj, 259, 381 

\bibitem[{{Sneden} {et~al.}(1986){Sneden}, {Pilachowski}, \&  {Vandenberg}}]{sneden86}
{Sneden}, C., {Pilachowski}, C.~A., \& {Vandenberg}, D.~A. 1986, \apj, 311, 826

\bibitem[{{Sneden}(1973)}]{sneden73}{Sneden}, C.~A. 1973, PhD thesis, Univ.\ Texas at Austin

\bibitem[{{Sousa} {et~al.}(2007){Sousa}, {Santos}, {Israelian},  {Mayor}, \&  {Monteiro}}]{sousa07}
{Sousa}, S.~G., {Santos}, N.~C., {Israelian}, G., {Mayor}, M., \& {Monteiro},  M.~J.~P.~F.~G. 2007, \aap, 469, 783

\bibitem[{{Strassmeier} {et~al.}(2000){Strassmeier}, {Washuettl}, {Granzer},  {Scheck}, \& {Weber}}]{strassmeier00}
{Strassmeier}, K., {Washuettl}, A., {Granzer}, T., {Scheck}, M., \& {Weber}, M.  2000, \aaps, 142, 275

\bibitem[{{Strassmeier} {et~al.}(1993){Strassmeier}, {Hall}, {Fekel}, \&  {Scheck}}]{strassmeier93}
{Strassmeier}, K.~G., {Hall}, D.~S., {Fekel}, F.~C., \& {Scheck}, M. 1993,  \aaps, 100, 173

\bibitem[{{Sweigart} {et~al.}(1989){Sweigart}, {Greggio}, \&  {Renzini}}]{sweigart89}
{Sweigart}, A.~V., {Greggio}, L., \& {Renzini}, A. 1989, \apjs, 69, 911

\bibitem[{{Sweigart} \& {Mengel}(1979)}]{sweigart79}{Sweigart}, A.~V., \& {Mengel}, J.~G. 1979, \apj, 229, 624

\bibitem[{{Uesugi} \& {Fukuda}(1982)}]{uesugi82}{Uesugi}, A., \& {Fukuda}, I. 1982, Catalogue of Stellar Rotational Velocities
   (Revised ed.; Kyoto: Univ.\ Kyoto, Department of Astronomy)

\bibitem[{{Vogt} {et~al.}(2000){Vogt}, {Marcy}, {Butler}, \& {Apps}}]{vogt00}{Vogt}, S.~S., {Marcy}, G.~W., {Butler}, R.~P., \& {Apps}, K. 2000, \apj, 536,  902

\bibitem[{{Wallerstein} \& {Sneden}(1982)}]{wallerstein82}{Wallerstein}, G., \& {Sneden}, C. 1982, \apj, 255, 577

\bibitem[{{Wasserburg} {et~al.}(1995){Wasserburg}, {Boothroyd}, \&  {Sackmann}}]{wasserburg95}
{Wasserburg}, G.~J., {Boothroyd}, A.~I., \& {Sackmann}, I. 1995, \apjl, 447,  L37

\bibitem[{{Wong} {et~al.}(2004){Wong}, {Mahaffy}, {Atreya}, {Niemann}, \&  {Owen}}]{wong04}
{Wong}, M.~H., {Mahaffy}, P.~R., {Atreya}, S.~K., {Niemann}, H.~B., \& {Owen},  T.~C. 2004, Icarus, 171, 153

\bibitem[{{Wright} {et~al.}(2011){Wright}, {Fakhouri}, {Marcy}, {Han}, {Feng},
  {Johnson}, {Howard}, {Fischer}, {Valenti}, {Anderson}, \&  {Piskunov}}]{wright11}
  {Wright}, J.~T., {Fakhouri}, O., {Marcy}, G.~W., {Han}, E., {Feng}, Y., {Johnson}, J.~A., {Howard}, A.~W., {Fischer}, D.~A., 
{Valenti}, J.~A., {Anderson}, J. \& {Piskunov}, N. 2011, \pasp, 123, 412

\bibitem[{{Wyller}(1966)}]{wyller66} Wyller, A.~A. 1966, \apj, 143, 828 

\end{thebibliography}
\end{document}